\documentclass[aip,jcp,preprint,letterpaper,onecolumn]{revtex4-1}
\usepackage[utf8]{inputenc}
\usepackage{caption}
\usepackage{subcaption}
\usepackage{natbib}
\usepackage{bm}
\usepackage{amsmath}
\usepackage{multirow}
\usepackage{cancel}
\usepackage{graphicx}
\usepackage{epstopdf}
\usepackage{xcolor}
\newcommand{\bra}[1]{\left\langle #1 \right|}
\newcommand{\ket}[1]{\left|#1\right\rangle}
\newcommand{\braket}[2]{\left\langle#1 |  #2\right\rangle}

\begin{document}
	\title
	{Accurate prediction of core-level spectra of radicals at density functional theory cost via square gradient minimization and recoupling of mixed configurations.}
	\author{Diptarka Hait}
	\email{diptarka@berkeley.edu}
	\author{Eric A. Haugen}
	\author{Zheyue Yang}
	\author{Katherine J. Oosterbaan}
	\affiliation
	{{Department of Chemistry, University of California, Berkeley, California 94720, USA.}}
	\affiliation{Chemical Sciences Division, Lawrence Berkeley National Laboratory, Berkeley, California 94720, USA.}
	\author{Stephen R. Leone}
	\affiliation
	{{Department of Chemistry, University of California, Berkeley, California 94720, USA.}}
	\affiliation{Chemical Sciences Division, Lawrence Berkeley National Laboratory, Berkeley, California 94720, USA.}
	\affiliation{Department of Physics, University of California, Berkeley, California 94720, USA.}
	\author{Martin Head-Gordon}
	\email{mhg@cchem.berkeley.edu}
	\affiliation
	{{Department of Chemistry, University of California, Berkeley, California 94720, USA.}}
	\affiliation{Chemical Sciences Division, Lawrence Berkeley National Laboratory, Berkeley, California 94720, USA.}
	\begin{abstract}
		State-specific orbital optimized approaches are more accurate at predicting core-level spectra than traditional linear-response protocols, but their utility had been restricted on account of the risk of `variational collapse' down to the ground state. We employ the recently developed square gradient minimization (SGM, J. Chem. Theory Comput. 16, 1699-1710, 2020) algorithm to reliably avoid variational collapse and study the effectiveness of orbital optimized density functional theory (DFT) at predicting second period element 1s core-level spectra of open-shell systems. Several density functionals (including SCAN, B3LYP and $\omega$B97X-D3) are found to predict excitation energies from the core to singly occupied levels to high accuracy ($\le 0.3$ eV RMS error), against available experimental data. Higher excited states are however more challenging by virtue of being intrinsically multiconfigurational. We thus present a CI inspired route to self-consistently recouple single determinant mixed configurations obtained from DFT, in order to obtain approximate doublet states. This recoupling scheme is used to predict the C K-edge spectra of the allyl radical, the O K-edge spectra of CO$^+$ and the N K-edge of NO$_2$ to high accuracy relative to experiment, indicating substantial promise in using this approach for computation of core-level spectra for doublet species (vs more traditional time dependent DFT, EOM-CCSD or using unrecoupled mixed configurations). We also present general guidelines for computing core-excited states from orbital optimized DFT. 
	\end{abstract}

	\maketitle
	
	\section{Introduction} \label{introduction}
	Linear-response time dependent density functional theory (TDDFT)\cite{runge1984density,casida1995time,dreuw2005single} is very widely used to model electronic excited states of chemical species. TDDFT is an appealing approach as it is computationally inexpensive ($O(N^{3-4})$ scaling vs number of basis functions $N$), nearly black-box and able to simultaneously compute a large number of excited states. However, the lack of explicit orbital relaxation renders it unsuitable for describing excitations that involve substantial reorganization of electron density, such as charge transfer\cite{dreuw2005single,peach2008excitation} or Rydberg excited states\cite{casida1998molecular,tozer2000determination}. Excitation of core electrons in particular involves a substantial relaxation of the core-hole (and an accompanying reorganization of valence electron density), which leads to substantial errors in excitation energies predicted by TDDFT by standard functionals. It is consequently not unusual to blue-shift TDDFT core-level spectra by $\sim10$ eV for alignment with experiment\cite{besley2010time,wenzel2014calculating,attar2017femtosecond,bhattacherjee2018photoinduced,chantzis2018ab,lestrange2015calibration}(though the qualitative nature of transitions is typically reasonably predicted). Some specialized short-range corrected functionals specifically trained to predict core-level spectra\cite{besley2009time} tend to fare better\cite{besley2020density,besley2011theoretical,robinson2010modelling,fogarty2018experimental,buckley2011theoretical,ljubic2016experimental,ljubic2018characterisation}, but the very strong sensitivity of TDDFT excitation energies on delocalization error\cite{perdew1982density,dreuw2005single} is troubling (as even small perturbations could have disproportionate impact on relative peak positions). 
	
	In contrast, linear-response based wave function theories like equation of motion coupled cluster singles and doubles (EOM-CCSD)\cite{sekino1984linear,stanton1993equation,krylov2008equation,peng2015energy,coriani2015communication,coriani2016molecular,carbone2019analysis,vidal2019new} tend to systematically overestimate core-excitation energies\cite{coriani2012coupled,frati2019coupled,tsuru2019time,peng2015energy} due to lack of explicit orbital relaxation, often necessitating empirical redshifting by 1-2 eV for alignment with experiment\cite{frati2019coupled,tsuru2019time,peng2015energy,coriani2015communication}. Encouragingly however, use of different core-valence separation\cite{cederbaum1980many} (CVS) schemes has been observed to reduce the magnitude of the shift required\cite{vidal2019new,lopez2020equation,vidal2020dyson}.
	A flavor of second order extended algebraic diagrammatic construction (ADC(2)-x\cite{wormit2014investigating}, specifically CVS-ADC(2)-x\cite{wenzel2014calculating}) is also often employed to calculate core-level spectra\cite{wenzel2014calculating,norman2018simulating,list2020probing,wenzel2016physical}. The accuracy of CVS-ADC(2)-x however owes a great deal to fortuitious cancellation between various sources of error\cite{wenzel2014calculating,wenzel2015analysis}, and performance actually worsens when third order ADC is employed\cite{wenzel2015analysis}. At any rate, the higher computational cost of these wave function theories ($O(N^6)$ for both EOM-CCSD and ADC(2)-x) and slower basis set convergence renders them impractical for large molecular systems or extended materials, relative to computationally inexpensive DFT approaches. Nonetheless, development of lower-scaling approximations to these wave function based methods is expected to broaden their applicability considerably\cite{myhre2016near,peng2015energy}.
	
	In contrast to these linear-response based protocols, state-specific orbital optimized (OO) methods have been much more successful at accurate prediction of core-level spectra even within the DFT paradigm\cite{besley2009self,derricotte2015simulation,michelitsch2019efficient,hait2020highly,ehlert2020psixas}. The main difficulty with these methods is the potential for `variational collapse' of the target excited state down to the ground state or another excited state, as it is challenging to optimize excited state orbitals (by virtue of excited states typically being saddle points of energy). The maximum overlap method (MOM)\cite{gilbert2008self,barca2018simple} was developed to address this problem for repeated Fock matrix diagonalization based methods like DIIS\cite{pulay1980convergence}, though convergence failures and variational collapse (via slow drifting of orbitals) are not always prevented\cite{mewes2014molecular,hait2020excited}. More recently, some of us have have proposed a square gradient minimization (SGM)\cite{hait2020excited} based direct minimization approach that appears to be robust against both modes of MOM failure. SGM has been employed in conjunction with the spin-pure restricted open-shell Kohn-Sham (ROKS) method\cite{filatov1999spin,kowalczyk2013excitation} to predict highly accurate ($<0.5$ eV error) core-level spectra of closed-shell molecules\cite{hait2020highly} at local DFT cost (using the modern SCAN\cite{SCAN} functional). It is also worth noting that there exist linear-response methods that incorporate partial OO character through relaxed core-ionized states, like Static Exchange (STEX)\cite{aagren1997direct} or Non-orthogonal Configuration Interaction Singles (NOCIS)\cite{oosterbaan2018non,oosterbaan2019non,oosterbaan2020generalized}, though such treatments are wave function based and $\sim1$ eV error remains common due to lack of dynamic correlation.   
	
	Stable open-shell molecules are fairly uncommon in nature and there is consequently a scarcity of static experimental spectra for such species. However open-shell systems are omnipresent in chemical dynamics experiments (either as fragments or excited states of closed-shell molecules) where transient X-ray absorption spectroscopy (XAS) is often employed\cite{chergui2017photoinduced,bhattacherjee2018ultrafast,schnorr2019tracing,yang2018electron}. It is consequently useful to have cheap and reliable theoretical techniques capable of modeling core-level spectra of such species. 
	The highly accurate ROKS method is however not applicable to most open-shell systems, as it is explicitly designed for singlet states with one broken electron pair.  In fact, open-shell systems pose additional challenges for many of the methods described above, as a spin-pure treatment of excited states necessitates inclusion of some double excitations\cite{maurice1996nature,oosterbaan2019non,oosterbaan2020generalized} even for states that conventionally appear to be single excitations breaking one electron pair. This is not too difficult for wave function approaches, as shown by the extended CIS (XCIS\cite{maurice1996nature}) and open-shell NOCIS\cite{oosterbaan2019non,oosterbaan2020generalized} methods. However, it is not at all straightforward to achieve this within TDDFT, which has no route for describing double excitations within the widely used adiabatic approximation\cite{maitra2004double,levine2006conical,dreuw2005single}. It is tempting to believe that missing such configurations would not be particularly significant if the unpaired electrons interact only weakly, but the failure of TDDFT in describing excited state single bond dissociations despite the unrestricted reference state being reasonable\cite{hait2019beyond} indicates some cause for caution.

	
	In this work, we apply OO excited state DFT in conjunction with SGM to study single core-excitations of open-shell systems. This entails investigation of excitations to both singly occupied levels (which can be well described by single determinants, in principle) and completely unoccupied levels (which result in intrinsically multiconfigurational states). We present a scheme for recoupling multiple configurations to obtain an approximate doublet state for the latter class of excitations and demonstrate the utility of this protocol by considering the C K-edge spectra of the allyl radical, O K-edge of CO$^+$ and the N K-edge of NO$_2$. We also discuss general principles for reliably using these techniques to predict core-excitation spectra. Overall, we demonstrate that highly accurate DFT results can be obtained via orbital optimization with the modern local SCAN functional at low computational cost, similar to behavior observed for closed-shell systems. Low error can also be achieved via cam-B3LYP, TPSS and $\omega$B97X-D3 functionals (albeit at a somewhat higher asymptotic cost for the hybrid functionals).
	
	\section{Theory}
	\subsection{Single configurational states}\label{enas}
	Excitations from the core to singly occupied molecular orbitals (SOMOs) of open-shell systems result in states representable via a single Slater determinant, as there is no change in the number of unpaired electrons.
	The simplest approach for modeling such states is $\Delta$ Self-Consistent Field ($\Delta$SCF)\cite{ziegler1977calculation,gilbert2008self,kowalczyk2011assessment,besley2009self}, where the non-aufbau solution to the Hartree-Fock\cite{szabo2012modern} or Kohn-Sham\cite{kohn1965self} DFT equations is converged via an excited state solver like SGM or MOM. The resulting excited KS determinant would not necessarily be exactly orthogonal to the ground state determinant but this is generally of little concern since KS determinants are fictitious entities useful for finding densities and thus there exists no requirement that ground and excited state determinants be orthogonal.  Nonetheless, a significant ($>0.1$, for example) squared overlap between the ground and excited state configurations would be concerning but we have not observed such occurrences in our investigations and do not believe them to be likely without at least partial variational collapse of the core-hole. 
	
	The principal dilemma for such states is choosing between spin-restricted or unrestricted orbitals for $\Delta$SCF. Unrestricted orbitals are typically more suitable for DFT studies on open-shell systems, though some functionals are known to yield atypically unphysical behavior in certain limits away from equilibrium\cite{hait2019wellbehaved}. On the other hand, restricted open-shell (RO) orbitals artificially enforce a spin-symmetry that does not exist in radicals. As will be shown later (in Table \ref{tab:uvsro}), use of unrestricted orbitals appears to systematically lower the core-excitation energies (via extra stabilization of the core-excited state relative to the ground state). The best functionals for predicting spectra of closed-shell species yield lower errors for radicals when unrestricted orbitals are employed, and we thus recommend the use of unrestricted orbitals over RO orbitals for radicals. RO orbitals however should be employed for closed-shell systems (via ROKS or related methods)\cite{hait2020highly}, on account of the existence of spin-symmetry in such species.
	
	\subsection{Multiconfigurational states}\label{multi}
	Multiconfigurational DFT is a difficult challenge even outside the unique challenges of TDDFT for double excitations, as the Kohn-Sham (KS) exchange-correlation energy is defined for a single determinant reference. KS-DFT target states therefore should be single determinants, and directly recoupling them via configuration interaction (CI) would result in double counting of some electron-electron interactions through both the functional and the CI off-diagonal terms. This is quite undesirable, making modeling such states fairly challenging. 
	
	One very reasonable solution is to note that single determinants with both $\alpha$ and $\beta$ unpaired electrons are mixtures of different spin-states, and the highest spin-state within that ensemble can be well approximated by a single determinant by merely making all unpaired spins point in the same direction. Approximate spin-projection (AP)\cite{yamaguchi1988spin} can consequently be applied to remove this high spin contribution from a spin impure mixed determinant. This approach should be sufficient when there are only two eigenstates that significantly contribute to the mixed configuration, as is the case for single excitations out of closed-shell molecules (where only the singlet and triplet states contribute). ROKS in fact utilizes this very feature to ensure spin-purity. ROKS employs a mixed configuration that has one unpaired $\alpha$ spin and one unpaired $\beta$ spin (which has energy $E_M$) and a triplet configuration that has both unpaired spins as $\alpha$ (which has energy $E_T$). The use of RO orbitals forces the mixed configuration to be exactly halfway between singlet and triplet, indicating $E_M=\dfrac{E_S+E_T}{2}$ where $E_S$ is the true singlet energy. ROKS consequently optimizes the purified singlet energy $E_S=2E_M-E_T$. 
	
	Things are however substantially more challenging for doublet states. A mixed configuration with two unpaired $\alpha$ electrons and one unpaired $\beta$ electron is a mixture of three states---two doublets and a quartet. The quartet contribution can be easily removed using an AP protocol similar to ROKS, but disentangling the two doublet energies is nontrivial. 
	
	Looking at the pure wave function based CI approach however offers some hints as to how to proceed. If we consider restricted open-shell configurations with three unpaired electrons occupying three spin-restricted orbitals (labeled $1,2$ and $3$, respectively), eight possible configurations exist. Spin-inversion symmetry in the absence of magnetic fields however indicate that only four provide unique information:
	
	\begin{enumerate}
		\item $\ket{Q}=\ket{\uparrow\uparrow\uparrow}$: All three spins are $\alpha$. This is the pure quartet with energy $E_Q$.
		\item $\ket{M_1}=\ket{\downarrow\uparrow\uparrow}$: Only the spin at orbital $1$ is $\beta$. This is a mixed configuration with energy $E_{M_1}=E_Q+K_{12}+K_{13}$, where $K_{pq}$ is the exchange interaction $\braket{pq}{qp}$ between an electron in orbital $p$ and another in orbital $q$. The inversion of the spin in orbital $1$ relative to the quartet leads to a loss of exchange stabilization between this orbital and the other two, leading to the energy going up by $K_{12}+K_{13}$.
		\item $\ket{M_2}=\ket{\uparrow\downarrow\uparrow}$ Only the spin at orbital $2$ is $\beta$. Consequently $E_{M_2}=E_Q+K_{12}+K_{23}$.
		\item $\ket{M_3}=\ket{\uparrow\uparrow\downarrow}$ Only the spin at orbital $3$ is $\beta$. Consequently $E_{M_3}=E_Q+K_{13}+K_{23}$.   
	\end{enumerate}
	Having the single determinant energies $E_Q,E_{M_1},E_{M_2},E_{M_3}$ is sufficient to uniquely solve for the exchange interactions $K_{pq}$, with $K_{12}=\dfrac{E_{M_1}+E_{M_2}-E_Q-E_{M_3}}{2}$ etc. This is quite useful, as the off-diagonal CI coupling elements are $\bra{M_i}H\ket{M_j}=-K_{ij}$ from Slater-Condon rules for double excitations\cite{szabo2012modern}. This indicates that the knowledge of the single determinant energies is sufficient for solving the CI problem. With this, we find the eigenvalues of $H$ within the subspace spanned by $\ket{M_{1,2,3}}$ to be:
	\begin{align}
	E_1 &= E_Q\\
	E_2 &= \dfrac{1}{2}\left(E_{M_1}+E_{M_2}+E_{M_3}-E_{Q}- \sqrt{2\left[\left(E_{M_1}-E_{M_2}\right)^2+\left(E_{M_2}-E_{M_3}\right)^2+\left(E_{M_3}-E_{M_1}\right)^2\right]}\right)\label{d1}\\
	E_3 &= \dfrac{1}{2}\left(E_{M_1}+E_{M_2}+E_{M_3}-E_{Q}+ \sqrt{2\left[\left(E_{M_1}-E_{M_2}\right)^2+\left(E_{M_2}-E_{M_3}\right)^2+\left(E_{M_3}-E_{M_1}\right)^2\right]}\right)\label{d2}
	\end{align}
	The first eigenvalue corresponds to the quartet within the $M_S=\dfrac{1}{2}$ subspace (which is a linear combination of all three configurations with equal weights). The other two correspond to the energies of the two possible doublet states. 
	
	We propose that the same approach be employed for recoupling DFT configurations, with the KS energies of configurations $\ket{M_{1,2,3}}$ being employed instead of the HF ones used in the wave function theory approach. The risk of double counting should be greatly reduced as the effective off-diagonal elements are found directly from the KS energies versus Slater-Condon rules. Indeed, the off-diagonal elements should no longer be viewed as exchange interactions but rather effective spin-spin coupling elements. The entire approach is basically equivalent to solving for the eigenstates of the effective Ising like Hamiltonian $H^\prime=-2J_{12}\vec{S}_1\cdot\vec{S}_2-2J_{13}\vec{S}_1\cdot\vec{S}_3-2J_{23}\vec{S}_2\cdot\vec{S}_3$ for three interacting spins, where the couplings $J_{ij}$ are obtained from DFT (and are equivalent to the exchange interactions $K_{ij}$ if HF is used as the functional). Such approaches have been used within broken-symmetry DFT to calculate spin coupling constants of transition metal species to reasonable accuracy\cite{yamaguchi1979singlet,noodleman1981valence,noodleman1985models,sinnecker2004calculating,lovell2001femo,adams1997density,mouesca1995density,witzke2020bimetallic}, and it is hoped that similar behavior will transfer over. Furthermore, equivalent logic for the case of two unpaired spins yields ROKS, which is known to be quite accurate for singlet states with one broken electron pair\cite{kowalczyk2013excitation,hait2016prediction,hait2020excited}. These known instances of successful behavior encourages us to believe that this protocol is worthwhile to explore. We also note that Eqns \ref{d1}-\ref{d2} were reported in Ref \onlinecite{kowalczyk2011assessment} without an explicit description of the derivation, but these have not been actually applied to core-level spectroscopy (or any excited state problem) to the best of our knowledge.
	
	Having obtained $E_{2,3}$ as spin-purified energies, we next seek to determine how to obtain the optimal orbitals. It is tempting to directly optimize  $E_{2,3}$ in a manner analogous to ROKS but we have elected not to do so at present. This optimization is nontrivial due to the nonlinear nature of the energy expression (vs the simpler form for ROKS). In addition, the derived equation is only precisely true for restricted open-shell orbitals, while Sec \ref{enas} seems to suggest unrestricted orbitals are optimal. We therefore look to AP-$\Delta$SCF\cite{ziegler1977calculation,kowalczyk2011assessment} for singlet excited states for inspiration, where the mixed determinant and triplet determinants are individually optimized (resulting in two sets of orbitals) and the singlet energy is simply computed as $2E_M-E_T$ from the individually optimized energies, instead of optimizing a single set of orbitals as in ROKS. The resulting energies however are often not dramatically different from ROKS\cite{kowalczyk2013excitation} and so we choose to follow a similar protocol here to determine if there is sufficient utility in this route for recoupling mixed determinants to justify optimizing a single set of unrestricted orbitals for computing the doublet energies.
	We consequently optimize $\ket{Q}$ and $\ket{M_{1,2,3}}$ individually and compute $E_{2,3}$ from those optimized energies.
	
	One rather inconvenient detail is that individually optimized $\ket{M_{1,2,3}}$ configurations would thus not be strictly orthogonal to each other due to slight differences in the orbitals. However we do not consider any non-orthogonality derived terms arising from mixed configurations, as the KS determinants are fictitious constructs. On a more practical note, we ensure low overlap via  providing restricted open-shell quartet orbitals as the initial guess for SGM optimization of the mixed determinants. The initial guesses are thus orthogonal, and orbital relaxation to the closest stationary point (which SGM is supposed to achieve) in unrestricted space should not lead to significant non-orthogonality for cases where this model of three unpaired electrons is a good approximation. Further details about initial guesses are enumerated in Sec \ref{sec:recs}. 
	
	\subsection{Transition Dipole Moments}\label{tmu}
	The magnitude of the transition dipole moment between the ground and excited states is essential for computing oscillator strengths (and thus relative intensities in computed spectra). The fictitious nature of the KS determinant (which represents a wave function of noninteracting electrons subjected to a fictitious potential) is a significant obstacle here, as it implies there is no rigourous route for computing transition dipole moments. However, treating the KS determinants as real wave functions might be a reasonable approximation for computing this quantity, in the hope that the KS determinants (or superpositions thereof) would have a reasonably large overlap with the true wave functions to make this exercise worthwhile.  Indeed, spectra computed via this route show fairly good agreement with experiment (as can be seen from previous work\cite{hait2020highly} by some of us, for instance). Such a protocol can (and should) account for nonorthogonality between ground and mixed determinants as it is fairly simple to compute NOCI dipole matrix elements\cite{thom2009hartree}.
	
	There are some additional factors to consider for the recoupled multiconfigurational states. The wave function inspired approach indicates that transition dipole moments should be computed via a linear combination of the transition dipole moments of individual determinants, as weighted by their coefficients in the eigenvectors corresponding to $E_{2,3}$. The effect of non-orthogonality between mixed determinants $\ket{M_{1,2,3}}$ on eigenvector coefficients is neglected here both because such terms are relatively small (because the mixed determinants have fairly low overlap with each other) and because it is not straightforward to calculate these effects. The decision to not consider this form of nonorthogonality does not appear to have any significant deleterious impact, as shown by the spectra presented later. 
	
	The other important factor to consider is that the analysis in Sec \ref{multi} found off-diagonal coupling elements directly from the energies $E_{M_{1,2,3}}$ and thus did not account for phases of $\ket{M_{1,2,3}}$. These phases however are critical for estimating transition dipole moments, and thus must be obtained somehow. A protocol for estimating these phases via the formally ``quartet" state is supplied in the Appendix. 
	
	\section{Results and Discussion}
	
	\subsection{Excitations to the SOMO}
	The relative scarcity of experimental XAS data for radicals leaves us with a fairly small dataset of 17 excitations for assessing the performance of single determinant $\Delta$SCF. The precise statistical values here are thus less reliable than those obtained in Ref \onlinecite{hait2020highly} from 40 excitations out of closed-shell molecules, but general qualitative trends can be drawn even from this restricted amount of data. The experimental excitation energies for all the C K-edge excitations (save allyl and CO$^+$) were measured by some of us, via radicals obtained from the photodissociation of the corresponding iodide\cite{yang2018electron,yang2018}. These values should have an uncertainty of $\pm 0.1$ eV, although vibrational excitations induced by photodissociation could shift the values somewhat. However, the resulting excitation energy for CH$_3$ agrees well with vibrationally resolved spectra obtained from radicals generated from flash pyrolysis\cite{alagia2007probing}. Furthermore, (as can be seen from Table \ref{tab:somodata}), the experimental shifts between the C K-edge of the allyl radical (obtained by authors of Ref \onlinecite{alagia2013soft} on cold radicals generated via flash pyrolysis) and other C K-edges are very well reproduced by theoretical methods, suggesting that any vibrational excitation induced effect was small overall. A full Frank-Condon analysis could prove useful in quantifying any such effect, but was not pursued at present.
	
	We only consider a relatively small number of density functionals as a combination of large experimental uncertainty  (typically 0.1 eV) and limited number of data points would make precise rankings of many functionals meaningless. We think it is more useful to investigate the performance of some representative functionals and see if they are sufficiently accurate to justify wider use. We therefore consider the following functionals from various rungs of Jacob's ladder\cite{perdew1982density}:
	\begin{enumerate}
		\item Rung 1 (local spin-density approximation/LSDA\cite{Slater,VWN,PW92}): Not considered due to very large errors found in Ref \onlinecite{hait2020highly}.
		\item Rung 2 (generalized gradient approximation/GGA): BLYP\cite{b88,lyp}, PBE\cite{PBE}.
		\item Rung 3 (meta-GGA): TPSS\cite{tpss}, SCAN\cite{SCAN}.
		\item Rung 4 (hybrids): B3LYP\cite{b3lyp}, PBE0\cite{pbe0} (global hybrids); cam-B3LYP\cite{camb3lyp}, $\omega$B97X-D3\cite{wB97XD3}, $\omega$B97X-V\cite{wb97xv} (range separated hybrids).
		\item Rung 5 (double hybrids): Not considered due to significant computational expense. 
	\end{enumerate}
	The Hartree-Fock (HF) wave function method is also considered, in order to determine the impact of neglecting correlation entirely.  The choice of functionals here was guided by both a desire to compare against closed-shell results reported earlier\cite{hait2020highly} and a desire to examine the behavior of classic, minimally parameterized functionals like B3LYP. 
	
	
	\begin{table}[htb!]
		\begin{tabular}{llllllllllll}
			Radical	&	Expt.	&	BLYP	&	PBE	&	TPSS	&	SCAN	&	B3LYP	&	PBE0	&	cam-B3LYP	&	$\omega$B97X-D3	&	$\omega$B97X-V	&	HF	\\
			CH$_3$	&	281.4\cite{yang2018,alagia2007probing}	&	281.6	&	280.8	&	281.8	&	281.8	&	281.7	&	281.2	&	281.6	&	281.8	&	281.9	&	282.8	\\
			CH$_3$CH$_2$	&	281.7\cite{yang2018}	&	282.0	&	281.3	&	282.1	&	282.2	&	282.1	&	281.6	&	282.0	&	282.2	&	282.3	&	283.0	\\
			(CH$_3$)$_2$CH	&	282.2\cite{yang2018}	&	282.3	&	281.6	&	282.4	&	282.5	&	282.4	&	282.0	&	282.3	&	282.5	&	282.6	&	283.3	\\
			(CH$_3$)$_3$C	&	282.6\cite{yang2018}	&	282.6	&	281.9	&	282.6	&	282.8	&	282.7	&	282.3	&	282.6	&	282.8	&	282.9	&	283.5	\\
			Allyl	&	282.0\cite{alagia2013soft}	&	282.3	&	281.5	&	282.4	&	282.5	&	282.4	&	281.9	&	282.3	&	282.5	&	282.6	&	283.5	\\
			CO$^+$	&	282.0\cite{couto2020carbon}	&	282.2	&	281.3	&	282.2	&	282.3	&	282.3	&	281.8	&	282.2	&	282.4	&	282.5	&	283.3	\\
			CH$_2$Br	&	282.6\cite{yang2018}	&	282.8	&	282.0	&	282.9	&	282.9	&	282.8	&	282.4	&	282.7	&	282.9	&	283.1	&	283.8	\\
			CH$_2$Cl	&	282.8\cite{yang2018electron}	&	283.0	&	282.2	&	283.1	&	283.2	&	283.0	&	282.6	&	282.9	&	283.1	&	283.3	&	284.0	\\
			NH$_2$	&	394.3\cite{parent2009irradiation}	&	394.5	&	393.6	&	394.6	&	394.7	&	394.5	&	394.1	&	394.5	&	394.7	&	394.8	&	395.7	\\
			N$_2^+$	&	394.3\cite{lindblad2020x}	&	394.4	&	393.5	&	394.3	&	394.3	&	394.2	&	393.6	&	394.0	&	394.2	&	394.3	&	394.4	\\
			NH$_3^+$	&	395.2\cite{bari2019inner}	&	395.0	&	394.2	&	395.2	&	395.3	&	395.1	&	394.6	&	395.0	&	395.2	&	395.4	&	396.4	\\
			NO$_2$	&	401.0\cite{zhang1990inner}	&	401.0	&	400.2	&	401.0	&	401.2	&	401.0	&	400.6	&	401.1	&	401.3	&	401.5	&	402.1	\\
			OH	&	525.8\cite{stranges2002high}	&	525.8	&	524.9	&	526.0	&	526.0	&	525.8	&	525.3	&	525.8	&	525.9	&	526.1	&	527.0	\\
			HO$_2$	&	528.6\cite{lacombe2006radical}	&	528.5	&	527.6	&	528.6	&	528.5	&	528.3	&	527.8	&	528.3	&	528.4	&	528.6	&	528.9	\\
			NO$_2$	&	530.3\cite{zhang1990inner}	&	530.5	&	529.7	&	530.5	&	530.5	&	530.2	&	529.7	&	530.1	&	530.2	&	530.4	&	529.9	\\
			O$_2$	&	530.8\cite{coreno1999vibrationally}	&	530.8	&	530.0	&	530.8	&	530.8	&	530.6	&	530.1	&	530.6	&	530.8	&	531.0	&	530.6	\\
			CO$^+$	&	528.5\cite{couto2020carbon}	&	528.3	&	527.5	&	528.4	&	528.5	&	528.0	&	527.6	&	528.0	&	528.1	&	528.2	&	528.3	\\
			&	&	&	&	&	&	&	&	&	&	&	\\											
			&	RMSE	&	0.2	&	0.7	&	0.2	&	0.3	&	0.3	&	0.5	&	0.2	&	0.3	&	0.4	&	1.1	\\
			&	ME	&	0.1	&	-0.7	&	0.2	&	0.2	&	0.0	&	-0.4	&	0.0	&	0.2	&	0.3	&	0.9	\\
			&	MAX	&	0.3	&	1.0	&	0.4	&	0.5	&	0.5	&	0.9	&	0.5	&	0.5	&	0.6	&	1.5	
		\end{tabular}
		\caption{$\Delta$SCF/aug-cc-pCVTZ\cite{dunning1989gaussian,kendall1992electron,woon1995gaussian} core to SOMO excitation energies (in eV) for open-shell species, as predicted by various functionals. Unrestricted orbitals were used for both the ground and excited states. Root mean squared error (RMSE), mean error (ME) and maximum absolute error (MAX) are also reported.}
		\label{tab:somodata}
	\end{table}
	
	Table \ref{tab:somodata} presents the excitation energies calculated using the chosen approaches (using spin-unrestricted orbitals), along with statistical measures of error like the root mean squared error (RMSE). None of the density functional methods deviate from experiment by more than 1 eV, which is in sharp contrast to the typical behavior of TDDFT with the same functionals\cite{hait2020highly}. Even HF has only $< 2$ eV error despite complete absence of correlation. We specifically observe that the BLYP, TPSS, SCAN, B3LYP, cam-B3LYP and $\omega$B97X-D3 functionals yield 0.3 eV or lower RMSE, and do not deviate by more than 0.5 eV from the experimental reference values. The good performance by local functionals like BLYP, TPSS and SCAN is quite impressive, as these functionals are much more computationally efficient than hybrids. Of the trio, the performance of only SCAN has been characterized for closed-shell systems\cite{hait2020highly}, where it was also found to be similarly accurate. We consequently focus on the performance of SCAN in later sections of this work, as good performance in both the closed and open-shell limits is critical for prediction of transient X-ray absorption spectroscopy. However, we believe that good performance can be obtained from many functionals considered in this work (as partially demonstrated in the Supporting Information). Interestingly, PBE and PBE0 perform surprisingly poorly, especially relative to BLYP and B3LYP, respectively.
	

	Table \ref{tab:somodata} furthermore shows that the small errors for many functionals are mostly systematic, which appears to suggest that the change in excitation energy between two species (say between methyl and tert-butyl, for instance) would be reproduced fairly accurately by most functionals. This is also in principle true for TDDFT, although the massive ($\sim 10 $ eV) errors in the individual excitation energies mean that even a relatively small variation in absolute error could have significant impact on relative peak positions (made more likely by very high sensitivity of TDDFT results to delocalization error\cite{perdew1982density,dreuw2005single}). Most functionals (including SCAN) appear to systematically overestimate energies, while PBE and PBE0 systematically underestimate (which might be the reason for their poor overall performance). Inclusion of relativistic effects\cite{takahashi2017relativistic} (which systematically increase excitation energies by binding core electrons more tightly) would therefore degrade performance of many functionals, while improving the performance of PBE and PBE0. The atom specific relativistic corrections for C,N and O are however quite small\cite{takahashi2017relativistic} (0.1-0.3 eV) and therefore are often neglected in studies (such as by the SRC functionals trained for TDDFT spectra prediction\cite{besley2009time}, which has these effects implicitly baked into what is fundamentally a nonrelativistic theory). The impact of incorporating these corrections on the errors of various models is provided in the supporting information, which shows that the RMSE of functionals (other than PBE and PBE0) goes up by $0.1$ eV at most, suggesting that this is not a major issue in practice.
	We also note that HF systematically overestimates excitation energies by $\sim 1$ eV due to missing correlation, which indicates that simple models for dynamical correlation (such as perturbative approaches\cite{szabo2012modern,cremer2011moller}) might be adequate for substantially lowering error, albeit at higher computational cost than DFT. HF however has a strong propensity to spuriously spin-contaminate Slater determinants, and the performance of perturbative corrections to HF references could consequently be greatly degraded\cite{gill1988does,cremer2011moller}.

	\begin{table}[htb!]
		\begin{tabular}{llllll}
			Radical	&	Experiment	&	RO-SCAN	&	USCAN	&	RO-PBE0	&	UPBE0	\\
			CH$_3$	&	281.4	&	281.9	&	281.8	&	281.3	&	281.2	\\
			CH$_3$CH$_2$	&	281.7	&	282.3	&	282.2	&	281.7	&	281.6	\\
			(CH$_3$)$_2$CH	&	282.2	&	282.6	&	282.5	&	282.1	&	282.0	\\
			(CH$_3$)$_3$C	&	282.6	&	282.9	&	282.8	&	282.4	&	282.3	\\
			Allyl	&	282.0	&	282.5	&	282.5	&	281.9	&	281.9	\\
			CO$^+$	&	282.0	&	282.5	&	282.3	&	281.9	&	281.8	\\
			CH$_2$Br	&	282.6	&	283.0	&	282.9	&	282.5	&	282.4	\\
			CH$_2$Cl	&	282.8	&	283.3	&	283.2	&	282.7	&	282.6	\\
			NH$_2$	&	394.3	&	394.8	&	394.7	&	394.2	&	394.1	\\
			N$_2^+$	&	394.3	&	394.5	&	394.3	&	393.8	&	393.6	\\
			NH$_3^+$	&	395.2	&	395.4	&	395.3	&	394.7	&	394.6	\\
			NO$_2$	&	401.0	&	401.4	&	401.2	&	400.7	&	400.6	\\
			OH	&	525.8	&	526.2	&	526.0	&	525.4	&	525.3	\\
			HO$_2$	&	528.5	&	528.6	&	528.5	&	527.7	&	527.6	\\
			NO$_2$	&	528.6	&	528.7	&	528.5	&	527.9	&	527.8	\\
			O$_2$	&	530.3	&	530.7	&	530.5	&	529.8	&	529.7	\\
			CO$^+$	&	530.8	&	531.0	&	530.8	&	530.2	&	530.1	\\
			&	&	&	&	&	\\					
			&	RMSE	&	0.4	&	0.3	&	0.4	&	0.5	\\
			&	ME	&	0.4	&	0.2	&	-0.3	&	-0.4	\\
			&	MAX	&	0.6	&	0.5	&	0.8	&	0.9	
		\end{tabular}
		\caption{Comparison of $\Delta$SCF/aug-cc-pCVTZ core to SOMO excitation energies (in eV) for restricted open-shell (RO) and unrestricted (U) orbitals with SCAN and PBE0. Results for other functionals are provided in the supporting information.}
		\label{tab:uvsro}
	\end{table}
	
	We also consider whether there is any benefit to using restricted open-shell orbitals over unrestricted orbitals. Table \ref{tab:uvsro} indicates that use of unrestricted orbitals systematically lowers excitation energies by $\sim 0.1$ eV relative to restricted open-shell results. This consequently indicates that use of RO orbitals instead of U would degrade the performance of most of the studied functionals (as they systematically overestimate with U orbitals) and improve the behavior of PBE and PBE0.  Indeed, Table \ref{tab:uvsro} shows that both RO-PBE0 and RO-SCAN have the same RMSE of 0.4 eV. This potentially argues that RO-PBE0 is perhaps preferable to USCAN, as the small relativistic corrections furthers improve the RO-PBE0 RMSE to 0.2 eV (while degrading USCAN's RMSE to 0.4 eV, as shown in the Supporting Information). However, we believe that SCAN with unrestricted orbitals is still the preferred route, even aside from  greater asymptotic computational efficiency. Open-shell systems tend to often arise in transient absorption experiments starting from closed-shell species, and so it is important to use an approach that is effective at predicting the spectra for both types of systems. PBE0 is perceptibly inferior to SCAN when it comes to closed-shell systems\cite{hait2020highly} (irrespective of inclusion of relativistic effects), and the two are fairly close in predictive ability for open-shell systems, making SCAN with unrestricted orbitals the preferred choice. We also note that a comparison between aug-cc-pCVTZ and aug-cc-pCVQZ results shows that a small part ($\sim 0.1$ eV) of the systematic overestimation predicted by SCAN for Table \ref{tab:somodata} values stems from basis set incompleteness (as shown by a comparison in the Supporting Information), similar to behavior of closed-shell species\cite{hait2020highly}. 
	
	\begin{figure}[htb!]
		\includegraphics[width=0.5\textwidth]{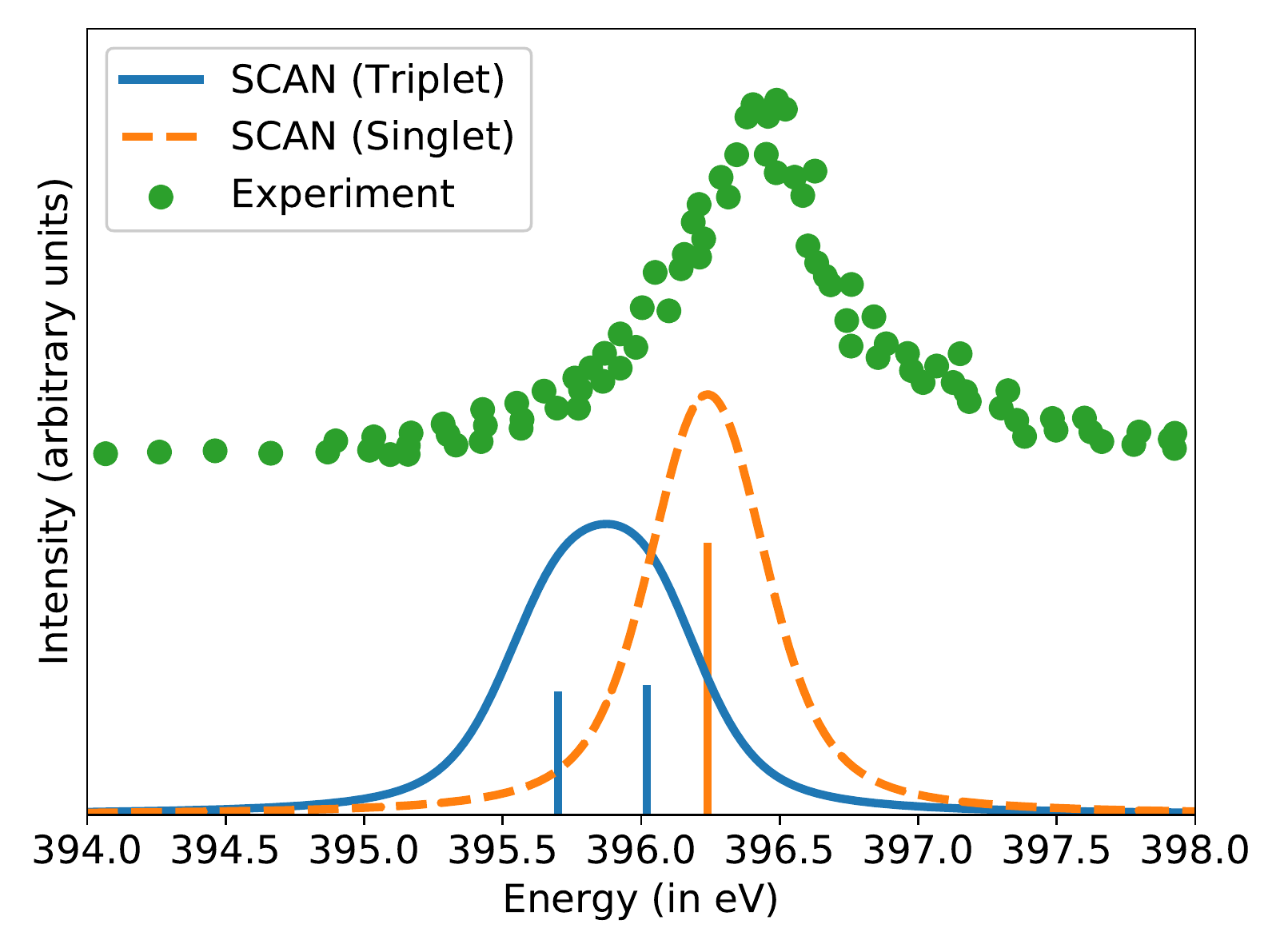}
		\caption{Comparison of experimental N K-edge spectrum of NH$_2^+$ (obtained from Ref \onlinecite{bari2019inner}) with those computed with SCAN/aug-cc-pCVTZ. 
			A Voigt profile with a Gaussian standard deviation of 0.2 eV and Lorentzian $\gamma=0.121$ eV was utilized for broadening the computed spectra. Bars are supplied to denote the location of the predicted excitation energies. The singlet and triplet spectra have been normalized by the same factor for a fair comparison.}
		\label{fig:nh2p}
	\end{figure}
	
	\subsubsection{The case of NH$_2^+$}
	
	\begin{table}[htb!]
		
		\begin{tabular}{lllll}
			Method          &       & Triplet &         & Singlet (ROKS) \\
			& Low   & High    & Average &                \\
			Experiment     &       &         & 396.4\cite{bari2019inner}   &                \\
			BLYP      & 395.7 & 396.0   & 395.8   & 396.1          \\
			PBE       & 394.9 & 395.2   & 395.0   & 395.3          \\
			TPSS      & 396.0 & 396.3   & 396.2   & 396.2          \\
			SCAN      & 395.7 & 396.0   & 395.9   & 396.2          \\
			B3LYP     & 395.7 & 396.0   & 395.9   & 396.1          \\
			PBE0      & 395.3 & 395.6   & 395.4   & 395.7          \\
			cam-B3LYP & 395.7 & 396.0   & 395.9   & 396.1          \\
			$\omega$B97X-D3  & 396.1 & 396.4   & 396.2   & 396.2          \\
			$\omega$B97X-V   & 396.3 & 396.6   & 396.4   & 396.3         
		\end{tabular}
		\caption{Comparison of $\Delta$SCF/aug-cc-pCVTZ core to SOMO excitation energies (in eV) for $^3B_1$ NH$_2^+$ with various functionals. The lowest core excitation energy for the $^1A_1$ singlet is also reported}
		\label{tab:nh2p}
	\end{table}
	
	The spectrum of NH$_2^+$ has been experimentally characterized\cite{bari2019inner}, but was not considered in Table \ref{tab:somodata} as the two possible excitations to singly occupied levels are  unresolved experimentally (assuming the radical cation is in the $^3B_1$ ground state). We used $\Delta$SCF to compute the two transitions separately, and report them in Table \ref{tab:nh2p}. These transitions have nearly the same oscillator strength and thus their average should roughly correspond to the experimental peak. The ROKS results for the lowest lying singlet $^1A_1$ excited state is also reported, in case it contributes to the experimental spectrum as well. Fig \ref{fig:nh2p} presents the representative case of the SCAN functional, with other methods yielding similar figures. 
	
	The computed average triplet excitation energies in Table \ref{fig:nh2p} agree fairly well with experiment, especially for good performers like SCAN, B3LYP or $\omega$B97X-D3. However, the values are somewhat red-shifted, in stark contrast to the general behavior seen in Table \ref{tab:somodata}. One possible explanation for this would be a blue-shifting of the experimental spectrum due to presence of singlet NH$_2^+$, since this state absorbs fairly strongly (roughly twice the oscillator strength than the individual triplet transitions) at slightly higher energies than the triplet, pushing the overall center of the band to higher energies (as hinted at by Fig \ref{fig:nh2p}). However, the computed triplet excitation average and the experimental maximum are not too far from each other (0.5 eV for SCAN), so it is not entirely impossible for DFT error to be the sole reason behind the discrepancies. $\omega$B97X-V for instance gives quite good agreement with experiment, without needing to invoke the singlet state.

	\subsection{Spectrum of the Allyl Radical}
	
	\begin{figure}[htb!]
		\begin{minipage}{0.48\textwidth}
			\centering
			\includegraphics[width=\linewidth]{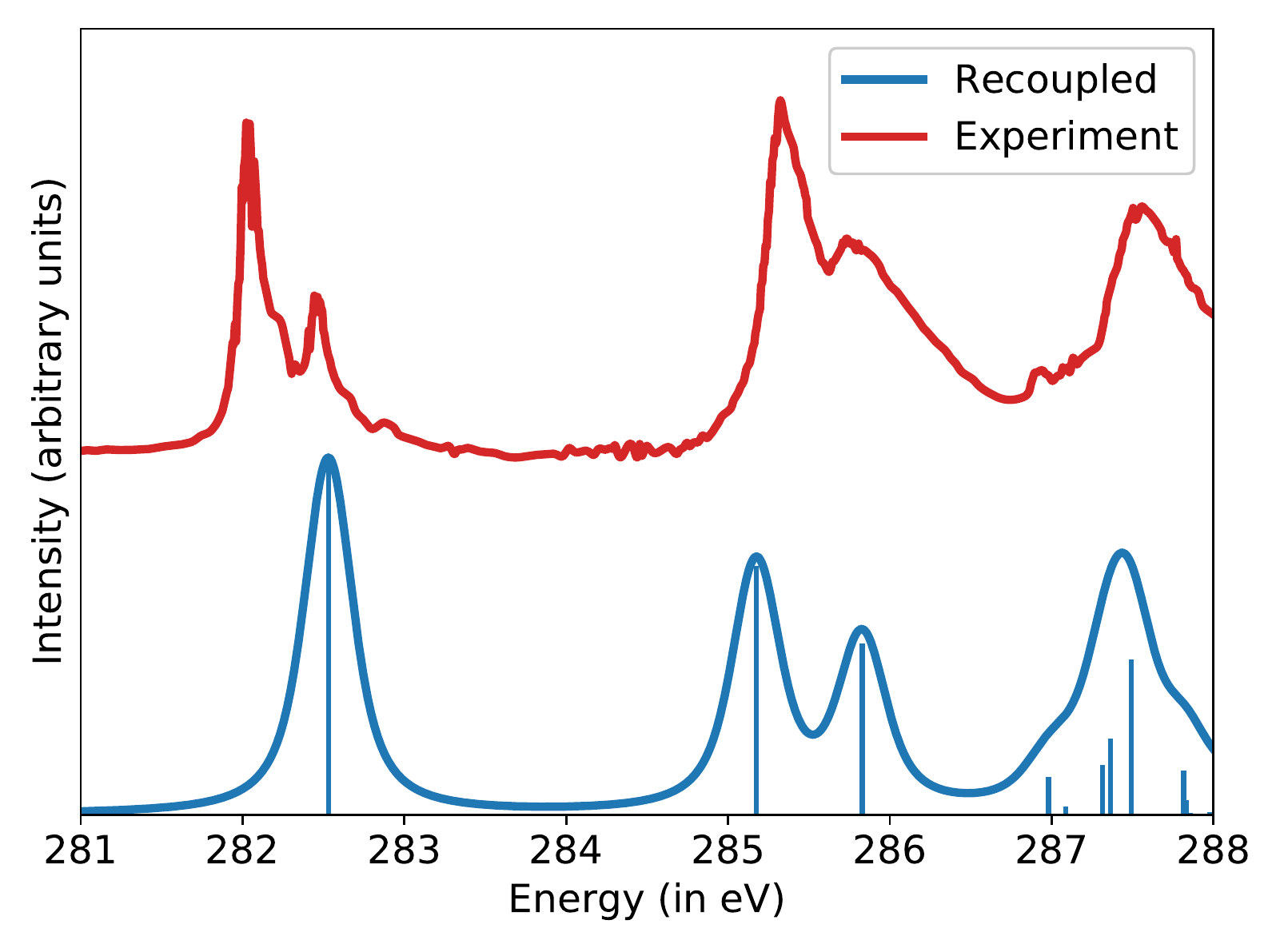}
			\subcaption{Recoupled configurations/SCAN}
			\label{fig:allylrec}
		\end{minipage}
		\begin{minipage}{0.48\textwidth}
			\centering
			\includegraphics[width=\linewidth]{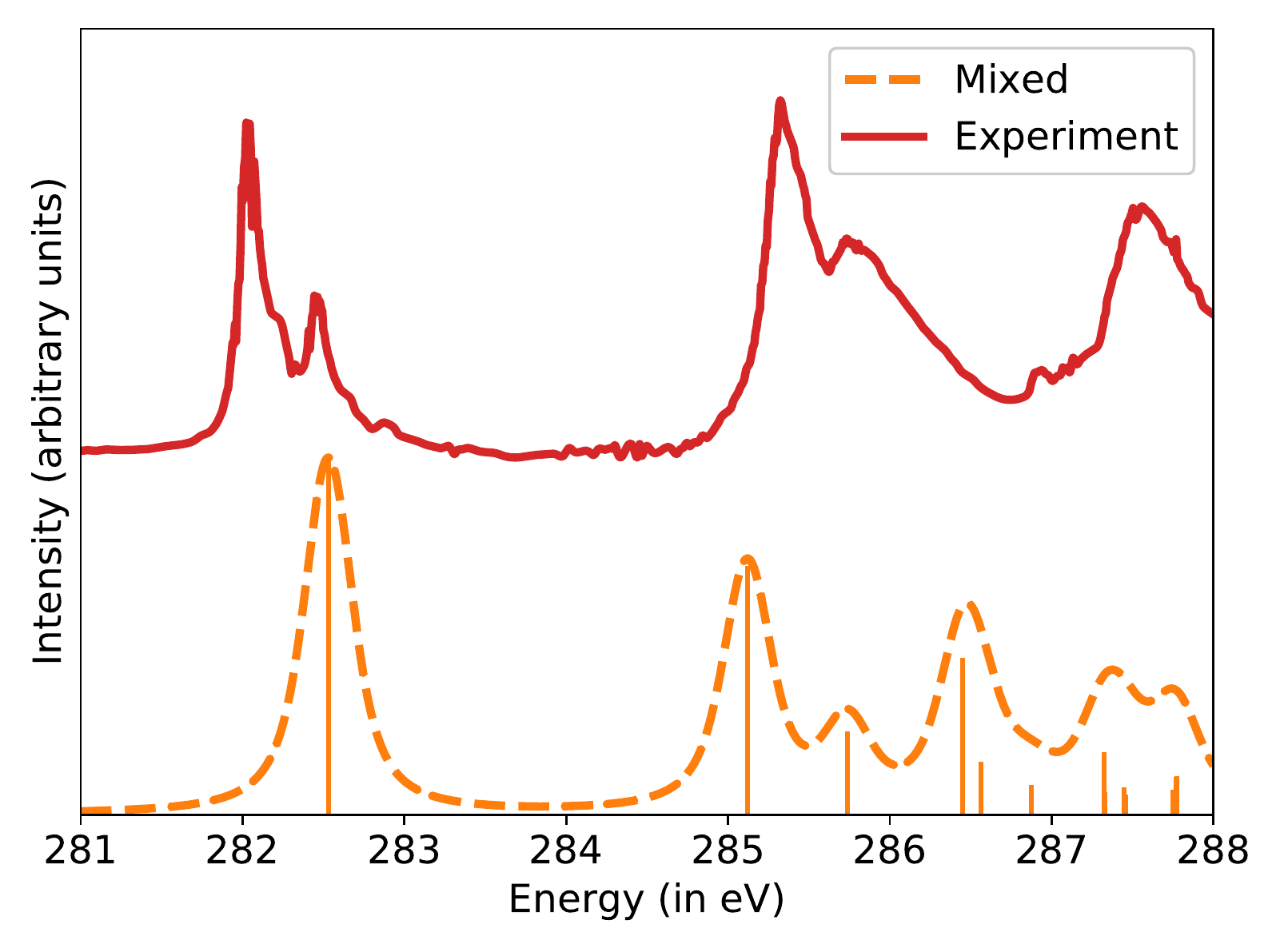}
			\subcaption{Mixed Configurations/SCAN}
			\label{fig:allylmix}
		\end{minipage}
		\begin{minipage}{0.48\textwidth}
			\centering
			\includegraphics[width=\linewidth]{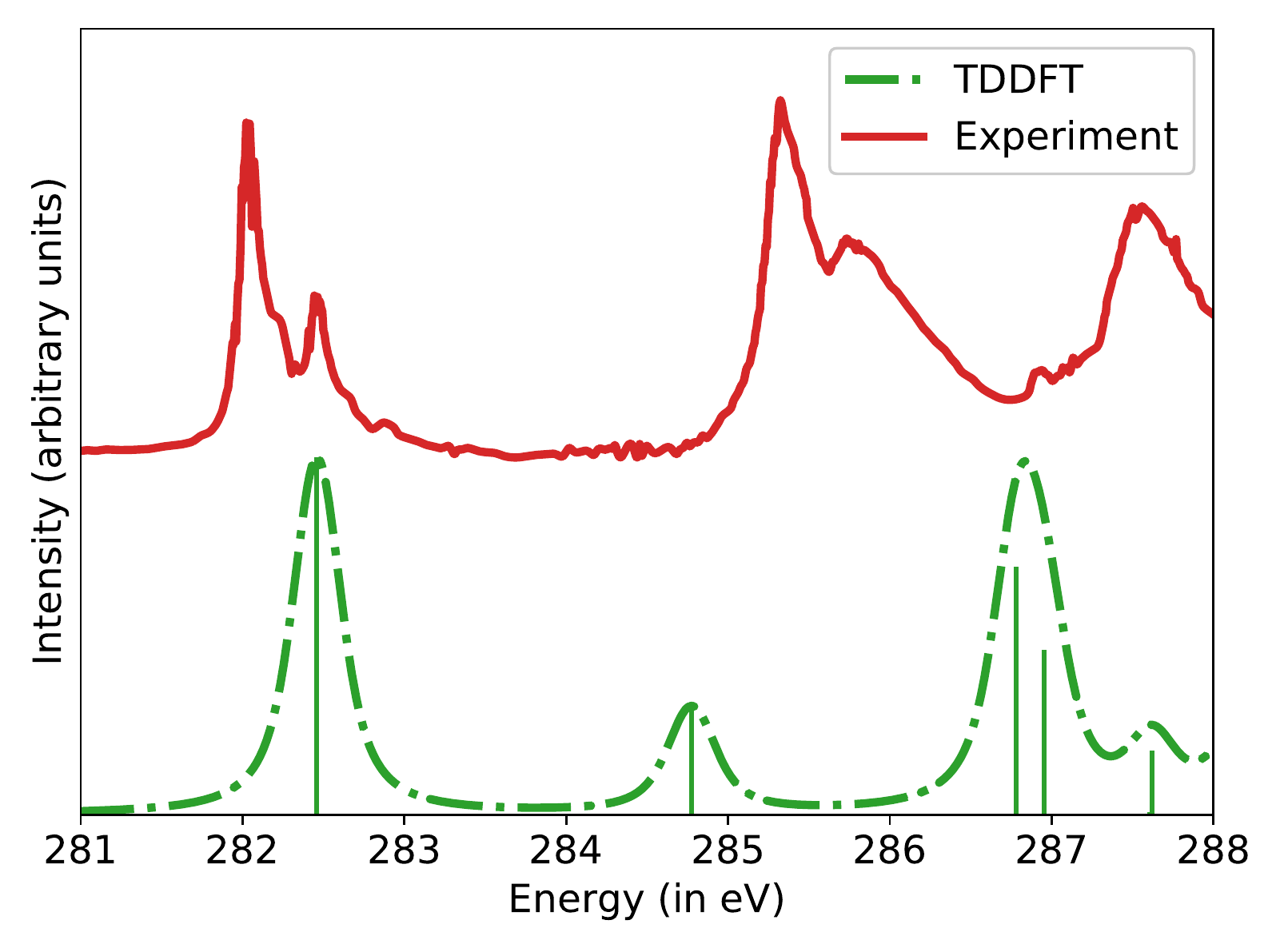}
			\subcaption{TDDFT/SRC2-R1}
			\label{fig:allyltddft}
		\end{minipage}
		\begin{minipage}{0.48\textwidth}
			\centering
			\includegraphics[width=\linewidth]{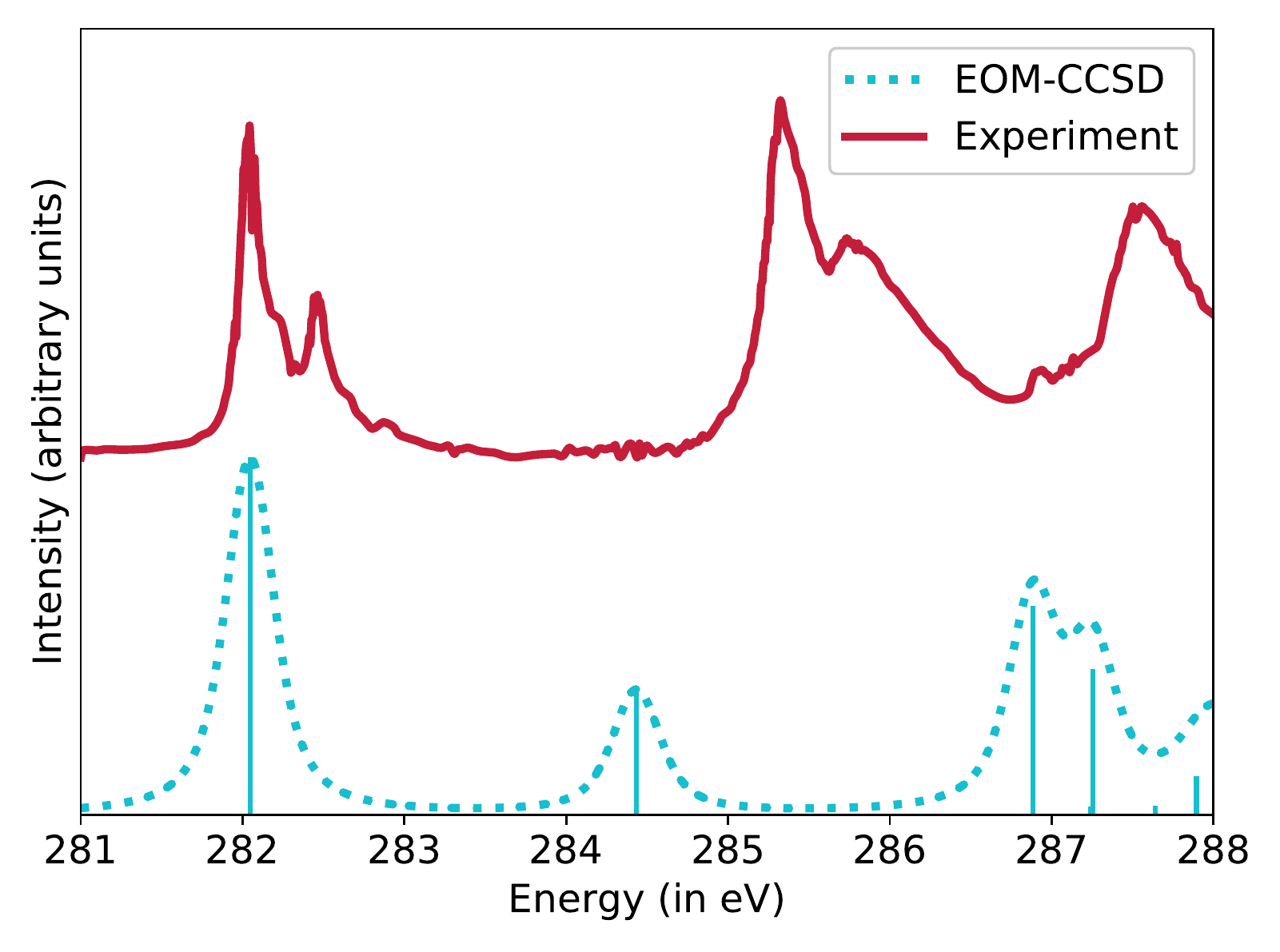}
			\subcaption{fc-CVS-EOM-CCSD\cite{vidal2019new}}
			\label{fig:allylccsd}
		\end{minipage}
		\caption{Comparison of experimental C K-edge spectrum of the allyl radical (obtained from Ref \onlinecite{alagia2013soft}) with those computed with DFT/aug-cc-pCVTZ and fc-CVS-EOM-CCSD/aug-cc-pCVTZ. The SRC2-R1 functional was employed for the TDDFT spectrum, while SCAN was utilized for both the recoupled and mixed configuration approaches. 
			A Voigt profile with a Gaussian standard deviation of 0.1 eV and Lorentzian $\gamma=0.121$ eV was utilized for broadening the computed spectra. Bars are supplied to denote the location of the predicted excitation energies.}
		\label{fig:allyl}
	\end{figure}
	
	Having explored the utility of $\Delta$SCF in predicting excitation energies to the SOMO, we next seek to investigate the utility of the theory described in Secs \ref{multi} and \ref{tmu} at predicting the full core-excitation spectrum. The recoupling approach described therein is expected to be most effective for excitations to unoccupied valence orbitals, as then all three unpaired spins (in the core, SOMO and valence excited levels) will be interacting strongly. The scarcity of experimental spectra to compare against is again a problem, and restricts us to only a few data points. Fortunately, the allyl radical has an experimentally characterized spectrum\cite{alagia2013soft} that is dominated by excitations to the unoccupied $\pi^*$ LUMO orbital, making it an excellent example for determining the utility of our recoupling approach, relative to simply using mixed configurations alone. 
	
	Fig \ref{fig:allyl} compares the performance of the orbital optimized methods in reproducing the C K-edge spectrum of the allyl radical. The performance of fc-CVS-EOM-CCSD and  TDDFT with the specialized short-range corrected SRC2-R1\cite{besley2009time} functional is also considered. All three DFT methods are reasonable at predicting the lowest energy allowed excitation (from the terminal C atoms to the SOMO, the corresponding transition from the central C atom being symmetry forbidden), though all systematically overestimate by approximately 0.5 eV, resulting in the computed peak aligning with the vibrational fine structure of the experimental band. This is potentially indicative of some multireference character of this excited state, though it is difficult to draw firm conclusions from density functional data alone (especially since it is possible to get better agreement via a functional that systematically underestimates 1s$\to$SOMO excitation energies, like PBE0). It is however worth noting that fc-CVS-EOM-CCSD is spot on for this excitation, without any need for empirical translation of spectrum (as can be seen from Fig \ref{fig:allylccsd}).
	
	Fig \ref{fig:allyltddft} lays bare the the failure of TDDFT at predicting excitations to the LUMO, as the peak positions are completely off. This is not a pecularity of the SRC2-R1 functional but rather a failure of the TDDFT family of methods, as translated TDDFT spectra from other functionals yield a similarly poor picture (as shown in the Supporting Information). Fig \ref{fig:allylccsd} also shows that fc-CVS-EOM-CCSD is unable to yield a qualitatively better spectrum than TDDFT, further highlighting the inadequacies of linear-response methods for this system. It is somewhat interesting that the inclusion of double excitations in fc-CVS-EOM-CCSD did not lead to any significant improvement over TDDFT (which is restricted to single excitations alone). The qualitative failure of both linear-response methods is likely a consequence of both spin-contamination and lack of orbital relaxation. Explicit inclusion of triple excitations should ameliorate both issues but the significant computational expense of full EOM-CCSDT would dramatically constrain practical use.

	The SCAN based orbital optimized approaches fare better, with both spin-contaminated mixed determinants and the recoupling approach yielding roughly qualitatively correct behavior. However, Fig \ref{fig:allylmix} shows that the mixed determinant approach fails to accurately predict the energy of the higher energy central C to LUMO transition, underestimating it by an eV. This substantially damages the quality of the predicted spectrum, by making this peak appear in an area where none are present experimentally.

	\begin{table}[htb!]
		\begin{tabular}{lllllll}
			Bright Transitions    & Experiment & MCSCF & Recoupled & Mixed & TDDFT &EOM-CCSD\\
			&  &  & SCAN & SCAN & SRC2-R1&\\
			C$_T\to$SOMO & 282.0      & 281.9 & 282.5     & 282.5 & 282.5 & 282.0\\
			C$_C\to$LUMO & 285.3      & 285.7 & 285.2     & 285.1 & 284.8 & 284.4\\
			C$_T\to$LUMO & 285.7      & 285.9 & 285.8     & 285.7 & 287.0 & 287.3\\
			C$_C\to$LUMO & 287.5      & 288.3 & 287.5     & 286.5 & 286.8& 286.9
		\end{tabular}
		\caption{Comparison of experimentally observed excitation energies (in eV) in the allyl core absorption spectrum with theoretical methods. The experimental values and MCSCF numbers were obtained from Ref \onlinecite{alagia2013soft}. C$_T$ stands for terminal carbon, while C$_C$ is central carbon.}.
		\label{tab:allyl}
	\end{table}
	
	The recoupling approach shifts this peak to the appropriate location and predicts a spectrum in excellent agreement with experiment (as can be seen from Fig \ref{fig:allylrec}). Indeed, Table \ref{tab:allyl} shows that the peaks predicted by recoupled SCAN agree better with experiment than MCSCF calculations reported in Ref \onlinecite{alagia2013soft} (though not too much should be inferred from this single data point). This good performance is not unique to SCAN alone, as several other functionals yield similar spectra in both the recoupled and mixed regimes (as shown in the Supporting Information). Specifically, we find that recoupled cam-B3LYP, PBE0 and TPSS give good predictions for the 1s$\to$ LUMO portion of the spectrum, while BLYP and PBE yield rather poor performance even after recoupling. SCAN and cam-B3LYP appear to give the best performance, while some of the higher energy peaks with PBE0 and TPSS are somewhat redshifted with respect to the experimental spectrum. This supports our decision of selecting SCAN as the principal functional for the manuscript, despite BLYP and TPSS having the same computational scaling and slightly lower RMSE for excitations to SOMO (as shown in Table \ref{tab:somodata}). The poor qualitative performance by BLYP and PBE also serves as a potential warning against attempting to use GGAs for prediction of core spectra, despite BLYP's excellent behavior for excitations to SOMO. Ultimately, the recoupling scheme cannot correct for deficient physics in the mixed configuration energies and a poor choice of functional could lead to poor results. Nonetheless, it is encouraging to see that all `advanced' functionals (Rungs 3 and 4) tested yield a reasonable spectrum after recoupling. 
	
	Overall, this example seems to suggest that orbital optimized approaches have an edge over TDDFT/EOM-CCSD when it comes to predicting core-excitation spectra of radicals. Furthermore, recoupling spin-contaminated mixed configurations to yield approximate doublets appears to not degrade performance and leads to some improvements.  The overall accuracy of recoupled SCAN at predicting the spectrum of allyl certainly appears to hint at the efficacy of using this approach for XAS studies of large carbon based polyradical systems, such as ones that might arise in soot formation during combustion\cite{johansson2018resonance}. 
	
	\subsection{O K-edge spectrum of CO$^+$}

	\begin{figure}[htb!]
		\begin{minipage}{0.48\textwidth}
			\centering
			\includegraphics[width=\linewidth]{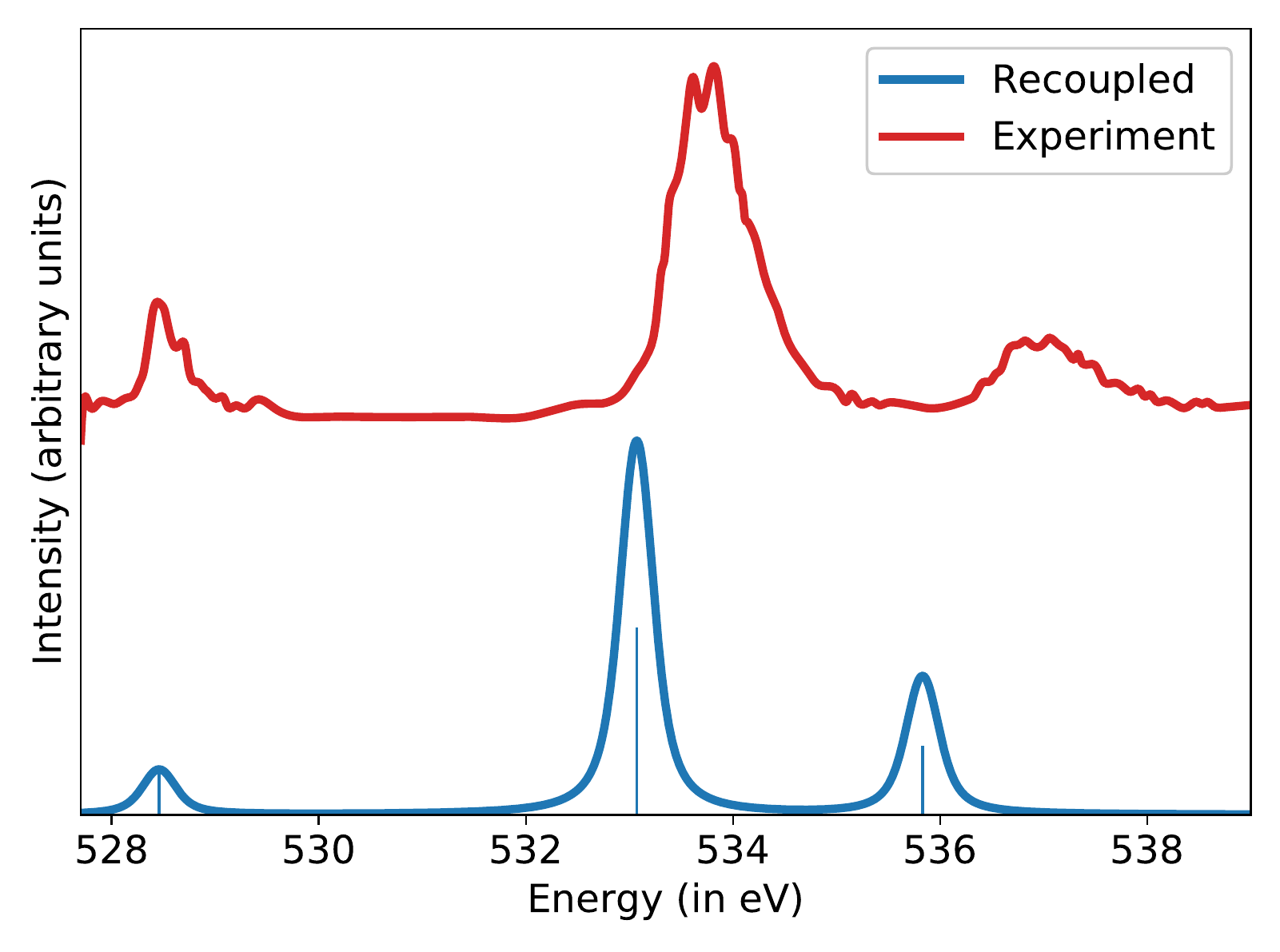}
			\subcaption{Recoupled configurations/SCAN}
			\label{fig:coprec}
		\end{minipage}
		\begin{minipage}{0.48\textwidth}
			\centering
			\includegraphics[width=\linewidth]{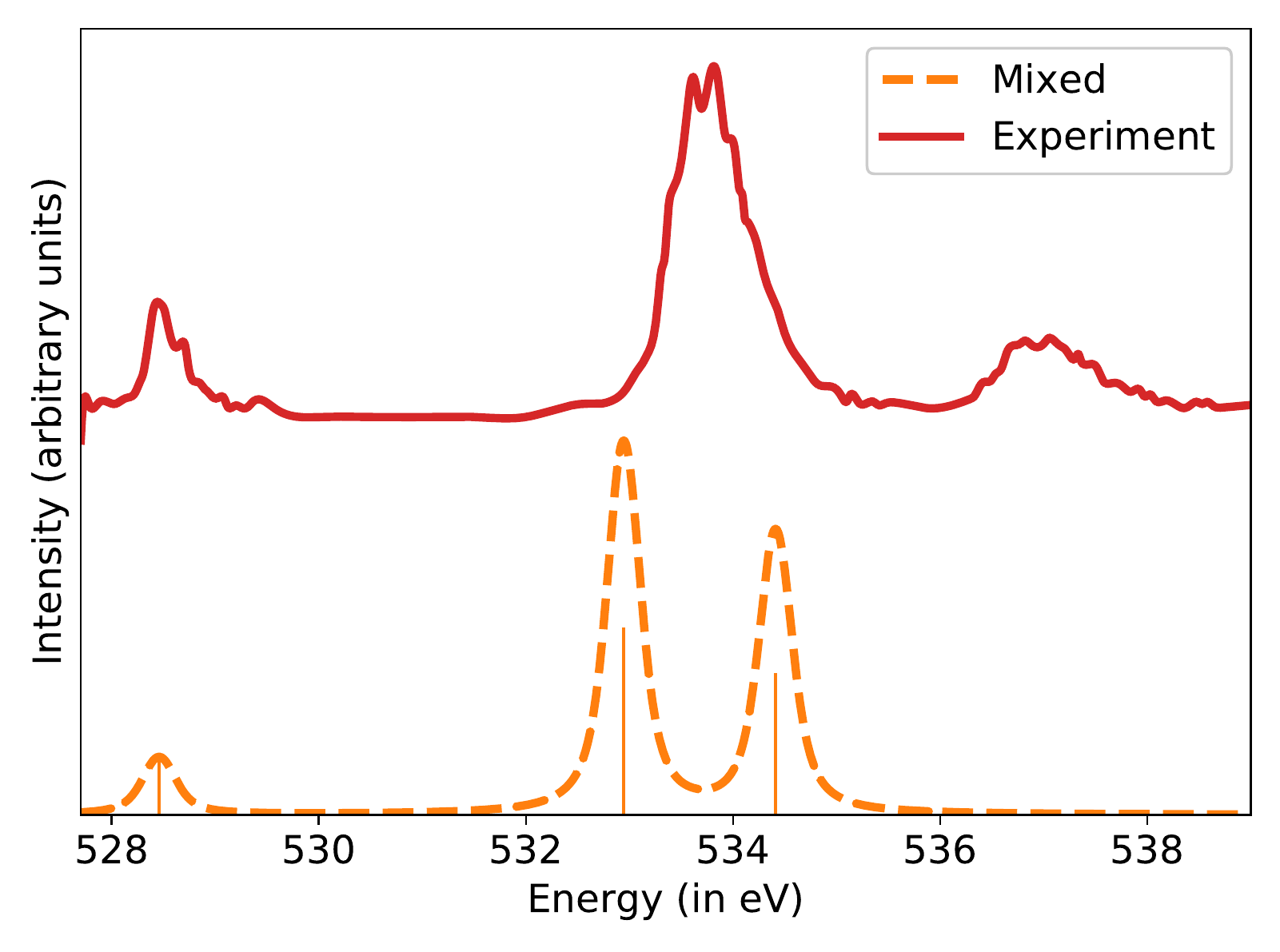}
			\subcaption{Mixed Configurations/SCAN}
			\label{fig:copmix}
		\end{minipage}
		\begin{minipage}{0.48\textwidth}
			\centering
			\includegraphics[width=\linewidth]{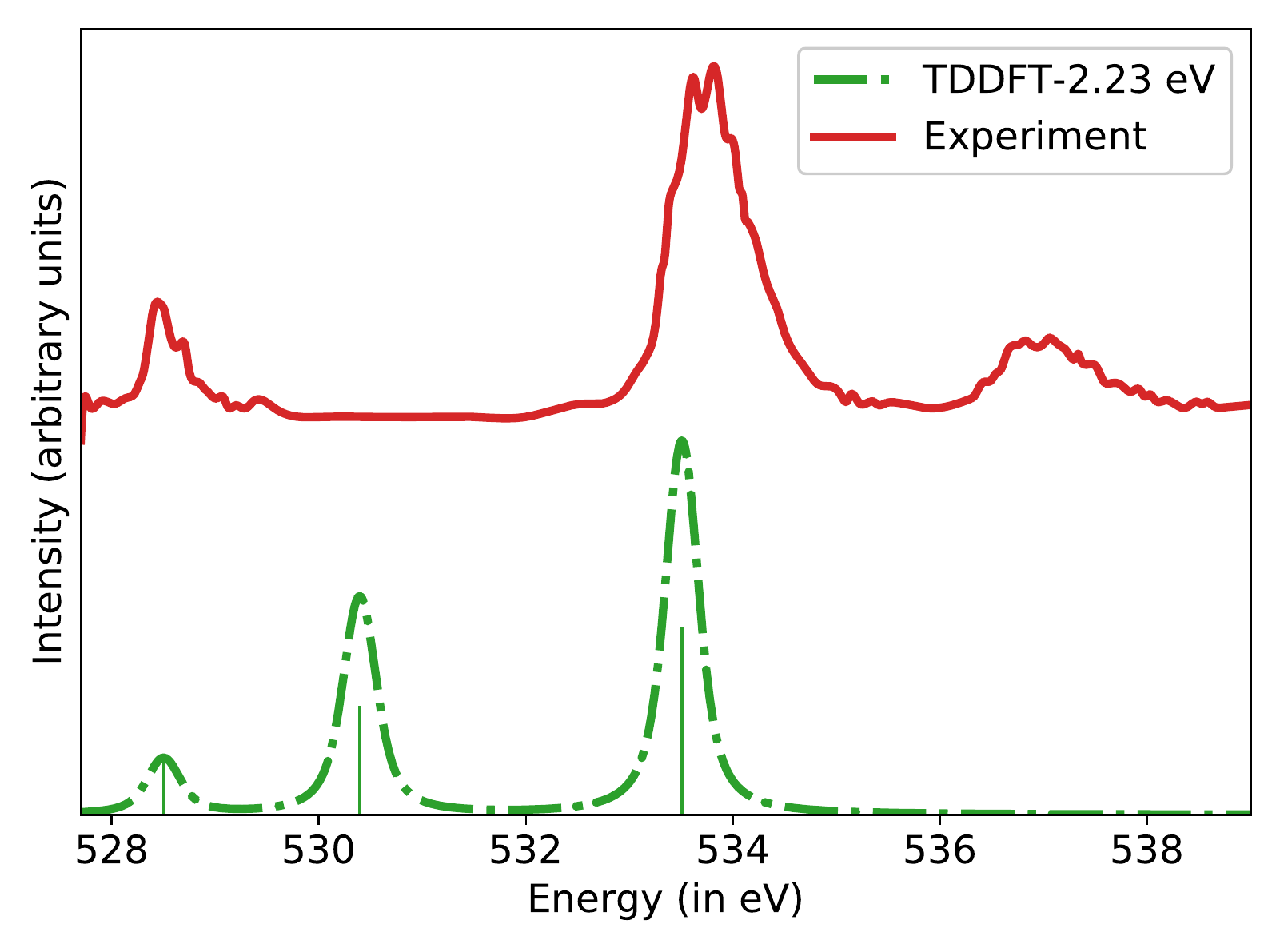}
			\subcaption{TDDFT/SRC2-R1}
			\label{fig:coptddft}
		\end{minipage}
		\begin{minipage}{0.48\textwidth}
			\centering
			\includegraphics[width=\linewidth]{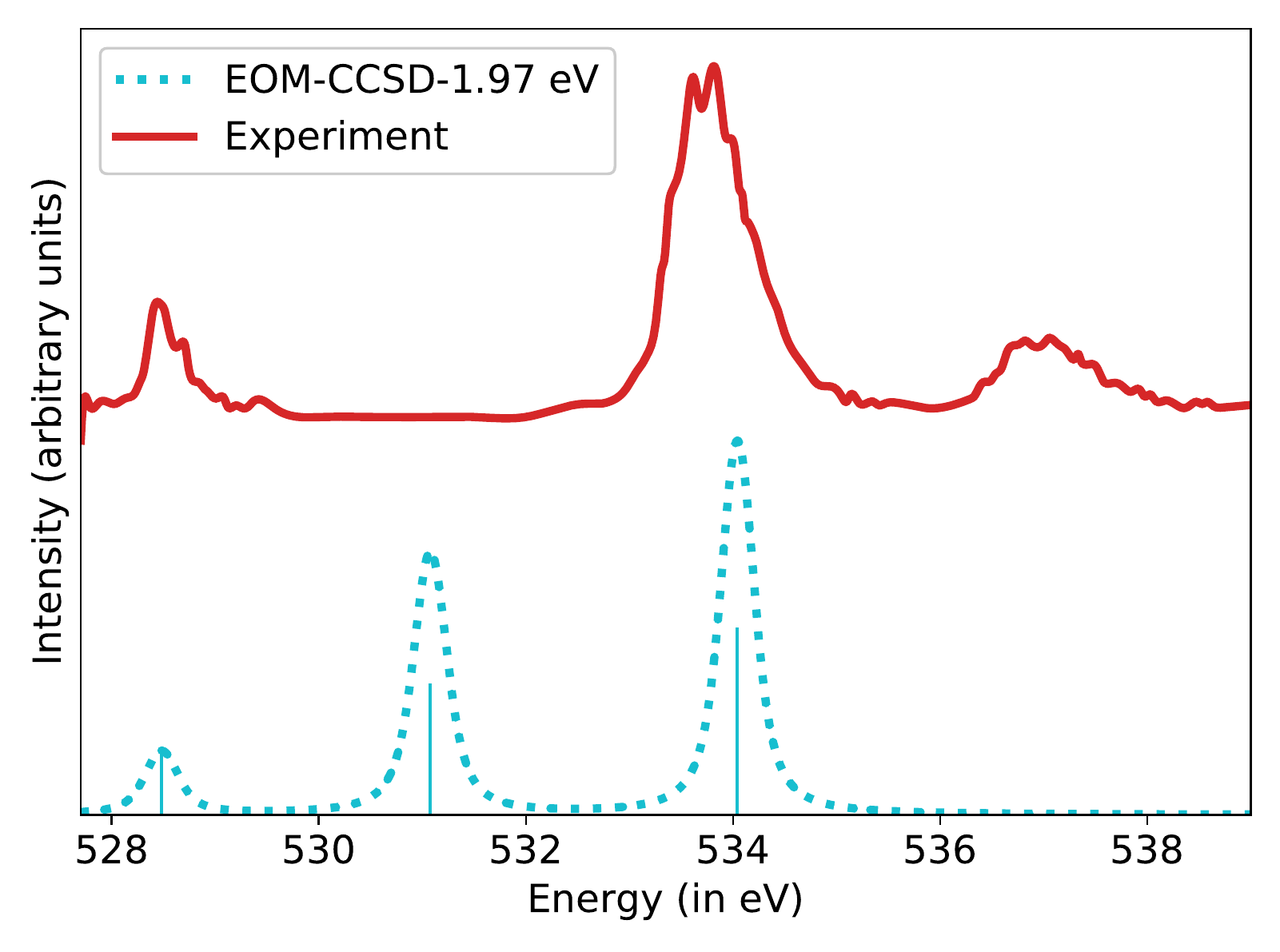}
			\subcaption{fc-CVS-EOM-CCSD\cite{vidal2019new}}
			\label{fig:copccsd}
		\end{minipage}
		\caption{Comparison of experimental O K-edge spectrum of CO$^+$ (obtained from Ref \onlinecite{couto2020carbon}) with those computed with DFT/aug-cc-pCVTZ and translated fc-CVS-EOM-CCSD/aug-cc-pCVTZ.
			A Voigt profile with a Gaussian standard deviation of 0.1 eV and Lorentzian $\gamma=0.121$ eV was utilized for broadening the computed spectra. Bars are supplied to denote the location of the predicted excitation energies.}
		\label{fig:cop}
	\end{figure}

	We next consider the rather challenging case of the CO$^+$ radical cation, whose experimental spectrum has been characterized very recently\cite{couto2020carbon}. We focus on the O K-edge as the two doublet states corresponding to the 1s$\to$LUMO excitation are experimentally well resolved, unlike the C K-edge (where vibrational fine structure of the lower energy excitation overlaps with the higher energy one). 
	
	Fig \ref{fig:cop} presents the orbital optimized SCAN spectrum (both recoupled and mixed), along with those from translated TDDFT and fc-CVS-EOM-CCSD. There are three peaks in all cases: the 1s$\to$ SOMO excitation (lowest in energy) and the two doublets arising from 1s$\to$LUMO excitations. We observe that the linear-response approaches yield a fairly poor picture. Both TD-SRC2-R1 (Fig \ref{fig:coptddft}) and EOM-CCSD (Fig \ref{fig:copccsd}) need to be redshifted by $\sim$ 2 eV to align the 1s $\to$ SOMO peak with experiment (vs the orbital optimized DFT spectra, which needs no such translation). The translated spectra are nonetheless greatly compressed relative to experiment and the relative intensities of the two 1s$\to$ LUMO peaks are incorrect.
	This is not merely a consequence of spin-contamination, as Fig \ref{fig:copmix} shows that SCAN using mixed configurations does better at reproducing the overall shape of the spectrum, despite having quartet contamination as well. Lack of orbital relaxation thus appears to be the critical factor that compromises the performance of TDDFT and EOM-CCSD for this system.

	Fig \ref{fig:copmix} however also shows that SCAN with mixed configurations has too small a spacing between the two 1s$\to$LUMO doublets (the two highest energy peaks). Fig \ref{fig:coprec} demonstrates that recoupling fixes this problem (and correctly reduces the intensity of the highest energy peak), yielding a spectrum is in decent agreement with experiment. The spacing between the two highest energy peaks remains somewhat small (2.8 eV) vs experiment ($\sim$3.4 eV but the unresolved broadness of the experimental second peak makes this hard to pinpoint). Other DFT functionals similarly underestimate this splitting (to varying extents), while reproducing the general shape of the spectrum (as can be seen from the Supporting Information). Nonetheless, it is undeniable that the spectrum quality is greatly improved by recoupling. We also note that the NOCIS method\cite{oosterbaan2018non,oosterbaan2019non,oosterbaan2020generalized} (which performs linear-response atop orbitals relaxed for the core-ionized state and is spin-pure in a manner analogous to XCIS\cite{maurice1996nature}) yields spectra in excellent agreement with experiment (as shown in the Supporting Information), further demonstrating the utility of orbital relaxation and configuration recoupling, in an unambiguous, wave function based manner. At any rate, the qualitative failure of TDDFT and EOM-CCSD seems to argue for the use of methods with explicit orbital relaxation and configuration recoupling (like the scheme presented here or NOCIS) for the computation of core-level spectra of open-shell systems, irrespective of whether the computed spectra is translated or not. 
	

	\subsection{N K-edge spectrum of NO$_2$}
	
	\begin{figure}[htb!]
		\begin{minipage}{0.48\textwidth}
			\centering
			\includegraphics[width=\linewidth]{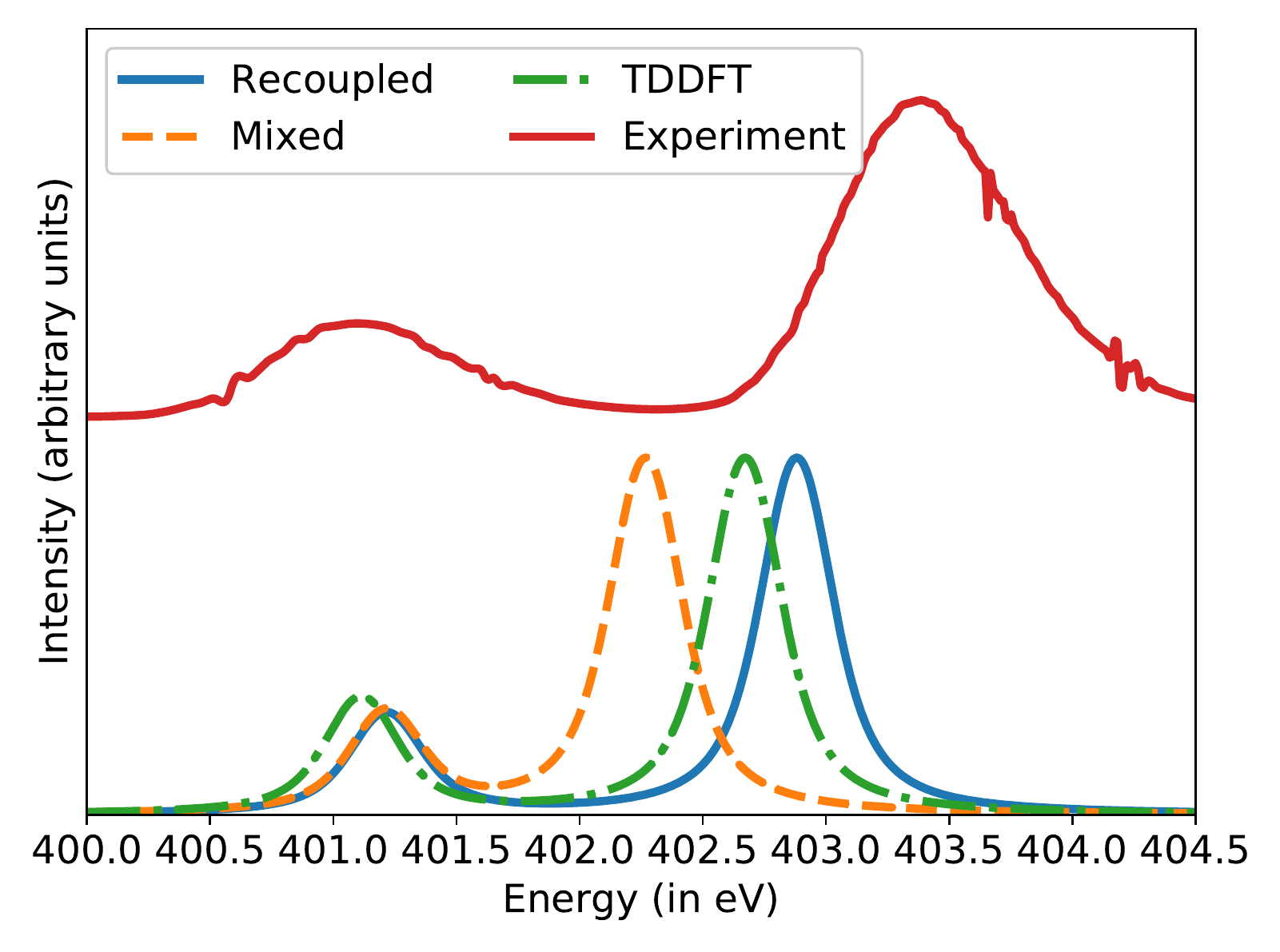}
			\subcaption{Valence excitations.}
			\label{fig:nkval}
		\end{minipage}
		\begin{minipage}{0.48\textwidth}
			\centering
			\includegraphics[width=\linewidth]{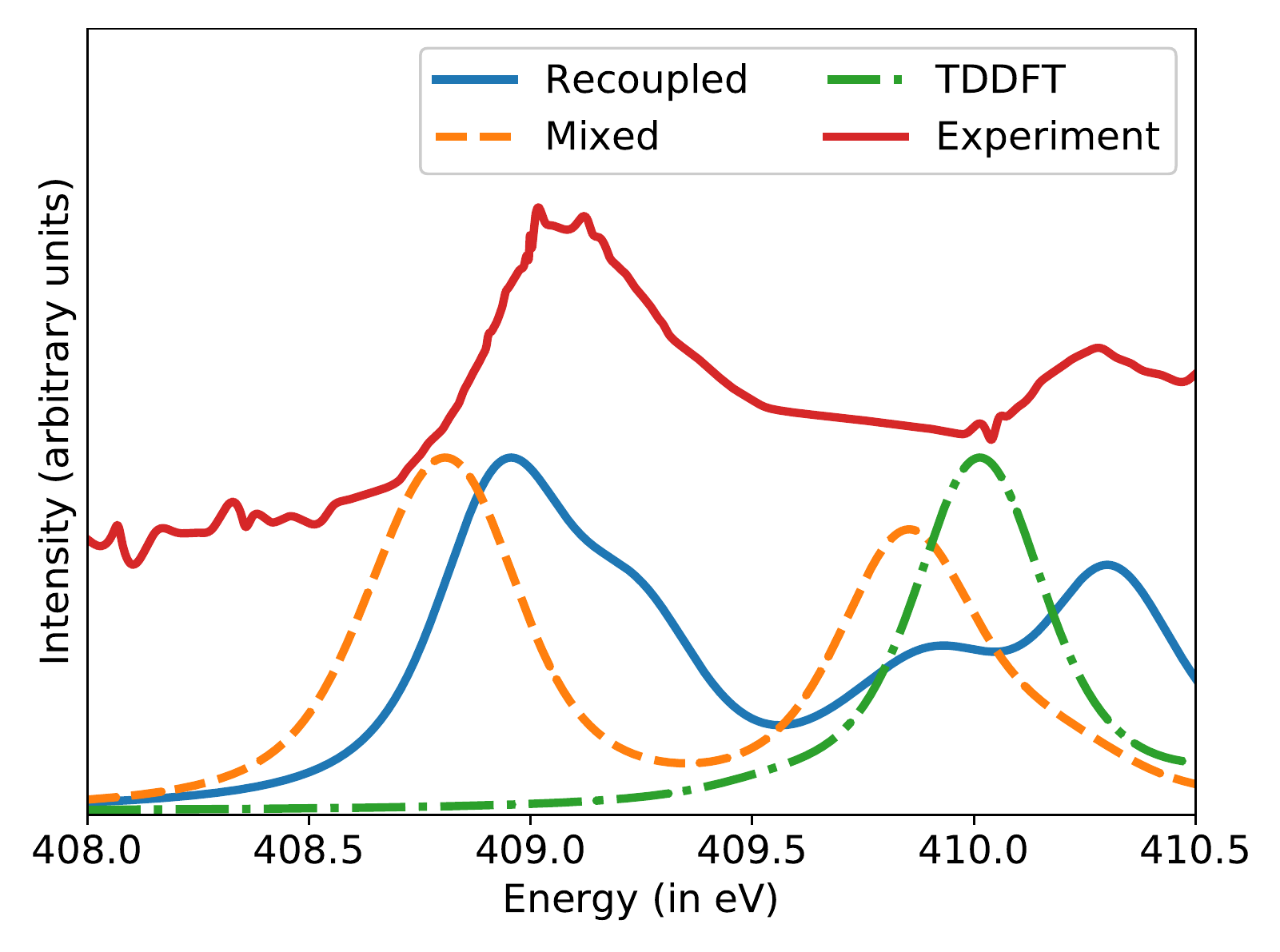}
			\subcaption{Rydberg excitations.}
			\label{fig:nkryd}
		\end{minipage}
		\caption{Comparison of experimental N K-edge spectrum of NO$_2$ (obtained from Ref \onlinecite{zhang1990inner}) with those computed with DFT/d-aug-cc-pCVTZ\cite{dunning1989gaussian,kendall1992electron,woon1995gaussian,woon1994gaussian} for both the valence (left) and Rydberg (right) regimes. The actual intensities of the Rydberg states are roughly an order of magnitude lower than that of the valence states, but have been magnified for easier comparison. The SRC2-R1 functional was employed for the TDDFT spectrum, while SCAN was utilized for both the recoupled and mixed configuration approaches.  
			A Voigt profile with a Gaussian standard deviation of 0.1 eV and Lorentzian $\gamma=0.121$ eV was utilized for broadening the computed spectra.}
		\label{fig:no2nkedge}
	\end{figure}
	
	NO$_2$ is another rare open-shell system with a known experimental high resolution core-level spectrum\cite{zhang1990inner}, by virtue of being quite stable for a radical. It is isoelectronic with allyl,  although the SOMO is not a $\pi^*$ orbital (but is rather a $\sigma$ orbital mostly localized on N). The spectrum is nonetheless dominated by the transitions to the SOMO and the $\pi^*$ LUMO levels. However, some Rydberg states have also been characterized, indicating that it could serve as an example to demonstrate whether our approach is balanced at predicting both valence and Rydberg excitations simultaneously. 
	
	Fig \ref{fig:no2nkedge} compares the experimental spectrum at the N K-edge with those predicted via DFT (employing the doubly augmented d-aug-cc-pCVTZ basis to properly converge Rydberg states). The valence regime spectrum in Fig \ref{fig:nkval} shows that all methods get the qualitative form right, though the 1s to $\pi^*$ LUMO transition is somewhat redshifted by all methods. The success of TDDFT here stands in contrast to the failure observed for the valence regime of the allyl radical, although the different symmetry of the SOMO ($\sigma$ vs $\pi^*$) may contribute to this. Recoupled SCAN performs better than mixed configuration SCAN for the second excitation by removing the quartet contribution to the energy. This blueshifts the 402.3 eV excitation energy predicted by the mixed configuration approach to 402.9 eV, which is much closer to the experimentally observed peak at 403.3 eV. This disagreement is not particularly small (and is in the opposite direction to the systematic overestimation exhibited by SCAN for excitations to the SOMO), but the recoupled DFT method gives best agreement with experiment. 
	
	The Rydberg regime depicted in Fig \ref{fig:nkryd} however shows somewhat surprising behavior. It was tempting to believe that the weak coupling between the excited electron and the other unpaired electrons would lead to good performance by all methods. However, TDDFT absolutely fails to reproduce the spectrum in this regime, significantly blueshifting the experimental peak at 408.9 eV to 410.0 eV. On the other hand, the mixed configurations are quartet contaminated, and are thus slightly redshifted from their optimal location. Our recoupling protocol eliminates this problem, giving excellent agreement with experiment. It is also worth noting that the recoupled approach appears to predict the shape of the curve better than individual mixed configurations, indicating that the protocol described in Sec \ref{tmu} was reasonably effective. This is however ultimately only one data point, and comparison against more high resolution experimental spectra would be useful in validating our observation. We therefore hope that spectra of more open-shell species in the Rydberg regime will be available in the near future. We note that high energy spectra for N$_2^+$ and CO$^+$ have been very recently reported\cite{lindblad2020x,couto2020carbon}, but the Rydberg region appears to also contain a large number of doubly excited states with significant multiconfigurational character (involving more than three orbitals), that DFT based methods are unlikely to successfully model. This is less likely to be the case for neutral species.
	
	\section{Recommendations for successful calculations}\label{sec:recs}
	The proposed protocol for recoupling mixed configurations appears to yield improved agreement with experiment, relative to simply using the two individual mixed configurations that correspond to single excitations. Nonetheless, it entails individual optimization of four configurations per excitation ($\ket{Q,M_{1,2,3}}$), to get two doublet state energies. We subsequently recommend the following protocol for ensuring maximum agreement between these configurations and minimizing computational cost. 
	
	\begin{enumerate}
		\item Optimize unrestricted KS ground state orbitals.
		\item Use these orbitals as initial guesses to optimize RO orbitals for the ground state. 
		\item Using the RO ground state orbitals as the initial guess, optimize the RO orbitals for the core-ionized state via SGM. This decouples the relaxation of the core-hole from the rest of the computations. 
		\item Using the RO core-ionized orbitals as the initial guess, optimize RO orbitals corresponding to the desired quartet state with SGM. The core-ionized orbitals can thus be computed only once, and repeatedly utilized for multiple excitations. Furthermore, the unoccupied orbitals for the core ionized state are much more representative of the optimized orbitals for the excited electron, than canonical ground state orbitals.  
		\item Using the RO core-excited quartet orbitals as initial guesses, find the unrestricted orbitals for the quartet $\ket{Q}$ and mixed configurations $\ket{M_{1,2,3}}$ with SGM. 
	\end{enumerate}
	Steps 1-3 also apply for excitations to the SOMO level, followed by use of the RO core ionized orbitals to initialize the excited state optimization for the core to SOMO excited configuration. They also apply for computation of core-excitations in closed-shell species via ROKS.
	We believe that the RO energies themselves are not particularly useful for radicals, but the RO orbitals act as useful intermediates to prevent the alpha and beta spatial orbitals from differing prior to the last optimization step (step 5). The RO orbital space in fact is much more tightly constrained and SGM is faster at those optimizations in practice. Difficult convergence cases in general could also be addressed via converging to the same state with a different (ideally, cheaper) functional and using the resulting orbitals as initial guesses.
	
	Three additional points regarding orbital optimized core-excitation calculations in general (for both closed and open-shell systems) are worth noting as well. 
	\begin{enumerate}
		\item Use of a localized core-hole is absolutely critical for systems where there are symmetry equivalent atoms (like the terminal carbons of allyl). Delocalized core-holes lead to substantial delocalization error\cite{perdew1982density,hait2018delocalization} driven underestimation of energy, as shown in Ref \onlinecite{hait2020highly}. Localization of core orbitals can be achieved via explict localization, or via weak electric fields that break symmetry. The mixed basis strategy described in the next point also leads to core orbital localizing symmetry breaking.
		\item It is absolutely essential to use at least a triple zeta level basis with split core functions (like cc-pCVTZ) at the local site of the core-excitation. The core-hole would otherwise not be able to adequately relax, and energies be systematically overestimated\cite{hait2020highly}. However, a smaller basis can be used for all other atoms, with cc-pVDZ being adequate in our experience\cite{hait2020highly} (though even smaller bases could potentially be fine). This mixed basis strategy helps bring down the computational cost considerably as well, as the overall computation cost is comparable to a double zeta basis DFT ground state calculation per iteration, though excited state orbital optimization does often require many more iterations than ground state computations.
		\item Many core-excited states possess significant Rydberg character. A good description of these states necessitates the presence of diffuse functions in the basis, and even double augmentation is sometimes necessary (such as the NO$_2$ spectrum presented in Fig \ref{fig:no2nkedge}, where singly augmented aug-cc-pCVTZ blueshifts the Rydberg peaks in Fig \ref{fig:nkryd} by 0.2 eV). This is easily the most onerous basis set requirement for such calculations but is functionally unavoidable for any electronic structure method seeking a correct description of Rydberg states. 
	\end{enumerate}
	
	\section{Conclusion}
	We have investigated orbital optimized density functional approaches to studying core-excitation spectra of open-shell systems, by employing the SGM approach for averting variational collapse. Lack of gas-phase experimental data proves to be a hindrance for assessing the performance of these methods, but existing data shows encouraging behavior. We firstly find that several density functionals like SCAN, TPSS, BLYP, B3LYP, cam-B3LYP and $\omega$B97X-D3 can be employed to predict excitation energies corresponding to 1s to SOMO transitions in radicals, to RMSE at or below 0.3 eV. The 1s$\to$ SOMO transitions are however not very challenging excitations as they do not result in a change in the total number of unpaired electrons and thus can be well approximated by single Slater determinants.
	
	Higher excitations entail breaking of electron pairs and thus are natively multiconfigurational. These states therefore cannot be described by single determinants, although somewhat reliable results can at times be obtained from symmetry broken mixed determinants in the limit of weak coupling between unpaired spins (analogous to how unrestricted HF/DFT being effective for single bond dissociations in closed-shell species). 
	For more general accuracy, we present a CI inspired approach for self-consistently recoupling these single determinant mixed configurations with unpaired spins to yield approximately spin-pure results corresponding to multiconfigurational doublet states.  The performance of this approach is compared against that of using unrecoupled mixed determinants alone and TDDFT/fc-CVS-EOM-CCSD for the core-level spectra of the allyl radical and CO$^+$ at the O K-edge. The N K-edge spectrum for NO$_2$ is also studied with both orbital optimized DFT and TDDFT. We find that the recoupling scheme leads to no degradation of performance and in fact consistently improves upon results obtained by merely using single mixed determinants (significantly so for the O K-edge of CO$^+$). It is nonetheless worth appreciating that unrecoupled determinants often yield fairly reasonable answers by themselves, especially relative to TDDFT/EOM-CCSD for the allyl radical and the O K-edge of CO$^+$. Our work therefore shows promise in using orbital optimized DFT approaches for predicting core-level spectra of radicals, where high accuracy can be obtained even from local functionals like SCAN, at low computational cost. Available evidence also appears to argue for recoupling mixed configurations, although this is roughly computationally twice as expensive (as four configurations need to be optimized as opposed to only two). The O K-edge of CO$^+$ also seems to suggest that our recoupling scheme somewhat underestimates doublet-doublet splitting in the strong coupling limit. More experimental spectra for open-shell systems (involving transitions to unoccupied valence orbitals) would however be immensely useful in fully characterizing the limitations of the recoupling approach. We will consequently continue to attempt to validate this approach via comparison to experiment as new data arises. 
	
	In future, we will also seek to develop approaches that optimize a single set of unrestricted orbitals for recoupling mixed configurations vs separately optimizing all four relevant states. This should reduce the computational cost of such calculations substantially, and enhance their utility. It would also be useful to generalize the recoupling approach to higher spin states like triplets, where there are more spins to recouple and a correspondingly larger number of coupling constants. Work along these directions is presently in progress.

	\section{Computational Methods}
	All calculations were performed with the Q-Chem 5.3 \cite{QCHEM4} package. 	Local exchange-correlation integrals were calculated over
	a radial grid with 99 points and an angular Lebedev grid with 590 points. Experimental geometries (from the NIST database\cite{johnson2015nist}) were used whenever possible, with MP2\cite{cremer2011moller}/cc-pVTZ\cite{dunning1989gaussian} optimized geometries being employed in their absence. The plots labeled `mixed' only used the two mixed configurations corresponding to single excitations from the ground state, as the third configuration is technically a double excitation that would not usually be considered due to formally zero (and in practice, typically small) oscillator strength. All TDDFT calculations employed the Tamm-Dancoff Approximation\cite{dreuw2005single,tamm1991relativistic,dancoff1950non,hirata1999time}. 
	\section{Acknowledgements}
	D.H., K.J.O. and M.H.-G. were supported by Director, Office of Science, Office of Basic Energy Sciences, of the U.S. Department of Energy under Contract No. DE-AC02-05CH11231, through the Atomic,
	Molecular, and Optical Sciences Program of the Chemical Sciences Division of Lawrence Berkeley National
	Laboratory. E.A.H., Z.Y. and S.R.L were funded by the Gas Phase Chemical Physics program, which operates
	under the same DOE-OS-BES contract number. We also thank Marta L. Vidal for helpful discussions regarding EOM-CCSD calculations.
	\section{Supporting Information}
	\noindent Additional spectra for the allyl radical and CO$^+$ (PDF)
	\newline Raw data (XLXS)
	\newline Molecular geometries (ZIP)
	\section{Data Availability}
	The data that supports the findings of this study are available within the article and its supplementary material.
	
	\appendix
	\section{Phase Estimation for Mixed Configurations}
	The phase convention chosen for $\ket{M_{1,2,3}}$ in Sec \ref{multi} ensures that the off-diagonal elements of the coupling matrix are $-J_{ij}$, where the couplings $J_{ij}$ are given by:
	\begin{align}
	J_{12}=\dfrac{E_{M_1}+E_{M_2}-E_Q-E_{M_3}}{2}\\
	J_{13}=\dfrac{E_{M_1}+E_{M_3}-E_Q-E_{M_2}}{2}\\
	J_{23}=\dfrac{E_{M_2}+E_{M_3}-E_Q-E_{M_1}}{2}
	\end{align}
	
	However, the determinants $\ket{M'_i}$ obtained from orbital optimization can differ from this ideal phase. Specifically, DFT can yield $\ket{M'_i}=p_i\ket{M_i}$ where $p_i=\pm 1$. This has no implication for the energies, but will affect properties like the transition dipole moment for which the relative phases of the configurations matter (as these properties depend on off-diagonal elements, and are computed from $\ket{M'_i}$ vs the idealized $\ket{M_i}$). 
	
	The easiest route for phase finding seems to be via the quartet state, which has the eigenvector $\begin{pmatrix}\dfrac{1}{\sqrt{3}}&\dfrac{1}{\sqrt{3}}&\dfrac{1}{\sqrt{3}}\end{pmatrix}^T$ in the $\ket{M_i}$ basis. This state should formally have zero transition dipole moment and thus could be employed to compute relative phases. 
	
	Specifically, let $\ket{M'_{i}}$ have transition dipole moments $\vec{\mu}_{i}=\bra{0}\hat{\vec{\mu}}\ket{M'_{i}}$ against the ground state determinant $\ket{0}$. Without loss of generality, we can set the phase $p_1$ of $\ket{M'_1}$ to $1$ (as only relative phases matter). Then the transition dipole moment of the ostensibly quartet state is $\vec{\mu}_Q=\dfrac{\vec{\mu}_1+p_2\vec{\mu}_2+p_3\vec{\mu}_3}{\sqrt{3}}$. Consequently, the signs of $p_{2,3}$ should be chosen to minimize this quantity. In practice, this protocol is often simplified on account of one of the three mixed determinants being a formal double excitation ($\ket{M_2}$ if the orbitals are ordered by energy), which would have typically have very low transition dipole moment (though generally not exactly zero on account of non-orthogonality between the ground and mixed determinant orbitals). The phase estimation problem here is thus often just finding whether $p_3$ (say) should be $1$ or $-1$ for $\vec{\mu}_Q$ to be smallest. 
	
	In fact, this is essentially an internal consistency check for determining the impact of neglecting non-orthogonality between mixed determinants and the overall quality of the optimized orbitals, as this ``quartet" transition dipole moment $\vec{\mu}_Q$ should be at least an order of magnitude smaller (and hopefully even less) than the largest transition dipole moment corresponding the two doublet states, after finding optimal phases. The oscillator strength scales as square of the transition dipole and thus any spurrious ``quartet" peak stemming from neglect of non-orthogonality etc. would be at least a hundred times weaker than the strongest doublet peak and thus the quality of the spectrum will be preserved.
	
	As an example, let us consider the N 1s$\to\pi^*$ transition in NO$_2$. The orbital optimized determinants $\ket{M'_{1,2,3}}$ we obtained had transition dipole moments (after ignoring terms smaller than $10^{-4}$). :
	\begin{align}
	\vec{\mu}_1&=-6.11\times 10^{-2}\hat{x}\\
	\vec{\mu}_2&=0\\
	\vec{\mu}_3&=5.98\times 10^{-2}	\hat{x}
	\end{align}
	$\vec{\mu}_Q$ is minimized if $p_3=1$, as then the dipoles will mostly cancel each other. 
	
	\bibliography{references}

\begin{thebibliography}{119}%
\makeatletter
\providecommand \@ifxundefined [1]{%
 \@ifx{#1\undefined}
}%
\providecommand \@ifnum [1]{%
 \ifnum #1\expandafter \@firstoftwo
 \else \expandafter \@secondoftwo
 \fi
}%
\providecommand \@ifx [1]{%
 \ifx #1\expandafter \@firstoftwo
 \else \expandafter \@secondoftwo
 \fi
}%
\providecommand \natexlab [1]{#1}%
\providecommand \enquote  [1]{``#1''}%
\providecommand \bibnamefont  [1]{#1}%
\providecommand \bibfnamefont [1]{#1}%
\providecommand \citenamefont [1]{#1}%
\providecommand \href@noop [0]{\@secondoftwo}%
\providecommand \href [0]{\begingroup \@sanitize@url \@href}%
\providecommand \@href[1]{\@@startlink{#1}\@@href}%
\providecommand \@@href[1]{\endgroup#1\@@endlink}%
\providecommand \@sanitize@url [0]{\catcode `\\12\catcode `\$12\catcode
  `\&12\catcode `\#12\catcode `\^12\catcode `\_12\catcode `\%12\relax}%
\providecommand \@@startlink[1]{}%
\providecommand \@@endlink[0]{}%
\providecommand \url  [0]{\begingroup\@sanitize@url \@url }%
\providecommand \@url [1]{\endgroup\@href {#1}{\urlprefix }}%
\providecommand \urlprefix  [0]{URL }%
\providecommand \Eprint [0]{\href }%
\providecommand \doibase [0]{http://dx.doi.org/}%
\providecommand \selectlanguage [0]{\@gobble}%
\providecommand \bibinfo  [0]{\@secondoftwo}%
\providecommand \bibfield  [0]{\@secondoftwo}%
\providecommand \translation [1]{[#1]}%
\providecommand \BibitemOpen [0]{}%
\providecommand \bibitemStop [0]{}%
\providecommand \bibitemNoStop [0]{.\EOS\space}%
\providecommand \EOS [0]{\spacefactor3000\relax}%
\providecommand \BibitemShut  [1]{\csname bibitem#1\endcsname}%
\let\auto@bib@innerbib\@empty
\bibitem [{\citenamefont {Runge}\ and\ \citenamefont
  {Gross}(1984)}]{runge1984density}%
  \BibitemOpen
  \bibfield  {author} {\bibinfo {author} {\bibfnamefont {E.}~\bibnamefont
  {Runge}}\ and\ \bibinfo {author} {\bibfnamefont {E.~K.~U.}\ \bibnamefont
  {Gross}},\ }\href@noop {} {\bibfield  {journal} {\bibinfo  {journal} {Phys.
  Rev. Lett.}\ }\textbf {\bibinfo {volume} {52}},\ \bibinfo {pages} {997}
  (\bibinfo {year} {1984})}\BibitemShut {NoStop}%
\bibitem [{\citenamefont {Casida}(1995)}]{casida1995time}%
  \BibitemOpen
  \bibfield  {author} {\bibinfo {author} {\bibfnamefont {M.~E.}\ \bibnamefont
  {Casida}},\ }in\ \href@noop {} {\emph {\bibinfo {booktitle} {Recent Advances
  In Density Functional Methods: (Part I)}}}\ (\bibinfo  {publisher} {World
  Scientific},\ \bibinfo {year} {1995})\ pp.\ \bibinfo {pages}
  {155--192}\BibitemShut {NoStop}%
\bibitem [{\citenamefont {Dreuw}\ and\ \citenamefont
  {Head-Gordon}(2005)}]{dreuw2005single}%
  \BibitemOpen
  \bibfield  {author} {\bibinfo {author} {\bibfnamefont {A.}~\bibnamefont
  {Dreuw}}\ and\ \bibinfo {author} {\bibfnamefont {M.}~\bibnamefont
  {Head-Gordon}},\ }\href@noop {} {\bibfield  {journal} {\bibinfo  {journal}
  {Chem. Rev.}\ }\textbf {\bibinfo {volume} {105}},\ \bibinfo {pages} {4009}
  (\bibinfo {year} {2005})}\BibitemShut {NoStop}%
\bibitem [{\citenamefont {Peach}\ \emph {et~al.}(2008)\citenamefont {Peach},
  \citenamefont {Benfield}, \citenamefont {Helgaker},\ and\ \citenamefont
  {Tozer}}]{peach2008excitation}%
  \BibitemOpen
  \bibfield  {author} {\bibinfo {author} {\bibfnamefont {M.~J.}\ \bibnamefont
  {Peach}}, \bibinfo {author} {\bibfnamefont {P.}~\bibnamefont {Benfield}},
  \bibinfo {author} {\bibfnamefont {T.}~\bibnamefont {Helgaker}}, \ and\
  \bibinfo {author} {\bibfnamefont {D.~J.}\ \bibnamefont {Tozer}},\ }\href@noop
  {} {\bibfield  {journal} {\bibinfo  {journal} {J. Chem. Phys.}\ }\textbf
  {\bibinfo {volume} {128}},\ \bibinfo {pages} {044118} (\bibinfo {year}
  {2008})}\BibitemShut {NoStop}%
\bibitem [{\citenamefont {Casida}\ \emph {et~al.}(1998)\citenamefont {Casida},
  \citenamefont {Jamorski}, \citenamefont {Casida},\ and\ \citenamefont
  {Salahub}}]{casida1998molecular}%
  \BibitemOpen
  \bibfield  {author} {\bibinfo {author} {\bibfnamefont {M.~E.}\ \bibnamefont
  {Casida}}, \bibinfo {author} {\bibfnamefont {C.}~\bibnamefont {Jamorski}},
  \bibinfo {author} {\bibfnamefont {K.~C.}\ \bibnamefont {Casida}}, \ and\
  \bibinfo {author} {\bibfnamefont {D.~R.}\ \bibnamefont {Salahub}},\
  }\href@noop {} {\bibfield  {journal} {\bibinfo  {journal} {J. Chem. Phys.}\
  }\textbf {\bibinfo {volume} {108}},\ \bibinfo {pages} {4439} (\bibinfo {year}
  {1998})}\BibitemShut {NoStop}%
\bibitem [{\citenamefont {Tozer}\ and\ \citenamefont
  {Handy}(2000)}]{tozer2000determination}%
  \BibitemOpen
  \bibfield  {author} {\bibinfo {author} {\bibfnamefont {D.~J.}\ \bibnamefont
  {Tozer}}\ and\ \bibinfo {author} {\bibfnamefont {N.~C.}\ \bibnamefont
  {Handy}},\ }\href@noop {} {\bibfield  {journal} {\bibinfo  {journal} {Phys.
  Chem. Chem. Phys.}\ }\textbf {\bibinfo {volume} {2}},\ \bibinfo {pages}
  {2117} (\bibinfo {year} {2000})}\BibitemShut {NoStop}%
\bibitem [{\citenamefont {Besley}\ and\ \citenamefont
  {Asmuruf}(2010)}]{besley2010time}%
  \BibitemOpen
  \bibfield  {author} {\bibinfo {author} {\bibfnamefont {N.~A.}\ \bibnamefont
  {Besley}}\ and\ \bibinfo {author} {\bibfnamefont {F.~A.}\ \bibnamefont
  {Asmuruf}},\ }\href@noop {} {\bibfield  {journal} {\bibinfo  {journal} {Phys.
  Chem. Chem. Phys.}\ }\textbf {\bibinfo {volume} {12}},\ \bibinfo {pages}
  {12024} (\bibinfo {year} {2010})}\BibitemShut {NoStop}%
\bibitem [{\citenamefont {Wenzel}, \citenamefont {Wormit},\ and\ \citenamefont
  {Dreuw}(2014)}]{wenzel2014calculating}%
  \BibitemOpen
  \bibfield  {author} {\bibinfo {author} {\bibfnamefont {J.}~\bibnamefont
  {Wenzel}}, \bibinfo {author} {\bibfnamefont {M.}~\bibnamefont {Wormit}}, \
  and\ \bibinfo {author} {\bibfnamefont {A.}~\bibnamefont {Dreuw}},\
  }\href@noop {} {\bibfield  {journal} {\bibinfo  {journal} {J. Chem. Theory
  Comput.}\ }\textbf {\bibinfo {volume} {10}},\ \bibinfo {pages} {4583}
  (\bibinfo {year} {2014})}\BibitemShut {NoStop}%
\bibitem [{\citenamefont {Attar}\ \emph {et~al.}(2017)\citenamefont {Attar},
  \citenamefont {Bhattacherjee}, \citenamefont {Pemmaraju}, \citenamefont
  {Schnorr}, \citenamefont {Closser}, \citenamefont {Prendergast},\ and\
  \citenamefont {Leone}}]{attar2017femtosecond}%
  \BibitemOpen
  \bibfield  {author} {\bibinfo {author} {\bibfnamefont {A.~R.}\ \bibnamefont
  {Attar}}, \bibinfo {author} {\bibfnamefont {A.}~\bibnamefont
  {Bhattacherjee}}, \bibinfo {author} {\bibfnamefont {C.}~\bibnamefont
  {Pemmaraju}}, \bibinfo {author} {\bibfnamefont {K.}~\bibnamefont {Schnorr}},
  \bibinfo {author} {\bibfnamefont {K.~D.}\ \bibnamefont {Closser}}, \bibinfo
  {author} {\bibfnamefont {D.}~\bibnamefont {Prendergast}}, \ and\ \bibinfo
  {author} {\bibfnamefont {S.~R.}\ \bibnamefont {Leone}},\ }\href@noop {}
  {\bibfield  {journal} {\bibinfo  {journal} {Science}\ }\textbf {\bibinfo
  {volume} {356}},\ \bibinfo {pages} {54} (\bibinfo {year} {2017})}\BibitemShut
  {NoStop}%
\bibitem [{\citenamefont {Bhattacherjee}\ \emph {et~al.}(2018)\citenamefont
  {Bhattacherjee}, \citenamefont {Schnorr}, \citenamefont {Oesterling},
  \citenamefont {Yang}, \citenamefont {Xue}, \citenamefont {de~Vivie-Riedle},\
  and\ \citenamefont {Leone}}]{bhattacherjee2018photoinduced}%
  \BibitemOpen
  \bibfield  {author} {\bibinfo {author} {\bibfnamefont {A.}~\bibnamefont
  {Bhattacherjee}}, \bibinfo {author} {\bibfnamefont {K.}~\bibnamefont
  {Schnorr}}, \bibinfo {author} {\bibfnamefont {S.}~\bibnamefont {Oesterling}},
  \bibinfo {author} {\bibfnamefont {Z.}~\bibnamefont {Yang}}, \bibinfo {author}
  {\bibfnamefont {T.}~\bibnamefont {Xue}}, \bibinfo {author} {\bibfnamefont
  {R.}~\bibnamefont {de~Vivie-Riedle}}, \ and\ \bibinfo {author} {\bibfnamefont
  {S.~R.}\ \bibnamefont {Leone}},\ }\href@noop {} {\bibfield  {journal}
  {\bibinfo  {journal} {J. Am. Chem. Soc.}\ }\textbf {\bibinfo {volume}
  {140}},\ \bibinfo {pages} {12538} (\bibinfo {year} {2018})}\BibitemShut
  {NoStop}%
\bibitem [{\citenamefont {Chantzis}\ \emph {et~al.}(2018)\citenamefont
  {Chantzis}, \citenamefont {Kowalska}, \citenamefont {Maganas}, \citenamefont
  {DeBeer},\ and\ \citenamefont {Neese}}]{chantzis2018ab}%
  \BibitemOpen
  \bibfield  {author} {\bibinfo {author} {\bibfnamefont {A.}~\bibnamefont
  {Chantzis}}, \bibinfo {author} {\bibfnamefont {J.~K.}\ \bibnamefont
  {Kowalska}}, \bibinfo {author} {\bibfnamefont {D.}~\bibnamefont {Maganas}},
  \bibinfo {author} {\bibfnamefont {S.}~\bibnamefont {DeBeer}}, \ and\ \bibinfo
  {author} {\bibfnamefont {F.}~\bibnamefont {Neese}},\ }\href@noop {}
  {\bibfield  {journal} {\bibinfo  {journal} {J. Chem. Theory Comput.}\
  }\textbf {\bibinfo {volume} {14}},\ \bibinfo {pages} {3686} (\bibinfo {year}
  {2018})}\BibitemShut {NoStop}%
\bibitem [{\citenamefont {Lestrange}, \citenamefont {Nguyen},\ and\
  \citenamefont {Li}(2015)}]{lestrange2015calibration}%
  \BibitemOpen
  \bibfield  {author} {\bibinfo {author} {\bibfnamefont {P.~J.}\ \bibnamefont
  {Lestrange}}, \bibinfo {author} {\bibfnamefont {P.~D.}\ \bibnamefont
  {Nguyen}}, \ and\ \bibinfo {author} {\bibfnamefont {X.}~\bibnamefont {Li}},\
  }\href@noop {} {\bibfield  {journal} {\bibinfo  {journal} {J. Chem. Theory
  Comput.}\ }\textbf {\bibinfo {volume} {11}},\ \bibinfo {pages} {2994}
  (\bibinfo {year} {2015})}\BibitemShut {NoStop}%
\bibitem [{\citenamefont {Besley}, \citenamefont {Peach},\ and\ \citenamefont
  {Tozer}(2009)}]{besley2009time}%
  \BibitemOpen
  \bibfield  {author} {\bibinfo {author} {\bibfnamefont {N.~A.}\ \bibnamefont
  {Besley}}, \bibinfo {author} {\bibfnamefont {M.~J.}\ \bibnamefont {Peach}}, \
  and\ \bibinfo {author} {\bibfnamefont {D.~J.}\ \bibnamefont {Tozer}},\
  }\href@noop {} {\bibfield  {journal} {\bibinfo  {journal} {Phys. Chem. Chem.
  Phys.}\ }\textbf {\bibinfo {volume} {11}},\ \bibinfo {pages} {10350}
  (\bibinfo {year} {2009})}\BibitemShut {NoStop}%
\bibitem [{\citenamefont {Besley}(2020)}]{besley2020density}%
  \BibitemOpen
  \bibfield  {author} {\bibinfo {author} {\bibfnamefont {N.~A.}\ \bibnamefont
  {Besley}},\ }\href@noop {} {\bibfield  {journal} {\bibinfo  {journal} {Acc.
  Chem. Res.}\ } (\bibinfo {year} {2020})}\BibitemShut {NoStop}%
\bibitem [{\citenamefont {Besley}\ and\ \citenamefont
  {Robinson}(2011)}]{besley2011theoretical}%
  \BibitemOpen
  \bibfield  {author} {\bibinfo {author} {\bibfnamefont {N.~A.}\ \bibnamefont
  {Besley}}\ and\ \bibinfo {author} {\bibfnamefont {D.}~\bibnamefont
  {Robinson}},\ }\href@noop {} {\bibfield  {journal} {\bibinfo  {journal}
  {Faraday Discuss.}\ }\textbf {\bibinfo {volume} {148}},\ \bibinfo {pages}
  {55} (\bibinfo {year} {2011})}\BibitemShut {NoStop}%
\bibitem [{\citenamefont {Robinson}\ and\ \citenamefont
  {Besley}(2010)}]{robinson2010modelling}%
  \BibitemOpen
  \bibfield  {author} {\bibinfo {author} {\bibfnamefont {D.}~\bibnamefont
  {Robinson}}\ and\ \bibinfo {author} {\bibfnamefont {N.~A.}\ \bibnamefont
  {Besley}},\ }\href@noop {} {\bibfield  {journal} {\bibinfo  {journal} {Phys.
  Chem. Chem. Phys.}\ }\textbf {\bibinfo {volume} {12}},\ \bibinfo {pages}
  {9667} (\bibinfo {year} {2010})}\BibitemShut {NoStop}%
\bibitem [{\citenamefont {Fogarty}\ \emph {et~al.}(2018)\citenamefont
  {Fogarty}, \citenamefont {Matthews}, \citenamefont {Ashworth}, \citenamefont
  {Brandt-Talbot}, \citenamefont {Palgrave}, \citenamefont {Bourne},
  \citenamefont {Vander~Hoogerstraete}, \citenamefont {Hunt},\ and\
  \citenamefont {Lovelock}}]{fogarty2018experimental}%
  \BibitemOpen
  \bibfield  {author} {\bibinfo {author} {\bibfnamefont {R.~M.}\ \bibnamefont
  {Fogarty}}, \bibinfo {author} {\bibfnamefont {R.~P.}\ \bibnamefont
  {Matthews}}, \bibinfo {author} {\bibfnamefont {C.~R.}\ \bibnamefont
  {Ashworth}}, \bibinfo {author} {\bibfnamefont {A.}~\bibnamefont
  {Brandt-Talbot}}, \bibinfo {author} {\bibfnamefont {R.~G.}\ \bibnamefont
  {Palgrave}}, \bibinfo {author} {\bibfnamefont {R.~A.}\ \bibnamefont
  {Bourne}}, \bibinfo {author} {\bibfnamefont {T.}~\bibnamefont
  {Vander~Hoogerstraete}}, \bibinfo {author} {\bibfnamefont {P.~A.}\
  \bibnamefont {Hunt}}, \ and\ \bibinfo {author} {\bibfnamefont {K.~R.}\
  \bibnamefont {Lovelock}},\ }\href@noop {} {\bibfield  {journal} {\bibinfo
  {journal} {J. Chem. Phys.}\ }\textbf {\bibinfo {volume} {148}},\ \bibinfo
  {pages} {193817} (\bibinfo {year} {2018})}\BibitemShut {NoStop}%
\bibitem [{\citenamefont {Buckley}\ and\ \citenamefont
  {Besley}(2011)}]{buckley2011theoretical}%
  \BibitemOpen
  \bibfield  {author} {\bibinfo {author} {\bibfnamefont {M.~W.}\ \bibnamefont
  {Buckley}}\ and\ \bibinfo {author} {\bibfnamefont {N.~A.}\ \bibnamefont
  {Besley}},\ }\href@noop {} {\bibfield  {journal} {\bibinfo  {journal} {Chem.
  Phys. Lett.}\ }\textbf {\bibinfo {volume} {501}},\ \bibinfo {pages} {540}
  (\bibinfo {year} {2011})}\BibitemShut {NoStop}%
\bibitem [{\citenamefont {Ljubi{\'c}}, \citenamefont {Kivim{\"a}ki},\ and\
  \citenamefont {Coreno}(2016)}]{ljubic2016experimental}%
  \BibitemOpen
  \bibfield  {author} {\bibinfo {author} {\bibfnamefont {I.}~\bibnamefont
  {Ljubi{\'c}}}, \bibinfo {author} {\bibfnamefont {A.}~\bibnamefont
  {Kivim{\"a}ki}}, \ and\ \bibinfo {author} {\bibfnamefont {M.}~\bibnamefont
  {Coreno}},\ }\href@noop {} {\bibfield  {journal} {\bibinfo  {journal} {Phys.
  Chem. Chem. Phys.}\ }\textbf {\bibinfo {volume} {18}},\ \bibinfo {pages}
  {10207} (\bibinfo {year} {2016})}\BibitemShut {NoStop}%
\bibitem [{\citenamefont {Ljubi{\'c}}\ \emph {et~al.}(2018)\citenamefont
  {Ljubi{\'c}}, \citenamefont {Kivim{\"a}ki}, \citenamefont {Coreno},
  \citenamefont {Kazazi{\'c}},\ and\ \citenamefont
  {Novak}}]{ljubic2018characterisation}%
  \BibitemOpen
  \bibfield  {author} {\bibinfo {author} {\bibfnamefont {I.}~\bibnamefont
  {Ljubi{\'c}}}, \bibinfo {author} {\bibfnamefont {A.}~\bibnamefont
  {Kivim{\"a}ki}}, \bibinfo {author} {\bibfnamefont {M.}~\bibnamefont
  {Coreno}}, \bibinfo {author} {\bibfnamefont {S.}~\bibnamefont {Kazazi{\'c}}},
  \ and\ \bibinfo {author} {\bibfnamefont {I.}~\bibnamefont {Novak}},\
  }\href@noop {} {\bibfield  {journal} {\bibinfo  {journal} {Phys. Chem. Chem.
  Phys.}\ }\textbf {\bibinfo {volume} {20}},\ \bibinfo {pages} {2480} (\bibinfo
  {year} {2018})}\BibitemShut {NoStop}%
\bibitem [{\citenamefont {Perdew}\ \emph {et~al.}(1982)\citenamefont {Perdew},
  \citenamefont {Parr}, \citenamefont {Levy},\ and\ \citenamefont
  {Balduz~Jr}}]{perdew1982density}%
  \BibitemOpen
  \bibfield  {author} {\bibinfo {author} {\bibfnamefont {J.~P.}\ \bibnamefont
  {Perdew}}, \bibinfo {author} {\bibfnamefont {R.~G.}\ \bibnamefont {Parr}},
  \bibinfo {author} {\bibfnamefont {M.}~\bibnamefont {Levy}}, \ and\ \bibinfo
  {author} {\bibfnamefont {J.~L.}\ \bibnamefont {Balduz~Jr}},\ }\href@noop {}
  {\bibfield  {journal} {\bibinfo  {journal} {Phys. Rev. Lett.}\ }\textbf
  {\bibinfo {volume} {49}},\ \bibinfo {pages} {1691} (\bibinfo {year}
  {1982})}\BibitemShut {NoStop}%
\bibitem [{\citenamefont {Sekino}\ and\ \citenamefont
  {Bartlett}(1984)}]{sekino1984linear}%
  \BibitemOpen
  \bibfield  {author} {\bibinfo {author} {\bibfnamefont {H.}~\bibnamefont
  {Sekino}}\ and\ \bibinfo {author} {\bibfnamefont {R.~J.}\ \bibnamefont
  {Bartlett}},\ }\href@noop {} {\bibfield  {journal} {\bibinfo  {journal} {Int.
  J. Quantum Chem}\ }\textbf {\bibinfo {volume} {26}},\ \bibinfo {pages} {255}
  (\bibinfo {year} {1984})}\BibitemShut {NoStop}%
\bibitem [{\citenamefont {Stanton}\ and\ \citenamefont
  {Bartlett}(1993)}]{stanton1993equation}%
  \BibitemOpen
  \bibfield  {author} {\bibinfo {author} {\bibfnamefont {J.~F.}\ \bibnamefont
  {Stanton}}\ and\ \bibinfo {author} {\bibfnamefont {R.~J.}\ \bibnamefont
  {Bartlett}},\ }\href@noop {} {\bibfield  {journal} {\bibinfo  {journal} {J.
  Chem. Phys.}\ }\textbf {\bibinfo {volume} {98}},\ \bibinfo {pages} {7029}
  (\bibinfo {year} {1993})}\BibitemShut {NoStop}%
\bibitem [{\citenamefont {Krylov}(2008)}]{krylov2008equation}%
  \BibitemOpen
  \bibfield  {author} {\bibinfo {author} {\bibfnamefont {A.~I.}\ \bibnamefont
  {Krylov}},\ }\href@noop {} {\bibfield  {journal} {\bibinfo  {journal} {Annu.
  Rev. Phys. Chem.}\ }\textbf {\bibinfo {volume} {59}},\ \bibinfo {pages} {433}
  (\bibinfo {year} {2008})}\BibitemShut {NoStop}%
\bibitem [{\citenamefont {Peng}\ \emph {et~al.}(2015)\citenamefont {Peng},
  \citenamefont {Lestrange}, \citenamefont {Goings}, \citenamefont {Caricato},\
  and\ \citenamefont {Li}}]{peng2015energy}%
  \BibitemOpen
  \bibfield  {author} {\bibinfo {author} {\bibfnamefont {B.}~\bibnamefont
  {Peng}}, \bibinfo {author} {\bibfnamefont {P.~J.}\ \bibnamefont {Lestrange}},
  \bibinfo {author} {\bibfnamefont {J.~J.}\ \bibnamefont {Goings}}, \bibinfo
  {author} {\bibfnamefont {M.}~\bibnamefont {Caricato}}, \ and\ \bibinfo
  {author} {\bibfnamefont {X.}~\bibnamefont {Li}},\ }\href@noop {} {\bibfield
  {journal} {\bibinfo  {journal} {J. Chem. Theory Comput.}\ }\textbf {\bibinfo
  {volume} {11}},\ \bibinfo {pages} {4146} (\bibinfo {year}
  {2015})}\BibitemShut {NoStop}%
\bibitem [{\citenamefont {Coriani}\ and\ \citenamefont
  {Koch}(2015)}]{coriani2015communication}%
  \BibitemOpen
  \bibfield  {author} {\bibinfo {author} {\bibfnamefont {S.}~\bibnamefont
  {Coriani}}\ and\ \bibinfo {author} {\bibfnamefont {H.}~\bibnamefont {Koch}},\
  }\href@noop {} {\bibfield  {journal} {\bibinfo  {journal} {J. Chem. Phys.}\
  }\textbf {\bibinfo {volume} {143}},\ \bibinfo {pages} {181103} (\bibinfo
  {year} {2015})}\BibitemShut {NoStop}%
\bibitem [{\citenamefont {Coriani}\ \emph {et~al.}(2016)\citenamefont
  {Coriani}, \citenamefont {Paw{\l}owski}, \citenamefont {Olsen},\ and\
  \citenamefont {J{\o}rgensen}}]{coriani2016molecular}%
  \BibitemOpen
  \bibfield  {author} {\bibinfo {author} {\bibfnamefont {S.}~\bibnamefont
  {Coriani}}, \bibinfo {author} {\bibfnamefont {F.}~\bibnamefont
  {Paw{\l}owski}}, \bibinfo {author} {\bibfnamefont {J.}~\bibnamefont {Olsen}},
  \ and\ \bibinfo {author} {\bibfnamefont {P.}~\bibnamefont {J{\o}rgensen}},\
  }\href@noop {} {\bibfield  {journal} {\bibinfo  {journal} {J. Chem. Phys.}\
  }\textbf {\bibinfo {volume} {144}},\ \bibinfo {pages} {024102} (\bibinfo
  {year} {2016})}\BibitemShut {NoStop}%
\bibitem [{\citenamefont {Carbone}\ \emph {et~al.}(2019)\citenamefont
  {Carbone}, \citenamefont {Cheng}, \citenamefont {Myhre}, \citenamefont
  {Matthews}, \citenamefont {Koch},\ and\ \citenamefont
  {Coriani}}]{carbone2019analysis}%
  \BibitemOpen
  \bibfield  {author} {\bibinfo {author} {\bibfnamefont {J.~P.}\ \bibnamefont
  {Carbone}}, \bibinfo {author} {\bibfnamefont {L.}~\bibnamefont {Cheng}},
  \bibinfo {author} {\bibfnamefont {R.~H.}\ \bibnamefont {Myhre}}, \bibinfo
  {author} {\bibfnamefont {D.}~\bibnamefont {Matthews}}, \bibinfo {author}
  {\bibfnamefont {H.}~\bibnamefont {Koch}}, \ and\ \bibinfo {author}
  {\bibfnamefont {S.}~\bibnamefont {Coriani}},\ }in\ \href@noop {} {\emph
  {\bibinfo {booktitle} {Advances in Quantum Chemistry}}},\ Vol.~\bibinfo
  {volume} {79}\ (\bibinfo  {publisher} {Elsevier},\ \bibinfo {year} {2019})\
  pp.\ \bibinfo {pages} {241--261}\BibitemShut {NoStop}%
\bibitem [{\citenamefont {Vidal}\ \emph {et~al.}(2019)\citenamefont {Vidal},
  \citenamefont {Feng}, \citenamefont {Epifanovsky}, \citenamefont {Krylov},\
  and\ \citenamefont {Coriani}}]{vidal2019new}%
  \BibitemOpen
  \bibfield  {author} {\bibinfo {author} {\bibfnamefont {M.~L.}\ \bibnamefont
  {Vidal}}, \bibinfo {author} {\bibfnamefont {X.}~\bibnamefont {Feng}},
  \bibinfo {author} {\bibfnamefont {E.}~\bibnamefont {Epifanovsky}}, \bibinfo
  {author} {\bibfnamefont {A.~I.}\ \bibnamefont {Krylov}}, \ and\ \bibinfo
  {author} {\bibfnamefont {S.}~\bibnamefont {Coriani}},\ }\href@noop {}
  {\bibfield  {journal} {\bibinfo  {journal} {J. Chem. Theory Comput.}\
  }\textbf {\bibinfo {volume} {15}},\ \bibinfo {pages} {3117} (\bibinfo {year}
  {2019})}\BibitemShut {NoStop}%
\bibitem [{\citenamefont {Coriani}\ \emph {et~al.}(2012)\citenamefont
  {Coriani}, \citenamefont {Christiansen}, \citenamefont {Fransson},\ and\
  \citenamefont {Norman}}]{coriani2012coupled}%
  \BibitemOpen
  \bibfield  {author} {\bibinfo {author} {\bibfnamefont {S.}~\bibnamefont
  {Coriani}}, \bibinfo {author} {\bibfnamefont {O.}~\bibnamefont
  {Christiansen}}, \bibinfo {author} {\bibfnamefont {T.}~\bibnamefont
  {Fransson}}, \ and\ \bibinfo {author} {\bibfnamefont {P.}~\bibnamefont
  {Norman}},\ }\href@noop {} {\bibfield  {journal} {\bibinfo  {journal} {Phys.
  Rev. A}\ }\textbf {\bibinfo {volume} {85}},\ \bibinfo {pages} {022507}
  (\bibinfo {year} {2012})}\BibitemShut {NoStop}%
\bibitem [{\citenamefont {Frati}\ \emph {et~al.}(2019)\citenamefont {Frati},
  \citenamefont {De~Groot}, \citenamefont {Cerezo}, \citenamefont {Santoro},
  \citenamefont {Cheng}, \citenamefont {Faber},\ and\ \citenamefont
  {Coriani}}]{frati2019coupled}%
  \BibitemOpen
  \bibfield  {author} {\bibinfo {author} {\bibfnamefont {F.}~\bibnamefont
  {Frati}}, \bibinfo {author} {\bibfnamefont {F.}~\bibnamefont {De~Groot}},
  \bibinfo {author} {\bibfnamefont {J.}~\bibnamefont {Cerezo}}, \bibinfo
  {author} {\bibfnamefont {F.}~\bibnamefont {Santoro}}, \bibinfo {author}
  {\bibfnamefont {L.}~\bibnamefont {Cheng}}, \bibinfo {author} {\bibfnamefont
  {R.}~\bibnamefont {Faber}}, \ and\ \bibinfo {author} {\bibfnamefont
  {S.}~\bibnamefont {Coriani}},\ }\href@noop {} {\bibfield  {journal} {\bibinfo
   {journal} {J. Chem. Phys.}\ }\textbf {\bibinfo {volume} {151}},\ \bibinfo
  {pages} {064107} (\bibinfo {year} {2019})}\BibitemShut {NoStop}%
\bibitem [{\citenamefont {Tsuru}\ \emph {et~al.}(2019)\citenamefont {Tsuru},
  \citenamefont {Vidal}, \citenamefont {P{\'a}pai}, \citenamefont {Krylov},
  \citenamefont {M{\o}ller},\ and\ \citenamefont {Coriani}}]{tsuru2019time}%
  \BibitemOpen
  \bibfield  {author} {\bibinfo {author} {\bibfnamefont {S.}~\bibnamefont
  {Tsuru}}, \bibinfo {author} {\bibfnamefont {M.~L.}\ \bibnamefont {Vidal}},
  \bibinfo {author} {\bibfnamefont {M.}~\bibnamefont {P{\'a}pai}}, \bibinfo
  {author} {\bibfnamefont {A.~I.}\ \bibnamefont {Krylov}}, \bibinfo {author}
  {\bibfnamefont {K.~B.}\ \bibnamefont {M{\o}ller}}, \ and\ \bibinfo {author}
  {\bibfnamefont {S.}~\bibnamefont {Coriani}},\ }\href@noop {} {\bibfield
  {journal} {\bibinfo  {journal} {J. Chem. Phys.}\ }\textbf {\bibinfo {volume}
  {151}},\ \bibinfo {pages} {124114} (\bibinfo {year} {2019})}\BibitemShut
  {NoStop}%
\bibitem [{\citenamefont {Cederbaum}, \citenamefont {Domcke},\ and\
  \citenamefont {Schirmer}(1980)}]{cederbaum1980many}%
  \BibitemOpen
  \bibfield  {author} {\bibinfo {author} {\bibfnamefont {L.~S.}\ \bibnamefont
  {Cederbaum}}, \bibinfo {author} {\bibfnamefont {W.}~\bibnamefont {Domcke}}, \
  and\ \bibinfo {author} {\bibfnamefont {J.}~\bibnamefont {Schirmer}},\
  }\href@noop {} {\bibfield  {journal} {\bibinfo  {journal} {Phys. Rev. A}\
  }\textbf {\bibinfo {volume} {22}},\ \bibinfo {pages} {206} (\bibinfo {year}
  {1980})}\BibitemShut {NoStop}%
\bibitem [{\citenamefont {Lopez~Vidal}\ \emph {et~al.}(2020)\citenamefont
  {Lopez~Vidal}, \citenamefont {Pokhilko}, \citenamefont {Krylov},\ and\
  \citenamefont {Coriani}}]{lopez2020equation}%
  \BibitemOpen
  \bibfield  {author} {\bibinfo {author} {\bibfnamefont {M.}~\bibnamefont
  {Lopez~Vidal}}, \bibinfo {author} {\bibfnamefont {P.}~\bibnamefont
  {Pokhilko}}, \bibinfo {author} {\bibfnamefont {A.}~\bibnamefont {Krylov}}, \
  and\ \bibinfo {author} {\bibfnamefont {S.}~\bibnamefont {Coriani}},\
  }\href@noop {} {\  (\bibinfo {year} {2020})}\BibitemShut {NoStop}%
\bibitem [{\citenamefont {Vidal}, \citenamefont {Krylov},\ and\ \citenamefont
  {Coriani}(2020)}]{vidal2020dyson}%
  \BibitemOpen
  \bibfield  {author} {\bibinfo {author} {\bibfnamefont {M.~L.}\ \bibnamefont
  {Vidal}}, \bibinfo {author} {\bibfnamefont {A.~I.}\ \bibnamefont {Krylov}}, \
  and\ \bibinfo {author} {\bibfnamefont {S.}~\bibnamefont {Coriani}},\
  }\href@noop {} {\bibfield  {journal} {\bibinfo  {journal} {Phys. Chem. Chem.
  Phys.}\ }\textbf {\bibinfo {volume} {22}},\ \bibinfo {pages} {2693} (\bibinfo
  {year} {2020})}\BibitemShut {NoStop}%
\bibitem [{\citenamefont {Wormit}\ \emph {et~al.}(2014)\citenamefont {Wormit},
  \citenamefont {Rehn}, \citenamefont {Harbach}, \citenamefont {Wenzel},
  \citenamefont {Krauter}, \citenamefont {Epifanovsky},\ and\ \citenamefont
  {Dreuw}}]{wormit2014investigating}%
  \BibitemOpen
  \bibfield  {author} {\bibinfo {author} {\bibfnamefont {M.}~\bibnamefont
  {Wormit}}, \bibinfo {author} {\bibfnamefont {D.~R.}\ \bibnamefont {Rehn}},
  \bibinfo {author} {\bibfnamefont {P.~H.}\ \bibnamefont {Harbach}}, \bibinfo
  {author} {\bibfnamefont {J.}~\bibnamefont {Wenzel}}, \bibinfo {author}
  {\bibfnamefont {C.~M.}\ \bibnamefont {Krauter}}, \bibinfo {author}
  {\bibfnamefont {E.}~\bibnamefont {Epifanovsky}}, \ and\ \bibinfo {author}
  {\bibfnamefont {A.}~\bibnamefont {Dreuw}},\ }\href@noop {} {\bibfield
  {journal} {\bibinfo  {journal} {Mol. Phys.}\ }\textbf {\bibinfo {volume}
  {112}},\ \bibinfo {pages} {774} (\bibinfo {year} {2014})}\BibitemShut
  {NoStop}%
\bibitem [{\citenamefont {Norman}\ and\ \citenamefont
  {Dreuw}(2018)}]{norman2018simulating}%
  \BibitemOpen
  \bibfield  {author} {\bibinfo {author} {\bibfnamefont {P.}~\bibnamefont
  {Norman}}\ and\ \bibinfo {author} {\bibfnamefont {A.}~\bibnamefont {Dreuw}},\
  }\href@noop {} {\bibfield  {journal} {\bibinfo  {journal} {Chem. Rev.}\
  }\textbf {\bibinfo {volume} {118}},\ \bibinfo {pages} {7208} (\bibinfo {year}
  {2018})}\BibitemShut {NoStop}%
\bibitem [{\citenamefont {List}\ \emph {et~al.}(2020)\citenamefont {List},
  \citenamefont {Dempwolff}, \citenamefont {Dreuw}, \citenamefont {Norman},\
  and\ \citenamefont {Mart{\'\i}nez}}]{list2020probing}%
  \BibitemOpen
  \bibfield  {author} {\bibinfo {author} {\bibfnamefont {N.~H.}\ \bibnamefont
  {List}}, \bibinfo {author} {\bibfnamefont {A.~L.}\ \bibnamefont {Dempwolff}},
  \bibinfo {author} {\bibfnamefont {A.}~\bibnamefont {Dreuw}}, \bibinfo
  {author} {\bibfnamefont {P.}~\bibnamefont {Norman}}, \ and\ \bibinfo {author}
  {\bibfnamefont {T.~J.}\ \bibnamefont {Mart{\'\i}nez}},\ }\href@noop {}
  {\bibfield  {journal} {\bibinfo  {journal} {Chem. Sci.}\ }\textbf {\bibinfo
  {volume} {11}},\ \bibinfo {pages} {4180} (\bibinfo {year}
  {2020})}\BibitemShut {NoStop}%
\bibitem [{\citenamefont {Wenzel}\ and\ \citenamefont
  {Dreuw}(2016)}]{wenzel2016physical}%
  \BibitemOpen
  \bibfield  {author} {\bibinfo {author} {\bibfnamefont {J.}~\bibnamefont
  {Wenzel}}\ and\ \bibinfo {author} {\bibfnamefont {A.}~\bibnamefont {Dreuw}},\
  }\href@noop {} {\bibfield  {journal} {\bibinfo  {journal} {J. Chem. Theory
  Comput.}\ }\textbf {\bibinfo {volume} {12}},\ \bibinfo {pages} {1314}
  (\bibinfo {year} {2016})}\BibitemShut {NoStop}%
\bibitem [{\citenamefont {Wenzel}\ \emph {et~al.}(2015)\citenamefont {Wenzel},
  \citenamefont {Holzer}, \citenamefont {Wormit},\ and\ \citenamefont
  {Dreuw}}]{wenzel2015analysis}%
  \BibitemOpen
  \bibfield  {author} {\bibinfo {author} {\bibfnamefont {J.}~\bibnamefont
  {Wenzel}}, \bibinfo {author} {\bibfnamefont {A.}~\bibnamefont {Holzer}},
  \bibinfo {author} {\bibfnamefont {M.}~\bibnamefont {Wormit}}, \ and\ \bibinfo
  {author} {\bibfnamefont {A.}~\bibnamefont {Dreuw}},\ }\href@noop {}
  {\bibfield  {journal} {\bibinfo  {journal} {J. Chem. Phys.}\ }\textbf
  {\bibinfo {volume} {142}},\ \bibinfo {pages} {214104} (\bibinfo {year}
  {2015})}\BibitemShut {NoStop}%
\bibitem [{\citenamefont {Myhre}, \citenamefont {Coriani},\ and\ \citenamefont
  {Koch}(2016)}]{myhre2016near}%
  \BibitemOpen
  \bibfield  {author} {\bibinfo {author} {\bibfnamefont {R.~H.}\ \bibnamefont
  {Myhre}}, \bibinfo {author} {\bibfnamefont {S.}~\bibnamefont {Coriani}}, \
  and\ \bibinfo {author} {\bibfnamefont {H.}~\bibnamefont {Koch}},\ }\href@noop
  {} {\bibfield  {journal} {\bibinfo  {journal} {J. Chem. Theory Comput.}\
  }\textbf {\bibinfo {volume} {12}},\ \bibinfo {pages} {2633} (\bibinfo {year}
  {2016})}\BibitemShut {NoStop}%
\bibitem [{\citenamefont {Besley}, \citenamefont {Gilbert},\ and\ \citenamefont
  {Gill}(2009)}]{besley2009self}%
  \BibitemOpen
  \bibfield  {author} {\bibinfo {author} {\bibfnamefont {N.~A.}\ \bibnamefont
  {Besley}}, \bibinfo {author} {\bibfnamefont {A.~T.}\ \bibnamefont {Gilbert}},
  \ and\ \bibinfo {author} {\bibfnamefont {P.~M.}\ \bibnamefont {Gill}},\
  }\href@noop {} {\bibfield  {journal} {\bibinfo  {journal} {J. Chem. Phys.}\
  }\textbf {\bibinfo {volume} {130}},\ \bibinfo {pages} {124308} (\bibinfo
  {year} {2009})}\BibitemShut {NoStop}%
\bibitem [{\citenamefont {Derricotte}\ and\ \citenamefont
  {Evangelista}(2015)}]{derricotte2015simulation}%
  \BibitemOpen
  \bibfield  {author} {\bibinfo {author} {\bibfnamefont {W.~D.}\ \bibnamefont
  {Derricotte}}\ and\ \bibinfo {author} {\bibfnamefont {F.~A.}\ \bibnamefont
  {Evangelista}},\ }\href@noop {} {\bibfield  {journal} {\bibinfo  {journal}
  {Phys. Chem. Chem. Phys.}\ }\textbf {\bibinfo {volume} {17}},\ \bibinfo
  {pages} {14360} (\bibinfo {year} {2015})}\BibitemShut {NoStop}%
\bibitem [{\citenamefont {Michelitsch}\ and\ \citenamefont
  {Reuter}(2019)}]{michelitsch2019efficient}%
  \BibitemOpen
  \bibfield  {author} {\bibinfo {author} {\bibfnamefont {G.~S.}\ \bibnamefont
  {Michelitsch}}\ and\ \bibinfo {author} {\bibfnamefont {K.}~\bibnamefont
  {Reuter}},\ }\href@noop {} {\bibfield  {journal} {\bibinfo  {journal} {J.
  Chem. Phys.}\ }\textbf {\bibinfo {volume} {150}},\ \bibinfo {pages} {074104}
  (\bibinfo {year} {2019})}\BibitemShut {NoStop}%
\bibitem [{\citenamefont {Hait}\ and\ \citenamefont
  {Head-Gordon}(2020{\natexlab{a}})}]{hait2020highly}%
  \BibitemOpen
  \bibfield  {author} {\bibinfo {author} {\bibfnamefont {D.}~\bibnamefont
  {Hait}}\ and\ \bibinfo {author} {\bibfnamefont {M.}~\bibnamefont
  {Head-Gordon}},\ }\href@noop {} {\bibfield  {journal} {\bibinfo  {journal}
  {J. Phys. Chem. Lett.}\ }\textbf {\bibinfo {volume} {11}},\ \bibinfo {pages}
  {775} (\bibinfo {year} {2020}{\natexlab{a}})}\BibitemShut {NoStop}%
\bibitem [{\citenamefont {Ehlert}\ and\ \citenamefont
  {Klamroth}(2020)}]{ehlert2020psixas}%
  \BibitemOpen
  \bibfield  {author} {\bibinfo {author} {\bibfnamefont {C.}~\bibnamefont
  {Ehlert}}\ and\ \bibinfo {author} {\bibfnamefont {T.}~\bibnamefont
  {Klamroth}},\ }\href@noop {} {\bibfield  {journal} {\bibinfo  {journal} {J.
  Comp. Chem.}\ }\textbf {\bibinfo {volume} {41}},\ \bibinfo {pages} {1781}
  (\bibinfo {year} {2020})}\BibitemShut {NoStop}%
\bibitem [{\citenamefont {Gilbert}, \citenamefont {Besley},\ and\ \citenamefont
  {Gill}(2008)}]{gilbert2008self}%
  \BibitemOpen
  \bibfield  {author} {\bibinfo {author} {\bibfnamefont {A.~T.}\ \bibnamefont
  {Gilbert}}, \bibinfo {author} {\bibfnamefont {N.~A.}\ \bibnamefont {Besley}},
  \ and\ \bibinfo {author} {\bibfnamefont {P.~M.}\ \bibnamefont {Gill}},\
  }\href@noop {} {\bibfield  {journal} {\bibinfo  {journal} {J. Phys. Chem A}\
  }\textbf {\bibinfo {volume} {112}},\ \bibinfo {pages} {13164} (\bibinfo
  {year} {2008})}\BibitemShut {NoStop}%
\bibitem [{\citenamefont {Barca}, \citenamefont {Gilbert},\ and\ \citenamefont
  {Gill}(2018)}]{barca2018simple}%
  \BibitemOpen
  \bibfield  {author} {\bibinfo {author} {\bibfnamefont {G.~M.}\ \bibnamefont
  {Barca}}, \bibinfo {author} {\bibfnamefont {A.~T.}\ \bibnamefont {Gilbert}},
  \ and\ \bibinfo {author} {\bibfnamefont {P.~M.}\ \bibnamefont {Gill}},\
  }\href@noop {} {\bibfield  {journal} {\bibinfo  {journal} {J. Chem. Theory
  Comput.}\ }\textbf {\bibinfo {volume} {14}},\ \bibinfo {pages} {1501}
  (\bibinfo {year} {2018})}\BibitemShut {NoStop}%
\bibitem [{\citenamefont {Pulay}(1980)}]{pulay1980convergence}%
  \BibitemOpen
  \bibfield  {author} {\bibinfo {author} {\bibfnamefont {P.}~\bibnamefont
  {Pulay}},\ }\href@noop {} {\bibfield  {journal} {\bibinfo  {journal} {Chem.
  Phys. Lett.}\ }\textbf {\bibinfo {volume} {73}},\ \bibinfo {pages} {393}
  (\bibinfo {year} {1980})}\BibitemShut {NoStop}%
\bibitem [{\citenamefont {Mewes}\ \emph {et~al.}(2014)\citenamefont {Mewes},
  \citenamefont {Jovanovi{\'c}}, \citenamefont {Marian},\ and\ \citenamefont
  {Dreuw}}]{mewes2014molecular}%
  \BibitemOpen
  \bibfield  {author} {\bibinfo {author} {\bibfnamefont {J.-M.}\ \bibnamefont
  {Mewes}}, \bibinfo {author} {\bibfnamefont {V.}~\bibnamefont
  {Jovanovi{\'c}}}, \bibinfo {author} {\bibfnamefont {C.~M.}\ \bibnamefont
  {Marian}}, \ and\ \bibinfo {author} {\bibfnamefont {A.}~\bibnamefont
  {Dreuw}},\ }\href@noop {} {\bibfield  {journal} {\bibinfo  {journal} {Phys.
  Chem. Chem. Phys.}\ }\textbf {\bibinfo {volume} {16}},\ \bibinfo {pages}
  {12393} (\bibinfo {year} {2014})}\BibitemShut {NoStop}%
\bibitem [{\citenamefont {Hait}\ and\ \citenamefont
  {Head-Gordon}(2020{\natexlab{b}})}]{hait2020excited}%
  \BibitemOpen
  \bibfield  {author} {\bibinfo {author} {\bibfnamefont {D.}~\bibnamefont
  {Hait}}\ and\ \bibinfo {author} {\bibfnamefont {M.}~\bibnamefont
  {Head-Gordon}},\ }\href@noop {} {\bibfield  {journal} {\bibinfo  {journal}
  {J. Chem. Theory Comput.}\ }\textbf {\bibinfo {volume} {16}},\ \bibinfo
  {pages} {1699} (\bibinfo {year} {2020}{\natexlab{b}})}\BibitemShut {NoStop}%
\bibitem [{\citenamefont {Filatov}\ and\ \citenamefont
  {Shaik}(1999)}]{filatov1999spin}%
  \BibitemOpen
  \bibfield  {author} {\bibinfo {author} {\bibfnamefont {M.}~\bibnamefont
  {Filatov}}\ and\ \bibinfo {author} {\bibfnamefont {S.}~\bibnamefont
  {Shaik}},\ }\href@noop {} {\bibfield  {journal} {\bibinfo  {journal} {Chem.
  Phys. Lett.}\ }\textbf {\bibinfo {volume} {304}},\ \bibinfo {pages} {429}
  (\bibinfo {year} {1999})}\BibitemShut {NoStop}%
\bibitem [{\citenamefont {Kowalczyk}\ \emph {et~al.}(2013)\citenamefont
  {Kowalczyk}, \citenamefont {Tsuchimochi}, \citenamefont {Chen}, \citenamefont
  {Top},\ and\ \citenamefont {Van~Voorhis}}]{kowalczyk2013excitation}%
  \BibitemOpen
  \bibfield  {author} {\bibinfo {author} {\bibfnamefont {T.}~\bibnamefont
  {Kowalczyk}}, \bibinfo {author} {\bibfnamefont {T.}~\bibnamefont
  {Tsuchimochi}}, \bibinfo {author} {\bibfnamefont {P.-T.}\ \bibnamefont
  {Chen}}, \bibinfo {author} {\bibfnamefont {L.}~\bibnamefont {Top}}, \ and\
  \bibinfo {author} {\bibfnamefont {T.}~\bibnamefont {Van~Voorhis}},\
  }\href@noop {} {\bibfield  {journal} {\bibinfo  {journal} {J. Chem. Phys.}\
  }\textbf {\bibinfo {volume} {138}},\ \bibinfo {pages} {164101} (\bibinfo
  {year} {2013})}\BibitemShut {NoStop}%
\bibitem [{\citenamefont {Sun}, \citenamefont {Ruzsinszky},\ and\ \citenamefont
  {Perdew}(2015)}]{SCAN}%
  \BibitemOpen
  \bibfield  {author} {\bibinfo {author} {\bibfnamefont {J.}~\bibnamefont
  {Sun}}, \bibinfo {author} {\bibfnamefont {A.}~\bibnamefont {Ruzsinszky}}, \
  and\ \bibinfo {author} {\bibfnamefont {J.~P.}\ \bibnamefont {Perdew}},\
  }\href@noop {} {\bibfield  {journal} {\bibinfo  {journal} {Phys. Rev. Lett.}\
  }\textbf {\bibinfo {volume} {115}},\ \bibinfo {pages} {036402} (\bibinfo
  {year} {2015})}\BibitemShut {NoStop}%
\bibitem [{\citenamefont {{\AA}gren}\ \emph {et~al.}(1997)\citenamefont
  {{\AA}gren}, \citenamefont {Carravetta}, \citenamefont {Vahtras},\ and\
  \citenamefont {Pettersson}}]{aagren1997direct}%
  \BibitemOpen
  \bibfield  {author} {\bibinfo {author} {\bibfnamefont {H.}~\bibnamefont
  {{\AA}gren}}, \bibinfo {author} {\bibfnamefont {V.}~\bibnamefont
  {Carravetta}}, \bibinfo {author} {\bibfnamefont {O.}~\bibnamefont {Vahtras}},
  \ and\ \bibinfo {author} {\bibfnamefont {L.~G.}\ \bibnamefont {Pettersson}},\
  }\href@noop {} {\bibfield  {journal} {\bibinfo  {journal} {Theo. Chem. Acc.}\
  }\textbf {\bibinfo {volume} {97}},\ \bibinfo {pages} {14} (\bibinfo {year}
  {1997})}\BibitemShut {NoStop}%
\bibitem [{\citenamefont {Oosterbaan}, \citenamefont {White},\ and\
  \citenamefont {Head-Gordon}(2018)}]{oosterbaan2018non}%
  \BibitemOpen
  \bibfield  {author} {\bibinfo {author} {\bibfnamefont {K.~J.}\ \bibnamefont
  {Oosterbaan}}, \bibinfo {author} {\bibfnamefont {A.~F.}\ \bibnamefont
  {White}}, \ and\ \bibinfo {author} {\bibfnamefont {M.}~\bibnamefont
  {Head-Gordon}},\ }\href@noop {} {\bibfield  {journal} {\bibinfo  {journal}
  {J. Chem. Phys.}\ }\textbf {\bibinfo {volume} {149}},\ \bibinfo {pages}
  {044116} (\bibinfo {year} {2018})}\BibitemShut {NoStop}%
\bibitem [{\citenamefont {Oosterbaan}, \citenamefont {White},\ and\
  \citenamefont {Head-Gordon}(2019)}]{oosterbaan2019non}%
  \BibitemOpen
  \bibfield  {author} {\bibinfo {author} {\bibfnamefont {K.~J.}\ \bibnamefont
  {Oosterbaan}}, \bibinfo {author} {\bibfnamefont {A.~F.}\ \bibnamefont
  {White}}, \ and\ \bibinfo {author} {\bibfnamefont {M.}~\bibnamefont
  {Head-Gordon}},\ }\href@noop {} {\bibfield  {journal} {\bibinfo  {journal}
  {J. Chem. Theory Comput.}\ }\textbf {\bibinfo {volume} {15}},\ \bibinfo
  {pages} {2966} (\bibinfo {year} {2019})}\BibitemShut {NoStop}%
\bibitem [{\citenamefont {Oosterbaan}\ \emph {et~al.}(2020)\citenamefont
  {Oosterbaan}, \citenamefont {White}, \citenamefont {Hait},\ and\
  \citenamefont {Head-Gordon}}]{oosterbaan2020generalized}%
  \BibitemOpen
  \bibfield  {author} {\bibinfo {author} {\bibfnamefont {K.~J.}\ \bibnamefont
  {Oosterbaan}}, \bibinfo {author} {\bibfnamefont {A.~F.}\ \bibnamefont
  {White}}, \bibinfo {author} {\bibfnamefont {D.}~\bibnamefont {Hait}}, \ and\
  \bibinfo {author} {\bibfnamefont {M.}~\bibnamefont {Head-Gordon}},\
  }\href@noop {} {\bibfield  {journal} {\bibinfo  {journal} {Phys. Chem. Chem.
  Phys.}\ }\textbf {\bibinfo {volume} {22}},\ \bibinfo {pages} {8182} (\bibinfo
  {year} {2020})}\BibitemShut {NoStop}%
\bibitem [{\citenamefont {Chergui}\ and\ \citenamefont
  {Collet}(2017)}]{chergui2017photoinduced}%
  \BibitemOpen
  \bibfield  {author} {\bibinfo {author} {\bibfnamefont {M.}~\bibnamefont
  {Chergui}}\ and\ \bibinfo {author} {\bibfnamefont {E.}~\bibnamefont
  {Collet}},\ }\href@noop {} {\bibfield  {journal} {\bibinfo  {journal} {Chem.
  Rev.}\ }\textbf {\bibinfo {volume} {117}},\ \bibinfo {pages} {11025}
  (\bibinfo {year} {2017})}\BibitemShut {NoStop}%
\bibitem [{\citenamefont {Bhattacherjee}\ and\ \citenamefont
  {Leone}(2018)}]{bhattacherjee2018ultrafast}%
  \BibitemOpen
  \bibfield  {author} {\bibinfo {author} {\bibfnamefont {A.}~\bibnamefont
  {Bhattacherjee}}\ and\ \bibinfo {author} {\bibfnamefont {S.~R.}\ \bibnamefont
  {Leone}},\ }\href@noop {} {\bibfield  {journal} {\bibinfo  {journal} {Acc.
  Chem. Res.}\ }\textbf {\bibinfo {volume} {51}},\ \bibinfo {pages} {3203}
  (\bibinfo {year} {2018})}\BibitemShut {NoStop}%
\bibitem [{\citenamefont {Schnorr}\ \emph {et~al.}(2019)\citenamefont
  {Schnorr}, \citenamefont {Bhattacherjee}, \citenamefont {Oosterbaan},
  \citenamefont {Delcey}, \citenamefont {Yang}, \citenamefont {Xue},
  \citenamefont {Attar}, \citenamefont {Chatterley}, \citenamefont
  {Head-Gordon}, \citenamefont {Leone},\ and\ \citenamefont
  {Gessner}}]{schnorr2019tracing}%
  \BibitemOpen
  \bibfield  {author} {\bibinfo {author} {\bibfnamefont {K.}~\bibnamefont
  {Schnorr}}, \bibinfo {author} {\bibfnamefont {A.}~\bibnamefont
  {Bhattacherjee}}, \bibinfo {author} {\bibfnamefont {K.~J.}\ \bibnamefont
  {Oosterbaan}}, \bibinfo {author} {\bibfnamefont {M.~G.}\ \bibnamefont
  {Delcey}}, \bibinfo {author} {\bibfnamefont {Z.}~\bibnamefont {Yang}},
  \bibinfo {author} {\bibfnamefont {T.}~\bibnamefont {Xue}}, \bibinfo {author}
  {\bibfnamefont {A.~R.}\ \bibnamefont {Attar}}, \bibinfo {author}
  {\bibfnamefont {A.~S.}\ \bibnamefont {Chatterley}}, \bibinfo {author}
  {\bibfnamefont {M.}~\bibnamefont {Head-Gordon}}, \bibinfo {author}
  {\bibfnamefont {S.~R.}\ \bibnamefont {Leone}}, \ and\ \bibinfo {author}
  {\bibfnamefont {O.}~\bibnamefont {Gessner}},\ }\href@noop {} {\bibfield
  {journal} {\bibinfo  {journal} {J. Phys. Chem Lett.}\ }\textbf {\bibinfo
  {volume} {10}},\ \bibinfo {pages} {1382} (\bibinfo {year}
  {2019})}\BibitemShut {NoStop}%
\bibitem [{\citenamefont {Yang}\ \emph {et~al.}(2018)\citenamefont {Yang},
  \citenamefont {Schnorr}, \citenamefont {Bhattacherjee}, \citenamefont
  {Lefebvre}, \citenamefont {Epshtein}, \citenamefont {Xue}, \citenamefont
  {Stanton},\ and\ \citenamefont {Leone}}]{yang2018electron}%
  \BibitemOpen
  \bibfield  {author} {\bibinfo {author} {\bibfnamefont {Z.}~\bibnamefont
  {Yang}}, \bibinfo {author} {\bibfnamefont {K.}~\bibnamefont {Schnorr}},
  \bibinfo {author} {\bibfnamefont {A.}~\bibnamefont {Bhattacherjee}}, \bibinfo
  {author} {\bibfnamefont {P.-L.}\ \bibnamefont {Lefebvre}}, \bibinfo {author}
  {\bibfnamefont {M.}~\bibnamefont {Epshtein}}, \bibinfo {author}
  {\bibfnamefont {T.}~\bibnamefont {Xue}}, \bibinfo {author} {\bibfnamefont
  {J.~F.}\ \bibnamefont {Stanton}}, \ and\ \bibinfo {author} {\bibfnamefont
  {S.~R.}\ \bibnamefont {Leone}},\ }\href@noop {} {\bibfield  {journal}
  {\bibinfo  {journal} {J. Am. Chem. Soc.}\ }\textbf {\bibinfo {volume}
  {140}},\ \bibinfo {pages} {13360} (\bibinfo {year} {2018})}\BibitemShut
  {NoStop}%
\bibitem [{\citenamefont {Maurice}\ and\ \citenamefont
  {Head-Gordon}(1996)}]{maurice1996nature}%
  \BibitemOpen
  \bibfield  {author} {\bibinfo {author} {\bibfnamefont {D.}~\bibnamefont
  {Maurice}}\ and\ \bibinfo {author} {\bibfnamefont {M.}~\bibnamefont
  {Head-Gordon}},\ }\href@noop {} {\bibfield  {journal} {\bibinfo  {journal}
  {J. Phys. Chem.}\ }\textbf {\bibinfo {volume} {100}},\ \bibinfo {pages}
  {6131} (\bibinfo {year} {1996})}\BibitemShut {NoStop}%
\bibitem [{\citenamefont {Maitra}\ \emph {et~al.}(2004)\citenamefont {Maitra},
  \citenamefont {Zhang}, \citenamefont {Cave},\ and\ \citenamefont
  {Burke}}]{maitra2004double}%
  \BibitemOpen
  \bibfield  {author} {\bibinfo {author} {\bibfnamefont {N.~T.}\ \bibnamefont
  {Maitra}}, \bibinfo {author} {\bibfnamefont {F.}~\bibnamefont {Zhang}},
  \bibinfo {author} {\bibfnamefont {R.~J.}\ \bibnamefont {Cave}}, \ and\
  \bibinfo {author} {\bibfnamefont {K.}~\bibnamefont {Burke}},\ }\href@noop {}
  {\bibfield  {journal} {\bibinfo  {journal} {J. Chem. Phys.}\ }\textbf
  {\bibinfo {volume} {120}},\ \bibinfo {pages} {5932} (\bibinfo {year}
  {2004})}\BibitemShut {NoStop}%
\bibitem [{\citenamefont {Levine}\ \emph {et~al.}(2006)\citenamefont {Levine},
  \citenamefont {Ko}, \citenamefont {Quenneville},\ and\ \citenamefont
  {Mart{\'i}nez}}]{levine2006conical}%
  \BibitemOpen
  \bibfield  {author} {\bibinfo {author} {\bibfnamefont {B.~G.}\ \bibnamefont
  {Levine}}, \bibinfo {author} {\bibfnamefont {C.}~\bibnamefont {Ko}}, \bibinfo
  {author} {\bibfnamefont {J.}~\bibnamefont {Quenneville}}, \ and\ \bibinfo
  {author} {\bibfnamefont {T.~J.}\ \bibnamefont {Mart{\'i}nez}},\ }\href@noop
  {} {\bibfield  {journal} {\bibinfo  {journal} {Mol. Phys.}\ }\textbf
  {\bibinfo {volume} {104}},\ \bibinfo {pages} {1039} (\bibinfo {year}
  {2006})}\BibitemShut {NoStop}%
\bibitem [{\citenamefont {Hait}, \citenamefont {Rettig},\ and\ \citenamefont
  {Head-Gordon}(2019{\natexlab{a}})}]{hait2019beyond}%
  \BibitemOpen
  \bibfield  {author} {\bibinfo {author} {\bibfnamefont {D.}~\bibnamefont
  {Hait}}, \bibinfo {author} {\bibfnamefont {A.}~\bibnamefont {Rettig}}, \ and\
  \bibinfo {author} {\bibfnamefont {M.}~\bibnamefont {Head-Gordon}},\
  }\href@noop {} {\bibfield  {journal} {\bibinfo  {journal} {Phys. Chem. Chem.
  Phys.}\ } (\bibinfo {year} {2019}{\natexlab{a}})}\BibitemShut {NoStop}%
\bibitem [{\citenamefont {Ziegler}, \citenamefont {Rauk},\ and\ \citenamefont
  {Baerends}(1977)}]{ziegler1977calculation}%
  \BibitemOpen
  \bibfield  {author} {\bibinfo {author} {\bibfnamefont {T.}~\bibnamefont
  {Ziegler}}, \bibinfo {author} {\bibfnamefont {A.}~\bibnamefont {Rauk}}, \
  and\ \bibinfo {author} {\bibfnamefont {E.~J.}\ \bibnamefont {Baerends}},\
  }\href@noop {} {\bibfield  {journal} {\bibinfo  {journal} {Theoretica chimica
  acta}\ }\textbf {\bibinfo {volume} {43}},\ \bibinfo {pages} {261} (\bibinfo
  {year} {1977})}\BibitemShut {NoStop}%
\bibitem [{\citenamefont {Kowalczyk}, \citenamefont {Yost},\ and\ \citenamefont
  {Voorhis}(2011)}]{kowalczyk2011assessment}%
  \BibitemOpen
  \bibfield  {author} {\bibinfo {author} {\bibfnamefont {T.}~\bibnamefont
  {Kowalczyk}}, \bibinfo {author} {\bibfnamefont {S.~R.}\ \bibnamefont {Yost}},
  \ and\ \bibinfo {author} {\bibfnamefont {T.~V.}\ \bibnamefont {Voorhis}},\
  }\href@noop {} {\bibfield  {journal} {\bibinfo  {journal} {J. Chem. Phys.}\
  }\textbf {\bibinfo {volume} {134}},\ \bibinfo {pages} {054128} (\bibinfo
  {year} {2011})}\BibitemShut {NoStop}%
\bibitem [{\citenamefont {Szabo}\ and\ \citenamefont
  {Ostlund}(1996)}]{szabo2012modern}%
  \BibitemOpen
  \bibfield  {author} {\bibinfo {author} {\bibfnamefont {A.}~\bibnamefont
  {Szabo}}\ and\ \bibinfo {author} {\bibfnamefont {N.~S.}\ \bibnamefont
  {Ostlund}},\ }\href@noop {} {\emph {\bibinfo {title} {{Modern Quantum
  Chemistry: Introduction to Advanced Electronic Structure Theory}}}}\
  (\bibinfo  {publisher} {Dover Publications, Inc.},\ \bibinfo {address}
  {Mineola, New York},\ \bibinfo {year} {1996})\ pp.\ \bibinfo {pages}
  {286--296}\BibitemShut {NoStop}%
\bibitem [{\citenamefont {Kohn}\ and\ \citenamefont
  {Sham}(1965)}]{kohn1965self}%
  \BibitemOpen
  \bibfield  {author} {\bibinfo {author} {\bibfnamefont {W.}~\bibnamefont
  {Kohn}}\ and\ \bibinfo {author} {\bibfnamefont {L.~J.}\ \bibnamefont
  {Sham}},\ }\href@noop {} {\bibfield  {journal} {\bibinfo  {journal} {Phys.
  Rev.}\ }\textbf {\bibinfo {volume} {140}},\ \bibinfo {pages} {A1133}
  (\bibinfo {year} {1965})}\BibitemShut {NoStop}%
\bibitem [{\citenamefont {Hait}, \citenamefont {Rettig},\ and\ \citenamefont
  {Head-Gordon}(2019{\natexlab{b}})}]{hait2019wellbehaved}%
  \BibitemOpen
  \bibfield  {author} {\bibinfo {author} {\bibfnamefont {D.}~\bibnamefont
  {Hait}}, \bibinfo {author} {\bibfnamefont {A.}~\bibnamefont {Rettig}}, \ and\
  \bibinfo {author} {\bibfnamefont {M.}~\bibnamefont {Head-Gordon}},\
  }\href@noop {} {\bibfield  {journal} {\bibinfo  {journal} {J. Chem. Phys.}\
  }\textbf {\bibinfo {volume} {150}},\ \bibinfo {pages} {094115} (\bibinfo
  {year} {2019}{\natexlab{b}})}\BibitemShut {NoStop}%
\bibitem [{\citenamefont {Yamaguchi}\ \emph {et~al.}(1988)\citenamefont
  {Yamaguchi}, \citenamefont {Jensen}, \citenamefont {Dorigo},\ and\
  \citenamefont {Houk}}]{yamaguchi1988spin}%
  \BibitemOpen
  \bibfield  {author} {\bibinfo {author} {\bibfnamefont {K.}~\bibnamefont
  {Yamaguchi}}, \bibinfo {author} {\bibfnamefont {F.}~\bibnamefont {Jensen}},
  \bibinfo {author} {\bibfnamefont {A.}~\bibnamefont {Dorigo}}, \ and\ \bibinfo
  {author} {\bibfnamefont {K.}~\bibnamefont {Houk}},\ }\href@noop {} {\bibfield
   {journal} {\bibinfo  {journal} {Chem. Phys. Lett.}\ }\textbf {\bibinfo
  {volume} {149}},\ \bibinfo {pages} {537} (\bibinfo {year}
  {1988})}\BibitemShut {NoStop}%
\bibitem [{\citenamefont {Yamaguchi}(1979)}]{yamaguchi1979singlet}%
  \BibitemOpen
  \bibfield  {author} {\bibinfo {author} {\bibfnamefont {K.}~\bibnamefont
  {Yamaguchi}},\ }\href@noop {} {\bibfield  {journal} {\bibinfo  {journal}
  {Chem. Phys. Lett.}\ }\textbf {\bibinfo {volume} {66}},\ \bibinfo {pages}
  {395} (\bibinfo {year} {1979})}\BibitemShut {NoStop}%
\bibitem [{\citenamefont {Noodleman}(1981)}]{noodleman1981valence}%
  \BibitemOpen
  \bibfield  {author} {\bibinfo {author} {\bibfnamefont {L.}~\bibnamefont
  {Noodleman}},\ }\href@noop {} {\bibfield  {journal} {\bibinfo  {journal} {J.
  Chem. Phys.}\ }\textbf {\bibinfo {volume} {74}},\ \bibinfo {pages} {5737}
  (\bibinfo {year} {1981})}\BibitemShut {NoStop}%
\bibitem [{\citenamefont {Noodleman}\ \emph {et~al.}(1985)\citenamefont
  {Noodleman}, \citenamefont {Norman~Jr}, \citenamefont {Osborne},
  \citenamefont {Aizman},\ and\ \citenamefont {Case}}]{noodleman1985models}%
  \BibitemOpen
  \bibfield  {author} {\bibinfo {author} {\bibfnamefont {L.}~\bibnamefont
  {Noodleman}}, \bibinfo {author} {\bibfnamefont {J.~G.}\ \bibnamefont
  {Norman~Jr}}, \bibinfo {author} {\bibfnamefont {J.~H.}\ \bibnamefont
  {Osborne}}, \bibinfo {author} {\bibfnamefont {A.}~\bibnamefont {Aizman}}, \
  and\ \bibinfo {author} {\bibfnamefont {D.~A.}\ \bibnamefont {Case}},\
  }\href@noop {} {\bibfield  {journal} {\bibinfo  {journal} {J. Am. Chem.
  Soc.}\ }\textbf {\bibinfo {volume} {107}},\ \bibinfo {pages} {3418} (\bibinfo
  {year} {1985})}\BibitemShut {NoStop}%
\bibitem [{\citenamefont {Sinnecker}\ \emph {et~al.}(2004)\citenamefont
  {Sinnecker}, \citenamefont {Neese}, \citenamefont {Noodleman},\ and\
  \citenamefont {Lubitz}}]{sinnecker2004calculating}%
  \BibitemOpen
  \bibfield  {author} {\bibinfo {author} {\bibfnamefont {S.}~\bibnamefont
  {Sinnecker}}, \bibinfo {author} {\bibfnamefont {F.}~\bibnamefont {Neese}},
  \bibinfo {author} {\bibfnamefont {L.}~\bibnamefont {Noodleman}}, \ and\
  \bibinfo {author} {\bibfnamefont {W.}~\bibnamefont {Lubitz}},\ }\href@noop {}
  {\bibfield  {journal} {\bibinfo  {journal} {J. Am. Chem. Soc.}\ }\textbf
  {\bibinfo {volume} {126}},\ \bibinfo {pages} {2613} (\bibinfo {year}
  {2004})}\BibitemShut {NoStop}%
\bibitem [{\citenamefont {Lovell}\ \emph {et~al.}(2001)\citenamefont {Lovell},
  \citenamefont {Li}, \citenamefont {Liu}, \citenamefont {Case},\ and\
  \citenamefont {Noodleman}}]{lovell2001femo}%
  \BibitemOpen
  \bibfield  {author} {\bibinfo {author} {\bibfnamefont {T.}~\bibnamefont
  {Lovell}}, \bibinfo {author} {\bibfnamefont {J.}~\bibnamefont {Li}}, \bibinfo
  {author} {\bibfnamefont {T.}~\bibnamefont {Liu}}, \bibinfo {author}
  {\bibfnamefont {D.~A.}\ \bibnamefont {Case}}, \ and\ \bibinfo {author}
  {\bibfnamefont {L.}~\bibnamefont {Noodleman}},\ }\href@noop {} {\bibfield
  {journal} {\bibinfo  {journal} {J. Am. Chem. Soc.}\ }\textbf {\bibinfo
  {volume} {123}},\ \bibinfo {pages} {12392} (\bibinfo {year}
  {2001})}\BibitemShut {NoStop}%
\bibitem [{\citenamefont {Adams}, \citenamefont {Noodleman},\ and\
  \citenamefont {Hendrickson}(1997)}]{adams1997density}%
  \BibitemOpen
  \bibfield  {author} {\bibinfo {author} {\bibfnamefont {D.~M.}\ \bibnamefont
  {Adams}}, \bibinfo {author} {\bibfnamefont {L.}~\bibnamefont {Noodleman}}, \
  and\ \bibinfo {author} {\bibfnamefont {D.~N.}\ \bibnamefont {Hendrickson}},\
  }\href@noop {} {\bibfield  {journal} {\bibinfo  {journal} {Inorg. Chem.}\
  }\textbf {\bibinfo {volume} {36}},\ \bibinfo {pages} {3966} (\bibinfo {year}
  {1997})}\BibitemShut {NoStop}%
\bibitem [{\citenamefont {Mouesca}, \citenamefont {Noodleman},\ and\
  \citenamefont {Case}(1995)}]{mouesca1995density}%
  \BibitemOpen
  \bibfield  {author} {\bibinfo {author} {\bibfnamefont {J.-M.}\ \bibnamefont
  {Mouesca}}, \bibinfo {author} {\bibfnamefont {L.}~\bibnamefont {Noodleman}},
  \ and\ \bibinfo {author} {\bibfnamefont {D.~A.}\ \bibnamefont {Case}},\
  }\href@noop {} {\bibfield  {journal} {\bibinfo  {journal} {Int. J. Quantum
  Chem}\ }\textbf {\bibinfo {volume} {56}},\ \bibinfo {pages} {95} (\bibinfo
  {year} {1995})}\BibitemShut {NoStop}%
\bibitem [{\citenamefont {Witzke}\ \emph {et~al.}(2020)\citenamefont {Witzke},
  \citenamefont {Hait}, \citenamefont {Chakarawet}, \citenamefont
  {Head-Gordon},\ and\ \citenamefont {Tilley}}]{witzke2020bimetallic}%
  \BibitemOpen
  \bibfield  {author} {\bibinfo {author} {\bibfnamefont {R.~J.}\ \bibnamefont
  {Witzke}}, \bibinfo {author} {\bibfnamefont {D.}~\bibnamefont {Hait}},
  \bibinfo {author} {\bibfnamefont {K.}~\bibnamefont {Chakarawet}}, \bibinfo
  {author} {\bibfnamefont {M.}~\bibnamefont {Head-Gordon}}, \ and\ \bibinfo
  {author} {\bibfnamefont {T.~D.}\ \bibnamefont {Tilley}},\ }\href@noop {}
  {\bibfield  {journal} {\bibinfo  {journal} {ACS Catalysis}\ } (\bibinfo
  {year} {2020})}\BibitemShut {NoStop}%
\bibitem [{\citenamefont {Hait}\ \emph {et~al.}(2016)\citenamefont {Hait},
  \citenamefont {Zhu}, \citenamefont {McMahon},\ and\ \citenamefont
  {Van~Voorhis}}]{hait2016prediction}%
  \BibitemOpen
  \bibfield  {author} {\bibinfo {author} {\bibfnamefont {D.}~\bibnamefont
  {Hait}}, \bibinfo {author} {\bibfnamefont {T.}~\bibnamefont {Zhu}}, \bibinfo
  {author} {\bibfnamefont {D.~P.}\ \bibnamefont {McMahon}}, \ and\ \bibinfo
  {author} {\bibfnamefont {T.}~\bibnamefont {Van~Voorhis}},\ }\href@noop {}
  {\bibfield  {journal} {\bibinfo  {journal} {J. Chem. Theory Comput.}\
  }\textbf {\bibinfo {volume} {12}},\ \bibinfo {pages} {3353} (\bibinfo {year}
  {2016})}\BibitemShut {NoStop}%
\bibitem [{\citenamefont {Thom}\ and\ \citenamefont
  {Head-Gordon}(2009)}]{thom2009hartree}%
  \BibitemOpen
  \bibfield  {author} {\bibinfo {author} {\bibfnamefont {A.~J.}\ \bibnamefont
  {Thom}}\ and\ \bibinfo {author} {\bibfnamefont {M.}~\bibnamefont
  {Head-Gordon}},\ }\href@noop {} {\bibfield  {journal} {\bibinfo  {journal}
  {J. Chem. Phys.}\ }\textbf {\bibinfo {volume} {131}},\ \bibinfo {pages}
  {124113} (\bibinfo {year} {2009})}\BibitemShut {NoStop}%
\bibitem [{\citenamefont {Yang}(2018)}]{yang2018}%
  \BibitemOpen
  \bibfield  {author} {\bibinfo {author} {\bibfnamefont {Z.}~\bibnamefont
  {Yang}},\ }\emph {\bibinfo {title} {Characterization of Substituted Radicals
  by Multi-Edge Femtosecond X-Ray Transient Absorption Spectroscopy}},\
  \href@noop {} {Ph.D. thesis},\ \bibinfo  {school} {UC Berkeley} (\bibinfo
  {year} {2018})\BibitemShut {NoStop}%
\bibitem [{\citenamefont {Alagia}\ \emph {et~al.}(2007)\citenamefont {Alagia},
  \citenamefont {Lavoll{\'e}e}, \citenamefont {Richter}, \citenamefont
  {Ekstr{\"o}m}, \citenamefont {Carravetta}, \citenamefont {Stranges},
  \citenamefont {Brunetti},\ and\ \citenamefont
  {Stranges}}]{alagia2007probing}%
  \BibitemOpen
  \bibfield  {author} {\bibinfo {author} {\bibfnamefont {M.}~\bibnamefont
  {Alagia}}, \bibinfo {author} {\bibfnamefont {M.}~\bibnamefont
  {Lavoll{\'e}e}}, \bibinfo {author} {\bibfnamefont {R.}~\bibnamefont
  {Richter}}, \bibinfo {author} {\bibfnamefont {U.}~\bibnamefont
  {Ekstr{\"o}m}}, \bibinfo {author} {\bibfnamefont {V.}~\bibnamefont
  {Carravetta}}, \bibinfo {author} {\bibfnamefont {D.}~\bibnamefont
  {Stranges}}, \bibinfo {author} {\bibfnamefont {B.}~\bibnamefont {Brunetti}},
  \ and\ \bibinfo {author} {\bibfnamefont {S.}~\bibnamefont {Stranges}},\
  }\href@noop {} {\bibfield  {journal} {\bibinfo  {journal} {Phys. Rev. A}\
  }\textbf {\bibinfo {volume} {76}},\ \bibinfo {pages} {022509} (\bibinfo
  {year} {2007})}\BibitemShut {NoStop}%
\bibitem [{\citenamefont {Alagia}\ \emph {et~al.}(2013)\citenamefont {Alagia},
  \citenamefont {Bodo}, \citenamefont {Decleva}, \citenamefont {Falcinelli},
  \citenamefont {Ponzi}, \citenamefont {Richter},\ and\ \citenamefont
  {Stranges}}]{alagia2013soft}%
  \BibitemOpen
  \bibfield  {author} {\bibinfo {author} {\bibfnamefont {M.}~\bibnamefont
  {Alagia}}, \bibinfo {author} {\bibfnamefont {E.}~\bibnamefont {Bodo}},
  \bibinfo {author} {\bibfnamefont {P.}~\bibnamefont {Decleva}}, \bibinfo
  {author} {\bibfnamefont {S.}~\bibnamefont {Falcinelli}}, \bibinfo {author}
  {\bibfnamefont {A.}~\bibnamefont {Ponzi}}, \bibinfo {author} {\bibfnamefont
  {R.}~\bibnamefont {Richter}}, \ and\ \bibinfo {author} {\bibfnamefont
  {S.}~\bibnamefont {Stranges}},\ }\href@noop {} {\bibfield  {journal}
  {\bibinfo  {journal} {Phys. Chem. Chem. Phys.}\ }\textbf {\bibinfo {volume}
  {15}},\ \bibinfo {pages} {1310} (\bibinfo {year} {2013})}\BibitemShut
  {NoStop}%
\bibitem [{\citenamefont {Dirac}(1931)}]{Slater}%
  \BibitemOpen
  \bibfield  {author} {\bibinfo {author} {\bibfnamefont {P.~A.}\ \bibnamefont
  {Dirac}},\ }in\ \href@noop {} {\emph {\bibinfo {booktitle} {{Proc. R. Soc.
  A}}}},\ Vol.\ \bibinfo {volume} {133}\ (\bibinfo {organization} {The Royal
  Society},\ \bibinfo {year} {1931})\ pp.\ \bibinfo {pages}
  {60--72}\BibitemShut {NoStop}%
\bibitem [{\citenamefont {Vosko}, \citenamefont {Wilk},\ and\ \citenamefont
  {Nusair}(1980)}]{VWN}%
  \BibitemOpen
  \bibfield  {author} {\bibinfo {author} {\bibfnamefont {S.~H.}\ \bibnamefont
  {Vosko}}, \bibinfo {author} {\bibfnamefont {L.}~\bibnamefont {Wilk}}, \ and\
  \bibinfo {author} {\bibfnamefont {M.}~\bibnamefont {Nusair}},\ }\href@noop {}
  {\bibfield  {journal} {\bibinfo  {journal} {Can. J. Phys.}\ }\textbf
  {\bibinfo {volume} {58}},\ \bibinfo {pages} {1200} (\bibinfo {year}
  {1980})}\BibitemShut {NoStop}%
\bibitem [{\citenamefont {Perdew}\ and\ \citenamefont {Wang}(1992)}]{PW92}%
  \BibitemOpen
  \bibfield  {author} {\bibinfo {author} {\bibfnamefont {J.~P.}\ \bibnamefont
  {Perdew}}\ and\ \bibinfo {author} {\bibfnamefont {Y.}~\bibnamefont {Wang}},\
  }\href@noop {} {\bibfield  {journal} {\bibinfo  {journal} {Phys. Rev. B}\
  }\textbf {\bibinfo {volume} {45}},\ \bibinfo {pages} {13244} (\bibinfo {year}
  {1992})}\BibitemShut {NoStop}%
\bibitem [{\citenamefont {Becke}(1988)}]{b88}%
  \BibitemOpen
  \bibfield  {author} {\bibinfo {author} {\bibfnamefont {A.~D.}\ \bibnamefont
  {Becke}},\ }\href@noop {} {\bibfield  {journal} {\bibinfo  {journal} {Phys.
  Rev. A}\ }\textbf {\bibinfo {volume} {38}},\ \bibinfo {pages} {3098}
  (\bibinfo {year} {1988})}\BibitemShut {NoStop}%
\bibitem [{\citenamefont {Lee}, \citenamefont {Yang},\ and\ \citenamefont
  {Parr}(1988)}]{lyp}%
  \BibitemOpen
  \bibfield  {author} {\bibinfo {author} {\bibfnamefont {C.}~\bibnamefont
  {Lee}}, \bibinfo {author} {\bibfnamefont {W.}~\bibnamefont {Yang}}, \ and\
  \bibinfo {author} {\bibfnamefont {R.~G.}\ \bibnamefont {Parr}},\ }\href@noop
  {} {\bibfield  {journal} {\bibinfo  {journal} {Phys. Rev. B}\ }\textbf
  {\bibinfo {volume} {37}},\ \bibinfo {pages} {785} (\bibinfo {year}
  {1988})}\BibitemShut {NoStop}%
\bibitem [{\citenamefont {Perdew}, \citenamefont {Burke},\ and\ \citenamefont
  {Ernzerhof}(1996)}]{PBE}%
  \BibitemOpen
  \bibfield  {author} {\bibinfo {author} {\bibfnamefont {J.~P.}\ \bibnamefont
  {Perdew}}, \bibinfo {author} {\bibfnamefont {K.}~\bibnamefont {Burke}}, \
  and\ \bibinfo {author} {\bibfnamefont {M.}~\bibnamefont {Ernzerhof}},\
  }\href@noop {} {\bibfield  {journal} {\bibinfo  {journal} {Phys. Rev. Lett.}\
  }\textbf {\bibinfo {volume} {77}},\ \bibinfo {pages} {3865} (\bibinfo {year}
  {1996})}\BibitemShut {NoStop}%
\bibitem [{\citenamefont {Tao}\ \emph {et~al.}(2003)\citenamefont {Tao},
  \citenamefont {Perdew}, \citenamefont {Staroverov},\ and\ \citenamefont
  {Scuseria}}]{tpss}%
  \BibitemOpen
  \bibfield  {author} {\bibinfo {author} {\bibfnamefont {J.}~\bibnamefont
  {Tao}}, \bibinfo {author} {\bibfnamefont {J.~P.}\ \bibnamefont {Perdew}},
  \bibinfo {author} {\bibfnamefont {V.~N.}\ \bibnamefont {Staroverov}}, \ and\
  \bibinfo {author} {\bibfnamefont {G.~E.}\ \bibnamefont {Scuseria}},\
  }\href@noop {} {\bibfield  {journal} {\bibinfo  {journal} {Phys. Rev. Lett.}\
  }\textbf {\bibinfo {volume} {91}},\ \bibinfo {pages} {146401} (\bibinfo
  {year} {2003})}\BibitemShut {NoStop}%
\bibitem [{\citenamefont {Becke}(1993)}]{b3lyp}%
  \BibitemOpen
  \bibfield  {author} {\bibinfo {author} {\bibfnamefont {A.~D.}\ \bibnamefont
  {Becke}},\ }\href@noop {} {\bibfield  {journal} {\bibinfo  {journal} {J.
  Chem. Phys.}\ }\textbf {\bibinfo {volume} {98}},\ \bibinfo {pages} {5648}
  (\bibinfo {year} {1993})}\BibitemShut {NoStop}%
\bibitem [{\citenamefont {Adamo}\ and\ \citenamefont {Barone}(1999)}]{pbe0}%
  \BibitemOpen
  \bibfield  {author} {\bibinfo {author} {\bibfnamefont {C.}~\bibnamefont
  {Adamo}}\ and\ \bibinfo {author} {\bibfnamefont {V.}~\bibnamefont {Barone}},\
  }\href@noop {} {\bibfield  {journal} {\bibinfo  {journal} {J. Chem. Phys.}\
  }\textbf {\bibinfo {volume} {110}},\ \bibinfo {pages} {6158} (\bibinfo {year}
  {1999})}\BibitemShut {NoStop}%
\bibitem [{\citenamefont {Yanai}, \citenamefont {Tew},\ and\ \citenamefont
  {Handy}(2004)}]{camb3lyp}%
  \BibitemOpen
  \bibfield  {author} {\bibinfo {author} {\bibfnamefont {T.}~\bibnamefont
  {Yanai}}, \bibinfo {author} {\bibfnamefont {D.~P.}\ \bibnamefont {Tew}}, \
  and\ \bibinfo {author} {\bibfnamefont {N.~C.}\ \bibnamefont {Handy}},\
  }\href@noop {} {\bibfield  {journal} {\bibinfo  {journal} {Chem. Phys.
  Lett.}\ }\textbf {\bibinfo {volume} {393}},\ \bibinfo {pages} {51} (\bibinfo
  {year} {2004})}\BibitemShut {NoStop}%
\bibitem [{\citenamefont {Lin}\ \emph {et~al.}(2012)\citenamefont {Lin},
  \citenamefont {Li}, \citenamefont {Mao},\ and\ \citenamefont
  {Chai}}]{wB97XD3}%
  \BibitemOpen
  \bibfield  {author} {\bibinfo {author} {\bibfnamefont {Y.-S.}\ \bibnamefont
  {Lin}}, \bibinfo {author} {\bibfnamefont {G.-D.}\ \bibnamefont {Li}},
  \bibinfo {author} {\bibfnamefont {S.-P.}\ \bibnamefont {Mao}}, \ and\
  \bibinfo {author} {\bibfnamefont {J.-D.}\ \bibnamefont {Chai}},\ }\href@noop
  {} {\bibfield  {journal} {\bibinfo  {journal} {J. Chem. Theory Comput.}\
  }\textbf {\bibinfo {volume} {9}},\ \bibinfo {pages} {263} (\bibinfo {year}
  {2012})}\BibitemShut {NoStop}%
\bibitem [{\citenamefont {Mardirossian}\ and\ \citenamefont
  {Head-Gordon}(2014)}]{wb97xv}%
  \BibitemOpen
  \bibfield  {author} {\bibinfo {author} {\bibfnamefont {N.}~\bibnamefont
  {Mardirossian}}\ and\ \bibinfo {author} {\bibfnamefont {M.}~\bibnamefont
  {Head-Gordon}},\ }\href@noop {} {\bibfield  {journal} {\bibinfo  {journal}
  {Phys. Chem. Chem. Phys.}\ }\textbf {\bibinfo {volume} {16}},\ \bibinfo
  {pages} {9904} (\bibinfo {year} {2014})}\BibitemShut {NoStop}%
\bibitem [{\citenamefont {Couto}\ \emph {et~al.}(2020)\citenamefont {Couto},
  \citenamefont {Kjellsson}, \citenamefont {{\AA}gren}, \citenamefont
  {Carravetta}, \citenamefont {Sorensen}, \citenamefont {Kubin}, \citenamefont
  {B{\"u}low}, \citenamefont {Timm}, \citenamefont {Zamudio-Bayer},
  \citenamefont {Issendorff} \emph {et~al.}}]{couto2020carbon}%
  \BibitemOpen
  \bibfield  {author} {\bibinfo {author} {\bibfnamefont {R.~C.}\ \bibnamefont
  {Couto}}, \bibinfo {author} {\bibfnamefont {L.}~\bibnamefont {Kjellsson}},
  \bibinfo {author} {\bibfnamefont {H.}~\bibnamefont {{\AA}gren}}, \bibinfo
  {author} {\bibfnamefont {V.}~\bibnamefont {Carravetta}}, \bibinfo {author}
  {\bibfnamefont {S.}~\bibnamefont {Sorensen}}, \bibinfo {author}
  {\bibfnamefont {M.}~\bibnamefont {Kubin}}, \bibinfo {author} {\bibfnamefont
  {C.}~\bibnamefont {B{\"u}low}}, \bibinfo {author} {\bibfnamefont
  {M.}~\bibnamefont {Timm}}, \bibinfo {author} {\bibfnamefont {V.}~\bibnamefont
  {Zamudio-Bayer}}, \bibinfo {author} {\bibfnamefont {B.~v.}\ \bibnamefont
  {Issendorff}},  \emph {et~al.},\ }\href@noop {} {\bibfield  {journal}
  {\bibinfo  {journal} {Phys. Chem. Chem. Phys.}\ } (\bibinfo {year}
  {2020})}\BibitemShut {NoStop}%
\bibitem [{\citenamefont {Parent}\ \emph {et~al.}(2009)\citenamefont {Parent},
  \citenamefont {Bournel}, \citenamefont {Lasne}, \citenamefont {Lacombe},
  \citenamefont {Strazzulla}, \citenamefont {Gardonio}, \citenamefont {Lizzit},
  \citenamefont {Kappler}, \citenamefont {Joly}, \citenamefont {Laffon} \emph
  {et~al.}}]{parent2009irradiation}%
  \BibitemOpen
  \bibfield  {author} {\bibinfo {author} {\bibfnamefont {P.}~\bibnamefont
  {Parent}}, \bibinfo {author} {\bibfnamefont {F.}~\bibnamefont {Bournel}},
  \bibinfo {author} {\bibfnamefont {J.}~\bibnamefont {Lasne}}, \bibinfo
  {author} {\bibfnamefont {S.}~\bibnamefont {Lacombe}}, \bibinfo {author}
  {\bibfnamefont {G.}~\bibnamefont {Strazzulla}}, \bibinfo {author}
  {\bibfnamefont {S.}~\bibnamefont {Gardonio}}, \bibinfo {author}
  {\bibfnamefont {S.}~\bibnamefont {Lizzit}}, \bibinfo {author} {\bibfnamefont
  {J.-P.}\ \bibnamefont {Kappler}}, \bibinfo {author} {\bibfnamefont
  {L.}~\bibnamefont {Joly}}, \bibinfo {author} {\bibfnamefont {C.}~\bibnamefont
  {Laffon}},  \emph {et~al.},\ }\href@noop {} {\bibfield  {journal} {\bibinfo
  {journal} {J. Chem. Phys.}\ }\textbf {\bibinfo {volume} {131}},\ \bibinfo
  {pages} {154308} (\bibinfo {year} {2009})}\BibitemShut {NoStop}%
\bibitem [{\citenamefont {Lindblad}\ \emph {et~al.}(2020)\citenamefont
  {Lindblad}, \citenamefont {Kjellsson}, \citenamefont {Couto}, \citenamefont
  {Timm}, \citenamefont {B{\"u}low}, \citenamefont {Zamudio-Bayer},
  \citenamefont {Lundberg}, \citenamefont {von Issendorff}, \citenamefont
  {Lau}, \citenamefont {Sorensen} \emph {et~al.}}]{lindblad2020x}%
  \BibitemOpen
  \bibfield  {author} {\bibinfo {author} {\bibfnamefont {R.}~\bibnamefont
  {Lindblad}}, \bibinfo {author} {\bibfnamefont {L.}~\bibnamefont {Kjellsson}},
  \bibinfo {author} {\bibfnamefont {R.~C.}\ \bibnamefont {Couto}}, \bibinfo
  {author} {\bibfnamefont {M.}~\bibnamefont {Timm}}, \bibinfo {author}
  {\bibfnamefont {C.}~\bibnamefont {B{\"u}low}}, \bibinfo {author}
  {\bibfnamefont {V.}~\bibnamefont {Zamudio-Bayer}}, \bibinfo {author}
  {\bibfnamefont {M.}~\bibnamefont {Lundberg}}, \bibinfo {author}
  {\bibfnamefont {B.}~\bibnamefont {von Issendorff}}, \bibinfo {author}
  {\bibfnamefont {J.}~\bibnamefont {Lau}}, \bibinfo {author} {\bibfnamefont
  {S.}~\bibnamefont {Sorensen}},  \emph {et~al.},\ }\href@noop {} {\bibfield
  {journal} {\bibinfo  {journal} {Phys. Rev. Lett.}\ }\textbf {\bibinfo
  {volume} {124}},\ \bibinfo {pages} {203001} (\bibinfo {year}
  {2020})}\BibitemShut {NoStop}%
\bibitem [{\citenamefont {Bari}\ \emph {et~al.}(2019)\citenamefont {Bari},
  \citenamefont {Inhester}, \citenamefont {Schubert}, \citenamefont {Mertens},
  \citenamefont {Schunck}, \citenamefont {D{\"o}rner}, \citenamefont {Deinert},
  \citenamefont {Schwob}, \citenamefont {Schippers}, \citenamefont {M{\"u}ller}
  \emph {et~al.}}]{bari2019inner}%
  \BibitemOpen
  \bibfield  {author} {\bibinfo {author} {\bibfnamefont {S.}~\bibnamefont
  {Bari}}, \bibinfo {author} {\bibfnamefont {L.}~\bibnamefont {Inhester}},
  \bibinfo {author} {\bibfnamefont {K.}~\bibnamefont {Schubert}}, \bibinfo
  {author} {\bibfnamefont {K.}~\bibnamefont {Mertens}}, \bibinfo {author}
  {\bibfnamefont {J.~O.}\ \bibnamefont {Schunck}}, \bibinfo {author}
  {\bibfnamefont {S.}~\bibnamefont {D{\"o}rner}}, \bibinfo {author}
  {\bibfnamefont {S.}~\bibnamefont {Deinert}}, \bibinfo {author} {\bibfnamefont
  {L.}~\bibnamefont {Schwob}}, \bibinfo {author} {\bibfnamefont
  {S.}~\bibnamefont {Schippers}}, \bibinfo {author} {\bibfnamefont
  {A.}~\bibnamefont {M{\"u}ller}},  \emph {et~al.},\ }\href@noop {} {\bibfield
  {journal} {\bibinfo  {journal} {Phys. Chem. Chem. Phys.}\ }\textbf {\bibinfo
  {volume} {21}},\ \bibinfo {pages} {16505} (\bibinfo {year}
  {2019})}\BibitemShut {NoStop}%
\bibitem [{\citenamefont {Zhang}\ \emph {et~al.}(1990)\citenamefont {Zhang},
  \citenamefont {Sze}, \citenamefont {Brion}, \citenamefont {Tong},\ and\
  \citenamefont {Li}}]{zhang1990inner}%
  \BibitemOpen
  \bibfield  {author} {\bibinfo {author} {\bibfnamefont {W.}~\bibnamefont
  {Zhang}}, \bibinfo {author} {\bibfnamefont {K.}~\bibnamefont {Sze}}, \bibinfo
  {author} {\bibfnamefont {C.}~\bibnamefont {Brion}}, \bibinfo {author}
  {\bibfnamefont {X.}~\bibnamefont {Tong}}, \ and\ \bibinfo {author}
  {\bibfnamefont {J.}~\bibnamefont {Li}},\ }\href@noop {} {\bibfield  {journal}
  {\bibinfo  {journal} {Chem. Phys.}\ }\textbf {\bibinfo {volume} {140}},\
  \bibinfo {pages} {265} (\bibinfo {year} {1990})}\BibitemShut {NoStop}%
\bibitem [{\citenamefont {Stranges}, \citenamefont {Richter},\ and\
  \citenamefont {Alagia}(2002)}]{stranges2002high}%
  \BibitemOpen
  \bibfield  {author} {\bibinfo {author} {\bibfnamefont {S.}~\bibnamefont
  {Stranges}}, \bibinfo {author} {\bibfnamefont {R.}~\bibnamefont {Richter}}, \
  and\ \bibinfo {author} {\bibfnamefont {M.}~\bibnamefont {Alagia}},\
  }\href@noop {} {\bibfield  {journal} {\bibinfo  {journal} {J. Chem. Phys.}\
  }\textbf {\bibinfo {volume} {116}},\ \bibinfo {pages} {3676} (\bibinfo {year}
  {2002})}\BibitemShut {NoStop}%
\bibitem [{\citenamefont {Lacombe}\ \emph {et~al.}(2006)\citenamefont
  {Lacombe}, \citenamefont {Bournel}, \citenamefont {Laffon},\ and\
  \citenamefont {Parent}}]{lacombe2006radical}%
  \BibitemOpen
  \bibfield  {author} {\bibinfo {author} {\bibfnamefont {S.}~\bibnamefont
  {Lacombe}}, \bibinfo {author} {\bibfnamefont {F.}~\bibnamefont {Bournel}},
  \bibinfo {author} {\bibfnamefont {C.}~\bibnamefont {Laffon}}, \ and\ \bibinfo
  {author} {\bibfnamefont {P.}~\bibnamefont {Parent}},\ }\href@noop {}
  {\bibfield  {journal} {\bibinfo  {journal} {Angew. Chem. Int. Ed.}\ }\textbf
  {\bibinfo {volume} {45}},\ \bibinfo {pages} {4159} (\bibinfo {year}
  {2006})}\BibitemShut {NoStop}%
\bibitem [{\citenamefont {Coreno}\ \emph {et~al.}(1999)\citenamefont {Coreno},
  \citenamefont {De~Simone}, \citenamefont {Prince}, \citenamefont {Richter},
  \citenamefont {Vondr{\'a}{\v{c}}ek}, \citenamefont {Avaldi},\ and\
  \citenamefont {Camilloni}}]{coreno1999vibrationally}%
  \BibitemOpen
  \bibfield  {author} {\bibinfo {author} {\bibfnamefont {M.}~\bibnamefont
  {Coreno}}, \bibinfo {author} {\bibfnamefont {M.}~\bibnamefont {De~Simone}},
  \bibinfo {author} {\bibfnamefont {K.}~\bibnamefont {Prince}}, \bibinfo
  {author} {\bibfnamefont {R.}~\bibnamefont {Richter}}, \bibinfo {author}
  {\bibfnamefont {M.}~\bibnamefont {Vondr{\'a}{\v{c}}ek}}, \bibinfo {author}
  {\bibfnamefont {L.}~\bibnamefont {Avaldi}}, \ and\ \bibinfo {author}
  {\bibfnamefont {R.}~\bibnamefont {Camilloni}},\ }\href@noop {} {\bibfield
  {journal} {\bibinfo  {journal} {Chem. Phys. Lett.}\ }\textbf {\bibinfo
  {volume} {306}},\ \bibinfo {pages} {269} (\bibinfo {year}
  {1999})}\BibitemShut {NoStop}%
\bibitem [{\citenamefont {Dunning~Jr}(1989)}]{dunning1989gaussian}%
  \BibitemOpen
  \bibfield  {author} {\bibinfo {author} {\bibfnamefont {T.~H.}\ \bibnamefont
  {Dunning~Jr}},\ }\href@noop {} {\bibfield  {journal} {\bibinfo  {journal} {J.
  Chem. Phys.}\ }\textbf {\bibinfo {volume} {90}},\ \bibinfo {pages} {1007}
  (\bibinfo {year} {1989})}\BibitemShut {NoStop}%
\bibitem [{\citenamefont {Kendall}, \citenamefont {Dunning~Jr},\ and\
  \citenamefont {Harrison}(1992)}]{kendall1992electron}%
  \BibitemOpen
  \bibfield  {author} {\bibinfo {author} {\bibfnamefont {R.~A.}\ \bibnamefont
  {Kendall}}, \bibinfo {author} {\bibfnamefont {T.~H.}\ \bibnamefont
  {Dunning~Jr}}, \ and\ \bibinfo {author} {\bibfnamefont {R.~J.}\ \bibnamefont
  {Harrison}},\ }\href@noop {} {\bibfield  {journal} {\bibinfo  {journal} {J.
  Chem. Phys.}\ }\textbf {\bibinfo {volume} {96}},\ \bibinfo {pages} {6796}
  (\bibinfo {year} {1992})}\BibitemShut {NoStop}%
\bibitem [{\citenamefont {Woon}\ and\ \citenamefont
  {Dunning~Jr}(1995)}]{woon1995gaussian}%
  \BibitemOpen
  \bibfield  {author} {\bibinfo {author} {\bibfnamefont {D.~E.}\ \bibnamefont
  {Woon}}\ and\ \bibinfo {author} {\bibfnamefont {T.~H.}\ \bibnamefont
  {Dunning~Jr}},\ }\href@noop {} {\bibfield  {journal} {\bibinfo  {journal} {J.
  Chem. Phys.}\ }\textbf {\bibinfo {volume} {103}},\ \bibinfo {pages} {4572}
  (\bibinfo {year} {1995})}\BibitemShut {NoStop}%
\bibitem [{\citenamefont {Takahashi}(2017)}]{takahashi2017relativistic}%
  \BibitemOpen
  \bibfield  {author} {\bibinfo {author} {\bibfnamefont {O.}~\bibnamefont
  {Takahashi}},\ }\href@noop {} {\bibfield  {journal} {\bibinfo  {journal}
  {Computational and Theoretical Chemistry}\ }\textbf {\bibinfo {volume}
  {1102}},\ \bibinfo {pages} {80} (\bibinfo {year} {2017})}\BibitemShut
  {NoStop}%
\bibitem [{\citenamefont {Cremer}(2011)}]{cremer2011moller}%
  \BibitemOpen
  \bibfield  {author} {\bibinfo {author} {\bibfnamefont {D.}~\bibnamefont
  {Cremer}},\ }\href@noop {} {\bibfield  {journal} {\bibinfo  {journal} {Wiley
  Interdisciplinary Reviews: Computational Molecular Science}\ }\textbf
  {\bibinfo {volume} {1}},\ \bibinfo {pages} {509} (\bibinfo {year}
  {2011})}\BibitemShut {NoStop}%
\bibitem [{\citenamefont {Gill}\ \emph {et~al.}(1988)\citenamefont {Gill},
  \citenamefont {Pople}, \citenamefont {Radom},\ and\ \citenamefont
  {Nobes}}]{gill1988does}%
  \BibitemOpen
  \bibfield  {author} {\bibinfo {author} {\bibfnamefont {P.~M.}\ \bibnamefont
  {Gill}}, \bibinfo {author} {\bibfnamefont {J.~A.}\ \bibnamefont {Pople}},
  \bibinfo {author} {\bibfnamefont {L.}~\bibnamefont {Radom}}, \ and\ \bibinfo
  {author} {\bibfnamefont {R.~H.}\ \bibnamefont {Nobes}},\ }\href@noop {}
  {\bibfield  {journal} {\bibinfo  {journal} {J. Chem. Phys.}\ }\textbf
  {\bibinfo {volume} {89}},\ \bibinfo {pages} {7307} (\bibinfo {year}
  {1988})}\BibitemShut {NoStop}%
\bibitem [{\citenamefont {Johansson}\ \emph {et~al.}(2018)\citenamefont
  {Johansson}, \citenamefont {Head-Gordon}, \citenamefont {Schrader},
  \citenamefont {Wilson},\ and\ \citenamefont
  {Michelsen}}]{johansson2018resonance}%
  \BibitemOpen
  \bibfield  {author} {\bibinfo {author} {\bibfnamefont {K.}~\bibnamefont
  {Johansson}}, \bibinfo {author} {\bibfnamefont {M.}~\bibnamefont
  {Head-Gordon}}, \bibinfo {author} {\bibfnamefont {P.}~\bibnamefont
  {Schrader}}, \bibinfo {author} {\bibfnamefont {K.}~\bibnamefont {Wilson}}, \
  and\ \bibinfo {author} {\bibfnamefont {H.}~\bibnamefont {Michelsen}},\
  }\href@noop {} {\bibfield  {journal} {\bibinfo  {journal} {Science}\ }\textbf
  {\bibinfo {volume} {361}},\ \bibinfo {pages} {997} (\bibinfo {year}
  {2018})}\BibitemShut {NoStop}%
\bibitem [{\citenamefont {Woon}\ and\ \citenamefont
  {Dunning~Jr}(1994)}]{woon1994gaussian}%
  \BibitemOpen
  \bibfield  {author} {\bibinfo {author} {\bibfnamefont {D.~E.}\ \bibnamefont
  {Woon}}\ and\ \bibinfo {author} {\bibfnamefont {T.~H.}\ \bibnamefont
  {Dunning~Jr}},\ }\href@noop {} {\bibfield  {journal} {\bibinfo  {journal} {J.
  Chem. Phys.}\ }\textbf {\bibinfo {volume} {100}},\ \bibinfo {pages} {2975}
  (\bibinfo {year} {1994})}\BibitemShut {NoStop}%
\bibitem [{\citenamefont {Hait}\ and\ \citenamefont
  {Head-Gordon}(2018)}]{hait2018delocalization}%
  \BibitemOpen
  \bibfield  {author} {\bibinfo {author} {\bibfnamefont {D.}~\bibnamefont
  {Hait}}\ and\ \bibinfo {author} {\bibfnamefont {M.}~\bibnamefont
  {Head-Gordon}},\ }\href@noop {} {\bibfield  {journal} {\bibinfo  {journal}
  {J. Phys. Chem. Lett.}\ }\textbf {\bibinfo {volume} {9}},\ \bibinfo {pages}
  {6280} (\bibinfo {year} {2018})}\BibitemShut {NoStop}%
\bibitem [{\citenamefont {Shao}\ \emph {et~al.}(2015)\citenamefont {Shao},
  \citenamefont {Gan}, \citenamefont {Epifanovsky}, \citenamefont {Gilbert},
  \citenamefont {Wormit}, \citenamefont {Kussmann}, \citenamefont {Lange},
  \citenamefont {Behn}, \citenamefont {Deng}, \citenamefont {Feng},
  \citenamefont {Ghosh}, \citenamefont {Goldey}, \citenamefont {Horn},
  \citenamefont {Jacobson}, \citenamefont {Kaliman}, \citenamefont
  {Khaliullin}, \citenamefont {K\'us}, \citenamefont {Landau}, \citenamefont
  {Liu}, \citenamefont {Proynov}, \citenamefont {Rhee}, \citenamefont
  {Richard}, \citenamefont {Rohrdanz}, \citenamefont {Steele}, \citenamefont
  {Sundstrom}, \citenamefont {{Woodcock III}}, \citenamefont {Zimmerman},
  \citenamefont {Zuev}, \citenamefont {Albrecht}, \citenamefont {Alguire},
  \citenamefont {Austin}, \citenamefont {Beran}, \citenamefont {Bernard},
  \citenamefont {Berquist}, \citenamefont {Brandhorst}, \citenamefont
  {Bravaya}, \citenamefont {Brown}, \citenamefont {Casanova}, \citenamefont
  {Chang}, \citenamefont {Chen}, \citenamefont {Chien}, \citenamefont
  {Closser}, \citenamefont {Crittenden}, \citenamefont {Diedenhofen},
  \citenamefont {{DiStasio Jr.}}, \citenamefont {Dop}, \citenamefont {Dutoi},
  \citenamefont {Edgar}, \citenamefont {Fatehi}, \citenamefont
  {{Fusti-Molnar}}, \citenamefont {Ghysels}, \citenamefont
  {{Golubeva-Zadorozhnaya}}, \citenamefont {Gomes}, \citenamefont
  {{Hanson-Heine}}, \citenamefont {Harbach}, \citenamefont {Hauser},
  \citenamefont {Hohenstein}, \citenamefont {Holden}, \citenamefont {Jagau},
  \citenamefont {Ji}, \citenamefont {Kaduk}, \citenamefont {Khistyaev},
  \citenamefont {Kim}, \citenamefont {Kim}, \citenamefont {King}, \citenamefont
  {Klunzinger}, \citenamefont {Kosenkov}, \citenamefont {Kowalczyk},
  \citenamefont {Krauter}, \citenamefont {Lao}, \citenamefont {Laurent},
  \citenamefont {Lawler}, \citenamefont {Levchenko}, \citenamefont {Lin},
  \citenamefont {Liu}, \citenamefont {Livshits}, \citenamefont {Lochan},
  \citenamefont {Luenser}, \citenamefont {Manohar}, \citenamefont {Manzer},
  \citenamefont {Mao}, \citenamefont {Mardirossian}, \citenamefont {Marenich},
  \citenamefont {Maurer}, \citenamefont {Mayhall}, \citenamefont {Oana},
  \citenamefont {{Olivares-Amaya}}, \citenamefont {O'Neill}, \citenamefont
  {Parkhill}, \citenamefont {Perrine}, \citenamefont {Peverati}, \citenamefont
  {Pieniazek}, \citenamefont {Prociuk}, \citenamefont {Rehn}, \citenamefont
  {Rosta}, \citenamefont {Russ}, \citenamefont {Sergueev}, \citenamefont
  {Sharada}, \citenamefont {Sharmaa}, \citenamefont {Small}, \citenamefont
  {Sodt}, \citenamefont {Stein}, \citenamefont {St\"uck}, \citenamefont {Su},
  \citenamefont {Thom}, \citenamefont {Tsuchimochi}, \citenamefont {Vogt},
  \citenamefont {Vydrov}, \citenamefont {Wang}, \citenamefont {Watson},
  \citenamefont {Wenzel}, \citenamefont {White}, \citenamefont {Williams},
  \citenamefont {Vanovschi}, \citenamefont {Yeganeh}, \citenamefont {Yost},
  \citenamefont {You}, \citenamefont {Zhang}, \citenamefont {Zhang},
  \citenamefont {Zhou}, \citenamefont {Brooks}, \citenamefont {Chan},
  \citenamefont {Chipman}, \citenamefont {Cramer}, \citenamefont {{Goddard
  III}}, \citenamefont {Gordon}, \citenamefont {Hehre}, \citenamefont {Klamt},
  \citenamefont {{Schaefer III}}, \citenamefont {Schmidt}, \citenamefont
  {Sherrill}, \citenamefont {Truhlar}, \citenamefont {Warshel}, \citenamefont
  {Xua}, \citenamefont {{Aspuru-Guzik}}, \citenamefont {Baer}, \citenamefont
  {Bell}, \citenamefont {Besley}, \citenamefont {Chai}, \citenamefont {Dreuw},
  \citenamefont {Dunietz}, \citenamefont {Furlani}, \citenamefont {Gwaltney},
  \citenamefont {Hsu}, \citenamefont {Jung}, \citenamefont {Kong},
  \citenamefont {Lambrecht}, \citenamefont {Liang}, \citenamefont {Ochsenfeld},
  \citenamefont {Rassolov}, \citenamefont {Slipchenko}, \citenamefont
  {Subotnik}, \citenamefont {{Van Voorhis}}, \citenamefont {Herbert},
  \citenamefont {Krylov}, \citenamefont {Gill},\ and\ \citenamefont
  {{Head-Gordon}}}]{QCHEM4}%
  \BibitemOpen
  \bibfield  {author} {\bibinfo {author} {\bibfnamefont {Y.}~\bibnamefont
  {Shao}}, \bibinfo {author} {\bibfnamefont {Z.}~\bibnamefont {Gan}}, \bibinfo
  {author} {\bibfnamefont {E.}~\bibnamefont {Epifanovsky}}, \bibinfo {author}
  {\bibfnamefont {A.~T.~B.}\ \bibnamefont {Gilbert}}, \bibinfo {author}
  {\bibfnamefont {M.}~\bibnamefont {Wormit}}, \bibinfo {author} {\bibfnamefont
  {J.}~\bibnamefont {Kussmann}}, \bibinfo {author} {\bibfnamefont {A.~W.}\
  \bibnamefont {Lange}}, \bibinfo {author} {\bibfnamefont {A.}~\bibnamefont
  {Behn}}, \bibinfo {author} {\bibfnamefont {J.}~\bibnamefont {Deng}}, \bibinfo
  {author} {\bibfnamefont {X.}~\bibnamefont {Feng}}, \bibinfo {author}
  {\bibfnamefont {D.}~\bibnamefont {Ghosh}}, \bibinfo {author} {\bibfnamefont
  {M.}~\bibnamefont {Goldey}}, \bibinfo {author} {\bibfnamefont {P.~R.}\
  \bibnamefont {Horn}}, \bibinfo {author} {\bibfnamefont {L.~D.}\ \bibnamefont
  {Jacobson}}, \bibinfo {author} {\bibfnamefont {I.}~\bibnamefont {Kaliman}},
  \bibinfo {author} {\bibfnamefont {R.~Z.}\ \bibnamefont {Khaliullin}},
  \bibinfo {author} {\bibfnamefont {T.}~\bibnamefont {K\'us}}, \bibinfo
  {author} {\bibfnamefont {A.}~\bibnamefont {Landau}}, \bibinfo {author}
  {\bibfnamefont {J.}~\bibnamefont {Liu}}, \bibinfo {author} {\bibfnamefont
  {E.~I.}\ \bibnamefont {Proynov}}, \bibinfo {author} {\bibfnamefont {Y.~M.}\
  \bibnamefont {Rhee}}, \bibinfo {author} {\bibfnamefont {R.~M.}\ \bibnamefont
  {Richard}}, \bibinfo {author} {\bibfnamefont {M.~A.}\ \bibnamefont
  {Rohrdanz}}, \bibinfo {author} {\bibfnamefont {R.~P.}\ \bibnamefont
  {Steele}}, \bibinfo {author} {\bibfnamefont {E.~J.}\ \bibnamefont
  {Sundstrom}}, \bibinfo {author} {\bibfnamefont {H.~L.}\ \bibnamefont
  {{Woodcock III}}}, \bibinfo {author} {\bibfnamefont {P.~M.}\ \bibnamefont
  {Zimmerman}}, \bibinfo {author} {\bibfnamefont {D.}~\bibnamefont {Zuev}},
  \bibinfo {author} {\bibfnamefont {B.}~\bibnamefont {Albrecht}}, \bibinfo
  {author} {\bibfnamefont {E.}~\bibnamefont {Alguire}}, \bibinfo {author}
  {\bibfnamefont {B.}~\bibnamefont {Austin}}, \bibinfo {author} {\bibfnamefont
  {G.~J.~O.}\ \bibnamefont {Beran}}, \bibinfo {author} {\bibfnamefont {Y.~A.}\
  \bibnamefont {Bernard}}, \bibinfo {author} {\bibfnamefont {E.}~\bibnamefont
  {Berquist}}, \bibinfo {author} {\bibfnamefont {K.}~\bibnamefont
  {Brandhorst}}, \bibinfo {author} {\bibfnamefont {K.~B.}\ \bibnamefont
  {Bravaya}}, \bibinfo {author} {\bibfnamefont {S.~T.}\ \bibnamefont {Brown}},
  \bibinfo {author} {\bibfnamefont {D.}~\bibnamefont {Casanova}}, \bibinfo
  {author} {\bibfnamefont {C.-M.}\ \bibnamefont {Chang}}, \bibinfo {author}
  {\bibfnamefont {Y.}~\bibnamefont {Chen}}, \bibinfo {author} {\bibfnamefont
  {S.~H.}\ \bibnamefont {Chien}}, \bibinfo {author} {\bibfnamefont {K.~D.}\
  \bibnamefont {Closser}}, \bibinfo {author} {\bibfnamefont {D.~L.}\
  \bibnamefont {Crittenden}}, \bibinfo {author} {\bibfnamefont
  {M.}~\bibnamefont {Diedenhofen}}, \bibinfo {author} {\bibfnamefont {R.~A.}\
  \bibnamefont {{DiStasio Jr.}}}, \bibinfo {author} {\bibfnamefont
  {H.}~\bibnamefont {Dop}}, \bibinfo {author} {\bibfnamefont {A.~D.}\
  \bibnamefont {Dutoi}}, \bibinfo {author} {\bibfnamefont {R.~G.}\ \bibnamefont
  {Edgar}}, \bibinfo {author} {\bibfnamefont {S.}~\bibnamefont {Fatehi}},
  \bibinfo {author} {\bibfnamefont {L.}~\bibnamefont {{Fusti-Molnar}}},
  \bibinfo {author} {\bibfnamefont {A.}~\bibnamefont {Ghysels}}, \bibinfo
  {author} {\bibfnamefont {A.}~\bibnamefont {{Golubeva-Zadorozhnaya}}},
  \bibinfo {author} {\bibfnamefont {J.}~\bibnamefont {Gomes}}, \bibinfo
  {author} {\bibfnamefont {M.~W.~D.}\ \bibnamefont {{Hanson-Heine}}}, \bibinfo
  {author} {\bibfnamefont {P.~H.~P.}\ \bibnamefont {Harbach}}, \bibinfo
  {author} {\bibfnamefont {A.~W.}\ \bibnamefont {Hauser}}, \bibinfo {author}
  {\bibfnamefont {E.~G.}\ \bibnamefont {Hohenstein}}, \bibinfo {author}
  {\bibfnamefont {Z.~C.}\ \bibnamefont {Holden}}, \bibinfo {author}
  {\bibfnamefont {T.-C.}\ \bibnamefont {Jagau}}, \bibinfo {author}
  {\bibfnamefont {H.}~\bibnamefont {Ji}}, \bibinfo {author} {\bibfnamefont
  {B.}~\bibnamefont {Kaduk}}, \bibinfo {author} {\bibfnamefont
  {K.}~\bibnamefont {Khistyaev}}, \bibinfo {author} {\bibfnamefont
  {J.}~\bibnamefont {Kim}}, \bibinfo {author} {\bibfnamefont {J.}~\bibnamefont
  {Kim}}, \bibinfo {author} {\bibfnamefont {R.~A.}\ \bibnamefont {King}},
  \bibinfo {author} {\bibfnamefont {P.}~\bibnamefont {Klunzinger}}, \bibinfo
  {author} {\bibfnamefont {D.}~\bibnamefont {Kosenkov}}, \bibinfo {author}
  {\bibfnamefont {T.}~\bibnamefont {Kowalczyk}}, \bibinfo {author}
  {\bibfnamefont {C.~M.}\ \bibnamefont {Krauter}}, \bibinfo {author}
  {\bibfnamefont {K.~U.}\ \bibnamefont {Lao}}, \bibinfo {author} {\bibfnamefont
  {A.}~\bibnamefont {Laurent}}, \bibinfo {author} {\bibfnamefont {K.~V.}\
  \bibnamefont {Lawler}}, \bibinfo {author} {\bibfnamefont {S.~V.}\
  \bibnamefont {Levchenko}}, \bibinfo {author} {\bibfnamefont {C.~Y.}\
  \bibnamefont {Lin}}, \bibinfo {author} {\bibfnamefont {F.}~\bibnamefont
  {Liu}}, \bibinfo {author} {\bibfnamefont {E.}~\bibnamefont {Livshits}},
  \bibinfo {author} {\bibfnamefont {R.~C.}\ \bibnamefont {Lochan}}, \bibinfo
  {author} {\bibfnamefont {A.}~\bibnamefont {Luenser}}, \bibinfo {author}
  {\bibfnamefont {P.}~\bibnamefont {Manohar}}, \bibinfo {author} {\bibfnamefont
  {S.~F.}\ \bibnamefont {Manzer}}, \bibinfo {author} {\bibfnamefont {S.-P.}\
  \bibnamefont {Mao}}, \bibinfo {author} {\bibfnamefont {N.}~\bibnamefont
  {Mardirossian}}, \bibinfo {author} {\bibfnamefont {A.~V.}\ \bibnamefont
  {Marenich}}, \bibinfo {author} {\bibfnamefont {S.~A.}\ \bibnamefont
  {Maurer}}, \bibinfo {author} {\bibfnamefont {N.~J.}\ \bibnamefont {Mayhall}},
  \bibinfo {author} {\bibfnamefont {C.~M.}\ \bibnamefont {Oana}}, \bibinfo
  {author} {\bibfnamefont {R.}~\bibnamefont {{Olivares-Amaya}}}, \bibinfo
  {author} {\bibfnamefont {D.~P.}\ \bibnamefont {O'Neill}}, \bibinfo {author}
  {\bibfnamefont {J.~A.}\ \bibnamefont {Parkhill}}, \bibinfo {author}
  {\bibfnamefont {T.~M.}\ \bibnamefont {Perrine}}, \bibinfo {author}
  {\bibfnamefont {R.}~\bibnamefont {Peverati}}, \bibinfo {author}
  {\bibfnamefont {P.~A.}\ \bibnamefont {Pieniazek}}, \bibinfo {author}
  {\bibfnamefont {A.}~\bibnamefont {Prociuk}}, \bibinfo {author} {\bibfnamefont
  {D.~R.}\ \bibnamefont {Rehn}}, \bibinfo {author} {\bibfnamefont
  {E.}~\bibnamefont {Rosta}}, \bibinfo {author} {\bibfnamefont {N.~J.}\
  \bibnamefont {Russ}}, \bibinfo {author} {\bibfnamefont {N.}~\bibnamefont
  {Sergueev}}, \bibinfo {author} {\bibfnamefont {S.~M.}\ \bibnamefont
  {Sharada}}, \bibinfo {author} {\bibfnamefont {S.}~\bibnamefont {Sharmaa}},
  \bibinfo {author} {\bibfnamefont {D.~W.}\ \bibnamefont {Small}}, \bibinfo
  {author} {\bibfnamefont {A.}~\bibnamefont {Sodt}}, \bibinfo {author}
  {\bibfnamefont {T.}~\bibnamefont {Stein}}, \bibinfo {author} {\bibfnamefont
  {D.}~\bibnamefont {St\"uck}}, \bibinfo {author} {\bibfnamefont {Y.-C.}\
  \bibnamefont {Su}}, \bibinfo {author} {\bibfnamefont {A.~J.~W.}\ \bibnamefont
  {Thom}}, \bibinfo {author} {\bibfnamefont {T.}~\bibnamefont {Tsuchimochi}},
  \bibinfo {author} {\bibfnamefont {L.}~\bibnamefont {Vogt}}, \bibinfo {author}
  {\bibfnamefont {O.}~\bibnamefont {Vydrov}}, \bibinfo {author} {\bibfnamefont
  {T.}~\bibnamefont {Wang}}, \bibinfo {author} {\bibfnamefont {M.~A.}\
  \bibnamefont {Watson}}, \bibinfo {author} {\bibfnamefont {J.}~\bibnamefont
  {Wenzel}}, \bibinfo {author} {\bibfnamefont {A.}~\bibnamefont {White}},
  \bibinfo {author} {\bibfnamefont {C.~F.}\ \bibnamefont {Williams}}, \bibinfo
  {author} {\bibfnamefont {V.}~\bibnamefont {Vanovschi}}, \bibinfo {author}
  {\bibfnamefont {S.}~\bibnamefont {Yeganeh}}, \bibinfo {author} {\bibfnamefont
  {S.~R.}\ \bibnamefont {Yost}}, \bibinfo {author} {\bibfnamefont {Z.-Q.}\
  \bibnamefont {You}}, \bibinfo {author} {\bibfnamefont {I.~Y.}\ \bibnamefont
  {Zhang}}, \bibinfo {author} {\bibfnamefont {X.}~\bibnamefont {Zhang}},
  \bibinfo {author} {\bibfnamefont {Y.}~\bibnamefont {Zhou}}, \bibinfo {author}
  {\bibfnamefont {B.~R.}\ \bibnamefont {Brooks}}, \bibinfo {author}
  {\bibfnamefont {G.~K.~L.}\ \bibnamefont {Chan}}, \bibinfo {author}
  {\bibfnamefont {D.~M.}\ \bibnamefont {Chipman}}, \bibinfo {author}
  {\bibfnamefont {C.~J.}\ \bibnamefont {Cramer}}, \bibinfo {author}
  {\bibfnamefont {W.~A.}\ \bibnamefont {{Goddard III}}}, \bibinfo {author}
  {\bibfnamefont {M.~S.}\ \bibnamefont {Gordon}}, \bibinfo {author}
  {\bibfnamefont {W.~J.}\ \bibnamefont {Hehre}}, \bibinfo {author}
  {\bibfnamefont {A.}~\bibnamefont {Klamt}}, \bibinfo {author} {\bibfnamefont
  {H.~F.}\ \bibnamefont {{Schaefer III}}}, \bibinfo {author} {\bibfnamefont
  {M.~W.}\ \bibnamefont {Schmidt}}, \bibinfo {author} {\bibfnamefont {C.~D.}\
  \bibnamefont {Sherrill}}, \bibinfo {author} {\bibfnamefont {D.~G.}\
  \bibnamefont {Truhlar}}, \bibinfo {author} {\bibfnamefont {A.}~\bibnamefont
  {Warshel}}, \bibinfo {author} {\bibfnamefont {X.}~\bibnamefont {Xua}},
  \bibinfo {author} {\bibfnamefont {A.}~\bibnamefont {{Aspuru-Guzik}}},
  \bibinfo {author} {\bibfnamefont {R.}~\bibnamefont {Baer}}, \bibinfo {author}
  {\bibfnamefont {A.~T.}\ \bibnamefont {Bell}}, \bibinfo {author}
  {\bibfnamefont {N.~A.}\ \bibnamefont {Besley}}, \bibinfo {author}
  {\bibfnamefont {J.-D.}\ \bibnamefont {Chai}}, \bibinfo {author}
  {\bibfnamefont {A.}~\bibnamefont {Dreuw}}, \bibinfo {author} {\bibfnamefont
  {B.~D.}\ \bibnamefont {Dunietz}}, \bibinfo {author} {\bibfnamefont {T.~R.}\
  \bibnamefont {Furlani}}, \bibinfo {author} {\bibfnamefont {S.~R.}\
  \bibnamefont {Gwaltney}}, \bibinfo {author} {\bibfnamefont {C.-P.}\
  \bibnamefont {Hsu}}, \bibinfo {author} {\bibfnamefont {Y.}~\bibnamefont
  {Jung}}, \bibinfo {author} {\bibfnamefont {J.}~\bibnamefont {Kong}}, \bibinfo
  {author} {\bibfnamefont {D.~S.}\ \bibnamefont {Lambrecht}}, \bibinfo {author}
  {\bibfnamefont {W.}~\bibnamefont {Liang}}, \bibinfo {author} {\bibfnamefont
  {C.}~\bibnamefont {Ochsenfeld}}, \bibinfo {author} {\bibfnamefont {V.~A.}\
  \bibnamefont {Rassolov}}, \bibinfo {author} {\bibfnamefont {L.~V.}\
  \bibnamefont {Slipchenko}}, \bibinfo {author} {\bibfnamefont {J.~E.}\
  \bibnamefont {Subotnik}}, \bibinfo {author} {\bibfnamefont {T.}~\bibnamefont
  {{Van Voorhis}}}, \bibinfo {author} {\bibfnamefont {J.~M.}\ \bibnamefont
  {Herbert}}, \bibinfo {author} {\bibfnamefont {A.~I.}\ \bibnamefont {Krylov}},
  \bibinfo {author} {\bibfnamefont {P.~M.~W.}\ \bibnamefont {Gill}}, \ and\
  \bibinfo {author} {\bibfnamefont {M.}~\bibnamefont {{Head-Gordon}}},\
  }\href@noop {} {\bibfield  {journal} {\bibinfo  {journal} {Mol.\ Phys.}\
  }\textbf {\bibinfo {volume} {113}},\ \bibinfo {pages} {184} (\bibinfo {year}
  {2015})}\BibitemShut {NoStop}%
\bibitem [{\citenamefont {Johnson~III}(2015)}]{johnson2015nist}%
  \BibitemOpen
  \bibfield  {author} {\bibinfo {author} {\bibfnamefont {R.~D.}\ \bibnamefont
  {Johnson~III}},\ }\href@noop {} {\bibfield  {journal} {\bibinfo  {journal}
  {http://cccbdb.nist.gov/}\ } (\bibinfo {year} {2015})}\BibitemShut {NoStop}%
\bibitem [{\citenamefont {Tamm}(1991)}]{tamm1991relativistic}%
  \BibitemOpen
  \bibfield  {author} {\bibinfo {author} {\bibfnamefont {I.}~\bibnamefont
  {Tamm}},\ }in\ \href@noop {} {\emph {\bibinfo {booktitle} {Selected
  Papers}}}\ (\bibinfo  {publisher} {Springer},\ \bibinfo {year} {1991})\ pp.\
  \bibinfo {pages} {157--174}\BibitemShut {NoStop}%
\bibitem [{\citenamefont {Dancoff}(1950)}]{dancoff1950non}%
  \BibitemOpen
  \bibfield  {author} {\bibinfo {author} {\bibfnamefont {S.}~\bibnamefont
  {Dancoff}},\ }\href@noop {} {\bibfield  {journal} {\bibinfo  {journal} {Phys.
  Rev.}\ }\textbf {\bibinfo {volume} {78}},\ \bibinfo {pages} {382} (\bibinfo
  {year} {1950})}\BibitemShut {NoStop}%
\bibitem [{\citenamefont {Hirata}\ and\ \citenamefont
  {Head-Gordon}(1999)}]{hirata1999time}%
  \BibitemOpen
  \bibfield  {author} {\bibinfo {author} {\bibfnamefont {S.}~\bibnamefont
  {Hirata}}\ and\ \bibinfo {author} {\bibfnamefont {M.}~\bibnamefont
  {Head-Gordon}},\ }\href@noop {} {\bibfield  {journal} {\bibinfo  {journal}
  {Chem. Phys. Lett.}\ }\textbf {\bibinfo {volume} {314}},\ \bibinfo {pages}
  {291} (\bibinfo {year} {1999})}\BibitemShut {NoStop}%
\end{thebibliography}%
\end{document}